\documentclass[twocolumn,apj,numberedappendix,appendixfloats,twocolappendix]{openjournal}

\usepackage{newtxtext,newtxmath,amsmath}
\usepackage[T1]{fontenc}
\usepackage{graphicx}
\usepackage{enumitem}
\usepackage{xcolor}
\usepackage{hyperref}
\usepackage{gensymb}
\usepackage{CJK}
\usepackage{tabularx}
\usepackage{orcidlink}
\hypersetup{
    unicode, 
    colorlinks=true,
    linkcolor=linkcolor,
    citecolor=linkcolor,
    filecolor=linkcolor,
    urlcolor=linkcolor,
}
\definecolor{linkcolor}{rgb}{0.0,0.3,0.5}

\newcommand{\adjusted}[1]{#1}
\newcolumntype{C}{>{\centering\arraybackslash}X}

\newcommand{\nihaoRmax}{$20\,\mathrm{kpc}$}

\newcommand{\nihaoAGEmax}{$0.5\,\mathrm{Gyr}$}

\begin{document}
\begin{CJK*}{UTF8}{gbsn}

%%%%%%%%%%%%%%%%%%%%%%%%%%%%%%%%%%%%%%%%%%%%%%%%%%
%%%%%%%%%%%%%%%%%% TITLE PAGE %%%%%%%%%%%%%%%%%%%%

\title{Local variations of the radial metallicity gradient in a simulated NIHAO-UHD Milky Way analogue and their implications for (extra-)galactic studies}

\author{\vspace{-1.6cm}Sven Buder$^{1,2,*}$\orcidlink{0000-0002-4031-8553}}
\author{Tobias Buck$^{3,4}$\orcidlink{0000-0003-2027-399X}}
\author{Qian-Hui Chen (陈千惠)$^{1,2}$\orcidlink{0000-0002-4382-1090}}
\author{Kathryn Grasha$^{1,2,*}$\orcidlink{0000-0002-3247-5321}}

\affiliation{$^{1}$Research School of Astronomy and Astrophysics, Australian National University, Canberra, ACT 2611, Australia}
\affiliation{$^{2}$ARC Centre of Excellence for All Sky Astrophysics in 3 Dimensions (ASTRO 3D), Australia}
\affiliation{$^{3}$Universit{\"a}t Heidelberg, Interdisziplin{\"a}res Zentrum f{\"u}r Wissenschaftliches Rechnen, Im Neuenheimer Feld 205, D-69120 Heidelberg, Germany}
\affiliation{$^{4}$Universit{\"a}t Heidelberg, Zentrum f{\"u}r Astronomie, Institut f{\"u}r Theoretische Astrophysik, Albert-Ueberle-Straße 2, D-69120 Heidelberg, Germany}

\thanks{E-mail: \href{mailto:sven.buder@anu.edu.au}{sven.buder@anu.edu.au}}
\thanks{$^*$Australian Research Council DECRA Fellow}

\begin{abstract}
Radial metallicity gradients are fundamental to understanding galaxy formation and evolution.
In our high-resolution simulation of a NIHAO-UHD Milky Way analogue, we analyze the linearity, scatter, spatial coherence, and age-related variations of metallicity gradients using young stars and gas.
While a global linear model generally captures the gradient, it ever so slightly overestimates metallicity in the inner galaxy and underestimates it in the outer regions of our simulated galaxy. Both a quadratic model, showing an initially steeper gradient that smoothly flattens outward, and a piecewise linear model with a break radius \adjusted{around 9.3-11.5~kpc (2.4-3.0} effective radii) fit the data equally better. The spread of [Fe/H] of young stars in the simulation increases by tenfold from the innermost to the outer galaxy at a radius of 20~kpc. We find that stars born at similar times along radial spirals drive this spread in the outer galaxy, with a chemical under- and over-enhancement of up to 0.1 dex at leading and trailing regions of such spirals, respectively.
This localised chemical variance highlights the need to examine radial and azimuthal selection effects for both Galactic and extragalactic observational studies. The arguably idealised but volume-complete simulations suggest that future studies should not only test linear and piecewise linear gradients, but also non-linear functions such as quadratic ones to test for a smooth gradient rather than one with a break radius. Either finding would help to determine the importance of different enrichment or mixing pathways and thus our understanding of galaxy formation and evolution scenarios.
\end{abstract}


\maketitle

%%%%%%%%%%%%%%%%%%%%%%%%%%%%%%%%%%%%%%%%%%%%%%%%%%
%%%%%%%%%%%%%%%%% BODY OF PAPER %%%%%%%%%%%%%%%%%%

%%%%%%%%%%%%%%%%%%%%%%%%%%%%%%%%%%%%%%%%%%%%%%%%%%
%%%%%%%%%%%%%%%%%%%%%%%%%%%%%%%%%%%%%%%%%%%%%%%%%%
\section{Introduction}
\label{sec:intro}
%%%%%%%%%%%%%%%%%%%%%%%%%%%%%%%%%%%%%%%%%%%%%%%%%%
%%%%%%%%%%%%%%%%%%%%%%%%%%%%%%%%%%%%%%%%%%%%%%%%%%

Understanding the radial metallicity gradient, defined as the change in heavy element abundance with galactocentric radius, in galaxies provides critical insights into their formation and evolutionary processes, such as inside-out formation, gas accretion, outflows, and radial migration \citep[e.g.][]{Quirk1973, Tinsley1980, Lacey1985, Wyse1989, Kauffman1996, Chiappini1997, Schoenrich2009b, Moran2012, Bird2013}. The decrease in metallicity with increasing distance from the Galactic centre is well-established both theoretically \citep{Larson1976, Tinsley1980, Chiosi1980} and observationally in the Milky Way \citep{Searle1971, Janes1979, Twarog1997} and other massive spiral galaxies \citep[e.g.][]{Tinsley1980, Zaritsky1994,Bresolin2012}. The Milky Way, being the only galaxy where we can resolve millions of stars, provides a unique opportunity to study these gradients and deviations from them in detail. Early evidence by \citet{Janes1979} suggested a linear gradient on the order of $\mathrm{d{[Fe/H]}} / \mathrm{d}R = -0.05 \pm 0.01\,\mathrm{dex\,kpc^{-1}}$ for the Milky Way which aligns very well with more recent measurements \citep{Anders2017, Hayden2015}. However, these gradients are accompanied by a significant spread in [Fe/H] of $0.1-0.15\,\mathrm{dex}$, as noted by \citet{Twarog1980}, which may imply a fine structure of the metallicity gradient \citep[see][]{Genovali2014}.

With increasing sample size and measurement precision, the specific shape and characteristics of this gradient remain somewhat unclear \citep{Chiappini2002}. Previous studies have for example claimed more intricate non-linear trends, bends, and flattening in the gradient of the Milky Way \citep[e.g.][]{Donor2020} and other galaxies \citep[e.g.][]{Pilyugin2003, Sanchez2014} or even sequences of shapes \citep{Pilyugin2017, Pilyugin2024}, which were fitted with different models \citep{RosalesOrtega2011, Bresolin2012}, such as piecewise linear ones \citep[e.g.][]{SanchezMenguiano2016} or non-linear ones \citep[e.g.][]{Scarano2013}.

Variations in the metal distribution, including breaks of the gradient at specific radii, give rise to a plethora of possible physical explanations, such as star formation efficiency variations and localised star formation bursts \citep{Sanchez2014, SanchezBlazquez2014, Ho2015}, gas accretion and dilution at different rates \citep{Bresolin2012, Sanchez2013, Belfiore2016, SanchezMenguiano2016}, gas outflows and feedback \citep{Lilly2013, Ma2017b}, as well as disk instabilities or local overdensities \citep{Grand2016, Ho2017c}. In particular, \citet{Scarano2013} suggested that gradient break radii coincided with the corotation radii of spiral arms.

Recent advancements in both computations and observations have significantly expanded our capabilities. For example, in terms of observational data in the Milky Way, the \textit{Gaia} mission \citep{Gaia-Collaboration2016} enables more detailed studies of these gradients. New suites of large-scale simulations now allow us to gain insights into radial metallicity gradients across a range of simulated galaxies, including Milky Way analogues. This presents opportunities to revisit outstanding challenges of the detailed shape of the radial metallicity gradient. For instance, \citet{Hogg2019} created an extensive metallicity map of the Milky Way using APOGEE and \textit{Gaia} data, while \citet{Poggio2022} mapped young stars and found metallicity variations around spiral arms \citep[see also][]{Zari2018, Zari2021, Poggio2021, Hackshaw2024}. Similarly, \citet[][among others]{Imig2023} traced gradients across different stellar populations and ages, emphasizing the importance of considering radial migration effects \citep{Binney2008, Frankel2018, Frankel2020}.

Historically, radial metallicity gradients have been measured using various stellar populations and gas tracers. Estimated gradients seem to be broadly consistent across different stellar tracers, such as young open clusters \citep[e.g.][]{Yong2012, Cunha2016, Magrini2017, Casamiquela2019, Donor2020, Spina2021,Myers2022}, young hot (OB-type) stars \citep{Zari2018, Zari2021, Poggio2021, Poggio2022}, field stars close to the Galactic plane \citep[e.g.][]{Bergemann2014} or Cepheids \citep{Andrievsky2002, Andrievsky2002b, Lemasle2007, Lemasle2013}.

Despite extensive observational efforts, several challenges persist for studies in the Milky Way. The completeness (or patchiness) of observed datasets remains a fundamental issue \citep{Bergemann2014}. The robustness of fits to the incomplete data is still contentious, including the need to actually fit two linear gradients with a break radius at corotation radius \citep[][and references therein]{Bresolin2012} or further out \citep{Yong2012, Donor2020} - or even more complicated functions \citep[see e.g.][]{Chiappini2001, Kubryk2015}. Furthermore, methodologies for fitting linear models to scattered data need re-evaluation \citep{Metha2021}. Different samples yield varying gradients, potentially due to biases in data or the inclusion of older stars \citep[e.g.][]{AllendePrieto2006, Hayden2014, Anders2014, Vickers2021, Willett2023}. The impact of spiral arm structures \citep{Poggio2021}, the Galactic warp \citep{Lemasle2022} or bar-driven mixing \citep{DiMatteo2013} on metallicity gradients is not fully understood.

Understanding these gradients in the Milky Way is also crucial for extragalactic studies, where spatial resolution limits observations in different ways. In extragalactic systems, metals are mainly traced via gas, because it provides a more direct measure of the ongoing enrichment processes, unlike stars, which primarily reflect the integrated chemical history of the past. Observationally, gas emission lines are typically brighter and more accessible across large distances than stellar absorption lines, allowing for broader spatial coverage, especially in distant galaxies. Consequently, extragalactic studies often focus on gas-phase metallicity as traced by oxygen, $\mathrm{A(O)} = 12 + \log(\mathrm{O/H})$, while Galactic studies typically use stellar iron abundance $\mathrm{[Fe/H]} = \mathrm{A(Fe)} - \mathrm{A(Fe)}_\odot$ as a metallicity tracer \citep[e.g.][]{Nicholls2017, FraserMcKelvie2022}.

New instruments like the MUSE integral field spectrograph have enabled a plethora of extragalactic studies to contrast the Milky Way and techniques like the spectroscopy of H~{\sc{ii}} regions and planetary nebulae have helped to infer gas metallicity distributions in external galaxies \citep{Shaver1983, Vilchez1996, Rolleston2000, Bresolin2012}. Recent examples include \citet{Sanchez2014} with CALIFA galaxy observations as well as \citet{Mun2024} and \citet{Chen2024} who use MAGPI observations to probe for example the effects of spiral arms. Notable is also the scatter that \citet{Chen2023} found for the gas metallicity across galactic radii with TYPHOON observations (see their. Figs. 4-6). \citet{Grasha2022} found that the gas metallicity gradient plateaus at a lower limit in their TYPHOON galaxies at the outermost radii - an observation replicated by IllustrisTNG simulations \citep{Hemler2021, Garcia2023}.

From a modelling perspective, galactic chemical evolution models can both test understanding of radial metallicity gradients and make predictions beyond the limited volumes and tracers tested by Milky Way and extragalactic studies. Such galactic chemical evolution models include \citet{Chiappini2001, Matteucci2001, Minchev2014b, Kubryk2015, Stanghellini2015, Rybizki2017, Spitoni2023, Johnson2024}. \citet{Sharda2021} even presented a model for gas phase metallicity gradients in galaxies and their evolution from first principles \citep[see also][]{Krumholz2018b}. 

In exploring radial metallicity gradients through simulations, we have better understood how different processes influence these gradients across galactic models and temporal scales. Studies such as \citet{Pilkington2012} in RaDES simulations reveal that gradients are typically established via inside-out galaxy formation. \citet{Khoperskov2023e} quantified the scatter of gas metallicity to $\approx 0.04-0.06\,\mathrm{dex}$ at a given galactocentric distance in their simulations. Meanwhile, the EAGLE simulations used by \citet{Tissera2019} provide insights into how these gradients vary with galaxy characteristics like stellar mass and merger history, emphasizing the dynamic nature of metallicity distributions. The plethora of simulations such as AURIGA \citep{Grand2016}, FIRE \citep[][see their Fig. 6]{Ma2017} or VINTERGATAN \citep[see their Fig. 9;][]{Agertz2021} also allow us to explore the gradient evolution of galactic timescales. \citet{Buck2023}, for example, found a link of major accretion events with periods of unexpected steepening in the metallicity gradient within NIHAO-UHD simulations -- closely resembling findings for the Milky Way by \citet{Lu2022} and \citet{Ratcliffe2023}. FIRE simulations, examined by \citet{Bellardini2021}, \citet{Bellardini2022}, \citet{Orr2023}, and \citet{Graf2025}, extend these findings by comparing radial metallicity gradients and their azimuthal scatter across gas and stellar components and illustrate the complex interplay between galactic structure and metal enrichment processes. Similarly, \citet{Grand2015,Grand2016} highlight temporal changes in metallicity gradients already within $120\,\mathrm{Myr}$, or roughly one galactic rotation. Such rapid changes underscore the impact of transient galactic events on the metal distribution, linking them to star formation patterns along spiral arms and the broader evolutionary history of the galaxy.

In this study, we analyze a high-resolution NIHAO-UHD simulation of a Milky Way analogue to bridge the observational gap between detailed studies of our galaxy and broader extragalactic surveys. We aim to reveal subtle features of the radial metallicity gradient, which may be obscured by observational constraints in both the Milky Way and distant galaxies, by testing the following properties within the observationally probed inner $R_\mathrm{Gal.} \leq 20\,\mathrm{kpc}$:
\begin{enumerate}
\item \textit{Linearity of the gradient:} Assess the extent to which the radial metallicity gradient of young stars is linear.
\item \textit{Scatter in the gradient:} Quantify the expected scatter in the radial metallicity gradient of young stars.
\item \textit{Coherence of the gradient with position:} Investigate the gradient's variation with radial coverage and azimuth.
\item \textit{Coherence of the gradient with age:} Test the reliability of stars as tracers of the gas disk over different ages.
\end{enumerate}

\begin{figure*}
    \centering
    \includegraphics[width=0.9\textwidth]{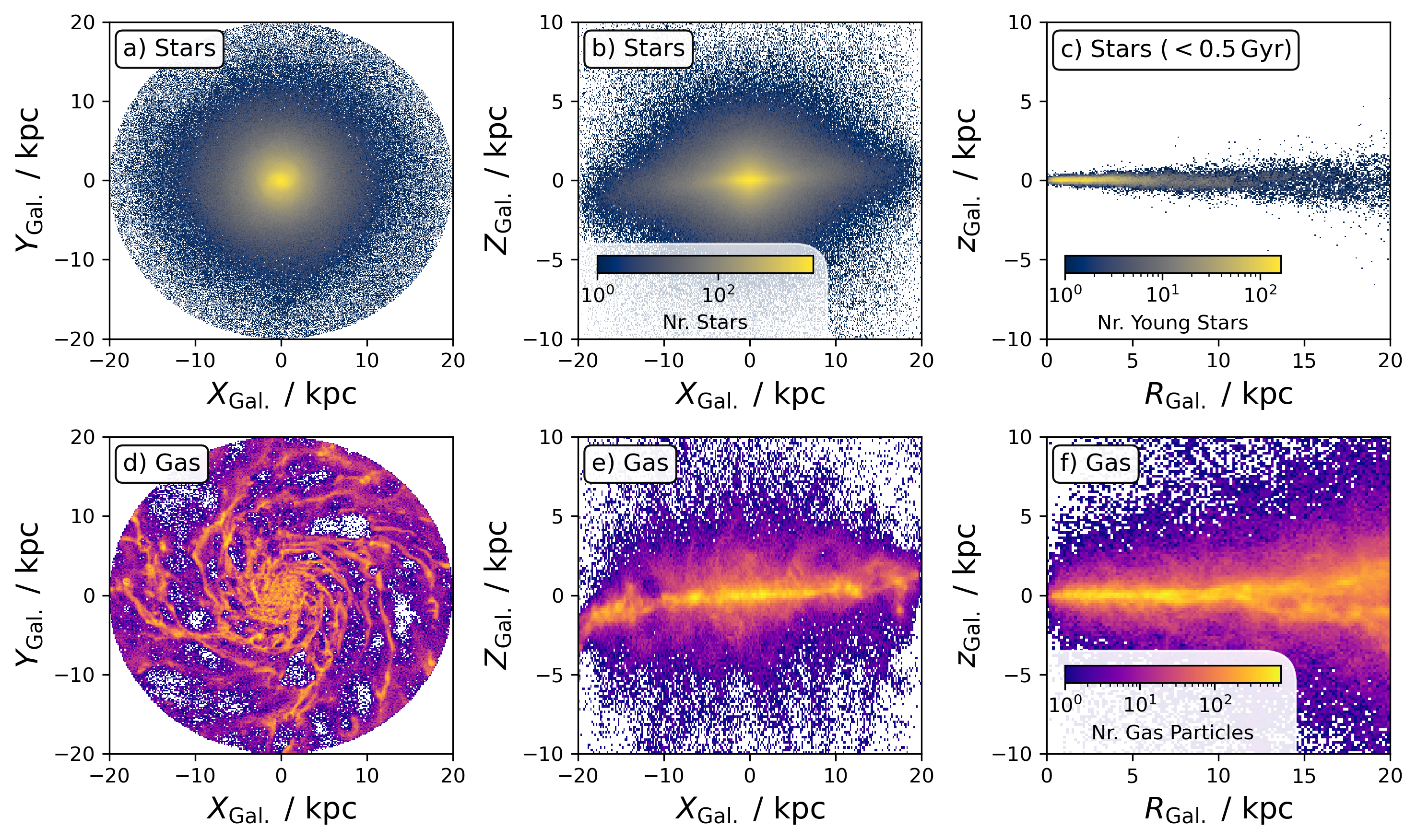}
    \caption{Logarithmic spatial density distribution of stars (upper panels) and gas (lower panels) within $R < 20\,\mathrm{kpc} \sim 5\,\mathrm{R_e}$ of the NIHAO-UHD Milky Way analogue \texttt{g8.26e11} in galactocentric cartesian and cylindrical coordinates. Panel c) shows the influence of selecting only young stars with ages below \nihaoAGEmax.}
    \label{fig:stars_and_gas_overview}
\end{figure*}

\begin{figure*}
    \centering
    \includegraphics[width=0.9\textwidth]{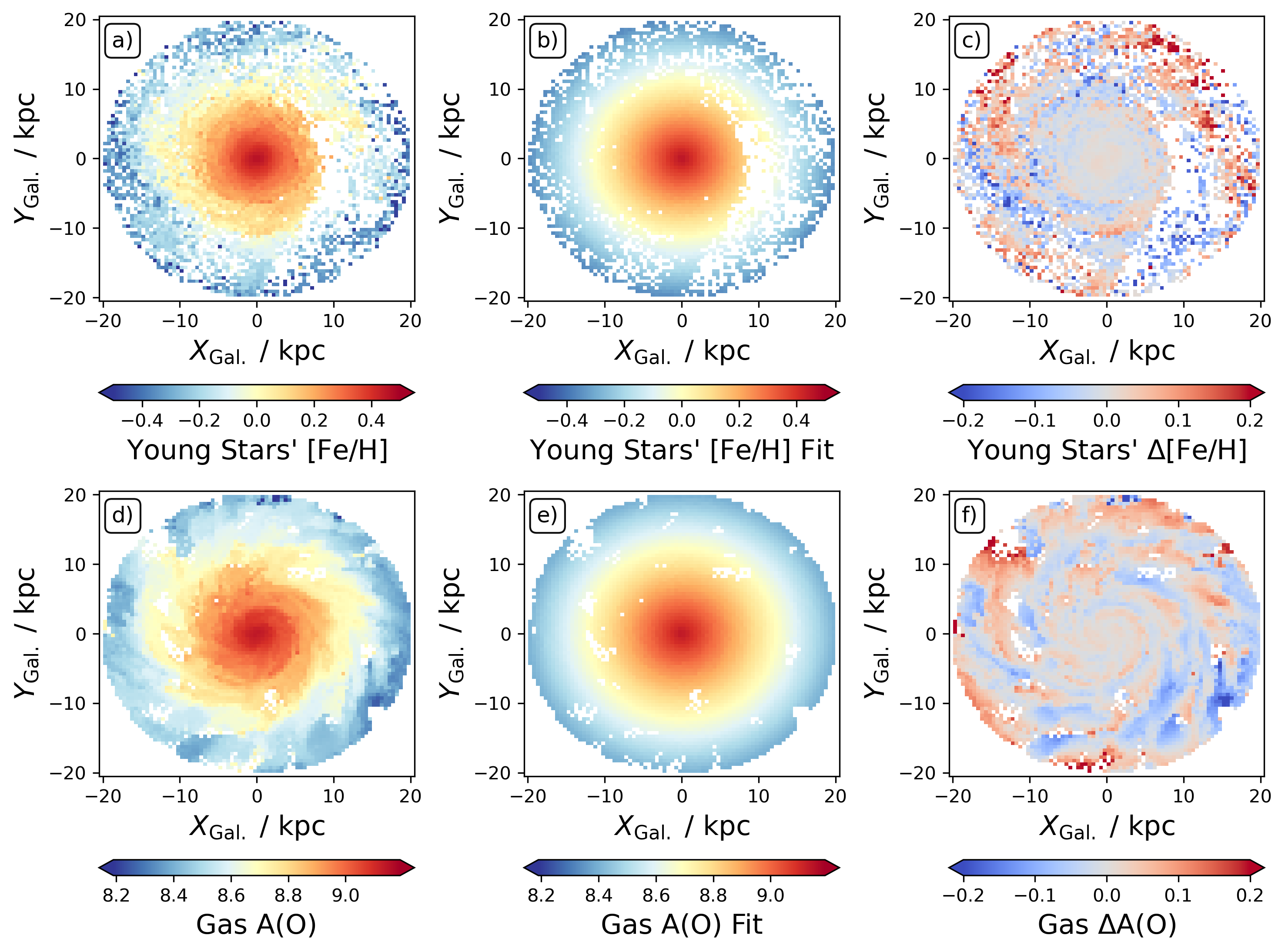}
    \caption{Face-on view of average simulated metallicity (left panels), a linear radial fit to it (middle panels, see Section~\ref{sec:global_fits} and Eq.~\ref{eq:function_linear}) and the fit residuals (right panels) in bins of $0.5\,\mathrm{kpc}$. Shown are metallicity as traced by young star iron abundances (top panels) and gas phase metallicity (bottom) panels.}
    \label{fig:stars_and_gas_2d_view}
\end{figure*}

The paper is structured as follows: Section~\ref{sec:data} describes the data of our Milky Way analogue NIHAO-UHD simulation. Section~\ref{sec:linear_radial_metallicity_gradients} analyses the linearity of the radial metallicity gradient of the simulation, the first of our four objectives. Section~\ref{sec:scatter_radial_metallicity_gradients} then analyses both the scatter and local deviations from the gradient as well as the coherence of the gradient with vertical and azimuthal position as well as age (the remaining three objectives). Section~\ref{sec:discussion} discusses them individually. We note that in this research we are mainly interested in the specific shape of the radial metallicity gradient for radii relevant to Galactic observations. However, we also discuss our results in the context of extragalactic results that probe beyond the inner $20\,\mathrm{kpc}$ of a galaxy. Section~\ref{sec:conc} bundles our results into an overarching conclusion. 

%%%%%%%%%%%%%%%%%%%%%%%%%%%%%%%%%%%%%%%%%%%%%%%%%%
%%%%%%%%%%%%%%%%%%%%%%%%%%%%%%%%%%%%%%%%%%%%%%%%%%
\section{Data: A NIHAO-UHD Milky Way analogue simulation} \label{sec:data}
%%%%%%%%%%%%%%%%%%%%%%%%%%%%%%%%%%%%%%%%%%%%%%%%%%
%%%%%%%%%%%%%%%%%%%%%%%%%%%%%%%%%%%%%%%%%%%%%%%%%%

For this project, we use a cosmological zoom-in simulation of a Milky Way analogue (\texttt{g8.26e11}) from the \textit{Numerical Investigation of a Hundred Astronomical Objects} \citep[NIHAO,][]{Wang2015} suite. This model galaxy was calculated as part of the NIHAO-UHD project \citep{Buck2020b} and has previously been used in various works studying Milky Way satellites \citep{Buck2019b}, Milky Way's dark halo spin \citep{Obreja2022}, inferring birth properties of stars with abundance clustering \citep{Ratcliffe2022}, as well as the evolution of the interstellar medium's radial metallicity gradient since redshift three \citep{Ratcliffe2024b}.

Simulations were carried out with the smoothed particle hydrodynamics code \texttt{Gasoline2} \citep{Wadsley2017}, including sub-grid turbulent mixing, using cosmological parameters from \citet{Planck2014} with initial conditions and energetic feedback descriptions from the NIHAO project \citep{Wang2015}. Zoom-in simulations were then performed as described in detail by \citet{Buck2021} with star formation following \citet{Stinson2006} and energetic feedback following \citet{Stinson2013}. We note that this is a slightly different rerun of the same simulation than the one studied by \citet{Buder2024}; in this work, we use a higher resolution version and updated chemical yields.

Because computational resources still limit the mass resolution of simulations, we are relying on tracer particles that represent simple stellar populations (SSPs) with the same age, overall metallicity and discrete initial mass function (IMF). \citet{Buck2021} have implemented the flexible chemical evolution code \textsc{chempy} \citep{Rybizki2017} to calculate the chemical yields for the SSPs. In particular, we use the alternative (\texttt{alt}) setup of \textsc{chempy} that assumes a \citet{Chabrier2003} IMF with high-mass slope of $\alpha_\text{IMF} = -2.3$ over a mass range of $0.1-100\,\mathrm{M_\odot}$ for SSPs across a metallicity range of $Z/Z_\odot \in [10^{-5},2]$. The code calculates the contribution from asymptotic giant branch (AGB) stars, CCSN across a mass range of $8-40\,\mathrm{M_\odot}$, and SNIa with an exponential function with exponent $-1.12$, a delay time of $40\,\mathrm{Myr}$, and a normalization of the SNIa rate of $\log_{10}(\mathrm{N_{Ia}}) = -2.9$. For each of these nucleosynthetic channels, yields from the following studies are used: \citet{Chieffi2004} for CCSN, \citet{Seitenzahl2013} for SNIa, and \citet{Karakas2016} for AGB stars \citep[\texttt{new\_fit} model in][]{Buck2021}. Contrary to a previous study by \citet{Buder2024}, we take the elemental abundances at face value and do not apply any shifts.

We limit the simulation data to the main halo by applying \textsc{pynbody}'s implementation of the Amiga Halo Finder \citep{Knollman2009} and then reposition and rotate this main halo to be face-on based on the angular momentum with \textsc{pynbody}'s \textsc{analysis.angmom.faceon} module \citep{pynbody}. We then further transform the resulting galactocentric Cartesian coordinate $(X,Y,Z)$ and velocities $(V_X,V_Y,V_Z)$ to Cylindrical ones as done in a previous study of this main halo by \citet{Buder2024}.

The model galaxy has a virial radius of $R_\mathrm{vir} = R_{200}=206\,\mathrm{kpc}$ and a total mass (gas, stars and dark matter) inside $R_{\rm vir}$ of $9.1 \cdot 10^{11}\,\mathrm{M_\odot}$. At redshift zero, it contains $8.2 \cdot 10^{11}\,\mathrm{M_\odot}$ dark matter, $6.4 \cdot 10^{10}\,\mathrm{M_\odot}$ gas mass and $2.3 \cdot 10^{10}\,\mathrm{M_\odot}$ stellar mass with a stellar mass resolution of around $7.5 \cdot 10^{3}\,\mathrm{M_\odot}$. When using a fifth of the virial radius as a reference to calculate total luminosity\footnote{Using \textsc{pb.analysis.luminosity.halo\_lum} \citep{pynbody}.} and mass, we estimate a half-light radius, that is, effective radius of $R_e = 3.79\,\mathrm{kpc}$ and a half-stellar-mass radius of $\mathrm{2.97\,\mathrm{kpc}}$. 

To achieve a roughly similar selection as the observational data of the Milky Way \citep{Genovali2014} and other galaxies \citep[e.g.][]{Chen2023}, we restrict the simulation data to a galactocentric radius of $R_\mathrm{Gal} \leq 20\,\mathrm{kpc}$ and $\vert z \vert \leq 10\,\mathrm{kpc}$, as shown in Fig.~\ref{fig:stars_and_gas_overview}. Similar to the Milky Way \citep{Poggio2018, Lemasle2022}, we note a warp of the stellar and gaseous disk (see Figs.~\ref{fig:stars_and_gas_overview}b and \ref{fig:stars_and_gas_overview}e, respectively) \adjusted{ leading to asymmetric deviations from the gas disk with a gas scale height of $h_{z\,\mathrm{gas}} = 1.25\,\mathrm{kpc}$}.

To avoid too strong effects of radial migration \citep{Binney2008, Frankel2018, Grand2016, Minchev2018} while maintaining a sufficiently large sample size we further enforce stars to be younger than \nihaoAGEmax, corresponding to roughly the time of four galactic rotations, and being half the value found by \citet{Minchev2018} for very limited migration in the Milky Way. This selection de-facto limits the vertical range of 99\% of stars to $\vert z \vert = 1.4\,\mathrm{kpc}$%
. The strong influence of this age cut on the vertical distribution of stars in the Milky Way analogue can best be appreciated from the difference of vertical density distributions of stars in Figs.~\ref{fig:stars_and_gas_overview}b and \ref{fig:stars_and_gas_overview}c. We are applying these cuts for all following analyses of the radial metallicity gradient in Section~\ref{sec:linear_radial_metallicity_gradients}.

%%%%%%%%%%%%%%%%%%%%%%%%%%%%%%%%%%%%%%%%%%%%%%%%%%
%%%%%%%%%%%%%%%%%%%%%%%%%%%%%%%%%%%%%%%%%%%%%%%%%%
\section{The linearity of the radial metallicity gradient in NIHAO-UHD}
\label{sec:linear_radial_metallicity_gradients}
%%%%%%%%%%%%%%%%%%%%%%%%%%%%%%%%%%%%%%%%%%%%%%%%%%
%%%%%%%%%%%%%%%%%%%%%%%%%%%%%%%%%%%%%%%%%%%%%%%%%%

In this section, we analyse the functional shape of the radial metallicity gradient. To get a first impression of possible shapes, we show the face-on view of the decreasing radial metallicity gradient of the simulation in Figs.~\ref{fig:stars_and_gas_2d_view}a (for young stars) and Fig.~\ref{fig:stars_and_gas_2d_view}d (for gas). Foreshadowing the later parts of this work on local variations, we also show a linear radial fit to either distribution in Fig.~\ref{fig:stars_and_gas_2d_view}b and \ref{fig:stars_and_gas_2d_view}e and show the fit residuals in Figs.~\ref{fig:stars_and_gas_2d_view}c and \ref{fig:stars_and_gas_2d_view}f.

At the moment, however, we focus on the linearity and thus the logarithmic density distribution of star particle iron abundances [Fe/H] across different galactocentric radii $R_\mathrm{Gal.}$. This distribution is shown in Fig.~\ref{fig:global_r_feh_fit}a and strongly suggests that the gradient is predominantly linear, similar to findings for the Milky Way. More complex shapes, such as piecewise linear ones have been suggested based on incomplete and limited data in the literature. We are thus also analysing these shapes with the complete and better-sampled data points of the NIHAO-UHD simulation. We firstly test different global fits in Section~\ref{sec:global_fits}, before testing the influence of binning and coverage in Sections~\ref{sec:binning} and \ref{sec:radial_coverage}, respectively.

\newpage
\subsection{Global gradient fits}
\label{sec:global_fits}

We fit three different functional forms to the global data: a linear function (used for Fig.~\ref{fig:stars_and_gas_2d_view}b)
\begin{equation}
f_{\text{lin}}(R_\mathrm{Gal.}) = c_1 \cdot R_\mathrm{Gal.} + c_2, \label{eq:function_linear}
\end{equation}
a piecewise linear with a break radius $R_\mathrm{break}$
\begin{equation}
f_{\text{piece}}(R_\mathrm{Gal.}) = 
\begin{cases} 
c_1 \cdot R_\mathrm{Gal.} + c_2 & \text{if } R_\mathrm{Gal.} \leq R_\mathrm{break} \\
c_3 \cdot R_\mathrm{Gal.} + c_4 & \text{if } R_\mathrm{Gal.} > R_\mathrm{break},  \label{eq:function_piecewise}
\end{cases}
\end{equation}
and a quadratic function
\begin{equation}
f_{\text{quad}}(R_\mathrm{Gal.}) = c_1 \cdot R_\mathrm{Gal.}^2 + c_2 \cdot R_\mathrm{Gal.} + c_3.  \label{eq:function_quadratic}
\end{equation}
The coefficients of the functions are fitted with the \textsc{scipy.optimize} function \textsc{curve\_fit} \citep{Scipy} and listed in Table~\ref{tab:global_fit_results_comparison}. To estimate the uncertainty of the break radius $R_\mathrm{break}$, we use the profile likelihood method to identify the radii at which the residual sum of squares (RSS) values are increased by $1\sigma$ from the best RSS radius in steps of $\Delta R_\mathrm{break} = 0.1\,\mathrm{kpc}$ and $0.5\,\mathrm{kpc}$ for the full and binned data set, respectively. We compute the RSS for each model $f_i$ (see Eqs.~\ref{eq:function_linear}-\ref{eq:function_quadratic}) based on the $N$ data points as 
\begin{equation} \label{eq:rss}
    \mathrm{RSS}_i = \sum_{n=1}^N \left( \mathrm{[Fe/H]}_n - f_i(R_{\mathrm{Gal.},n}) \right)^2.
\end{equation}

\begin{table}
\caption{Global linear gradient fit results with different methods. \textsc{LinearRegression} is part of the \textsc{sklearn} package.}
\label{tab:global_fit_results_per_method}
\begin{tabularx}{\columnwidth}{lCC}
\hline
Method & Intercept ($a_0 \pm \sigma_{a_0}$) & Slope ($a_1 \pm \sigma_{a_1}$) \\
\hline
\textsc{statsmodels.api.ODR} & $0.46266 \pm 0.00039$ & $-0.04109 \pm 0.00005$ \\
\textsc{scipy.odr} & $0.46268 \pm 0.00039$ & $-0.04109 \pm 0.00005$ \\
\textsc{np.polyfit} & $0.46266 \pm 0.00039$ & $-0.04109 \pm 0.00005$ \\
\textsc{LinearRegression} & $0.46286 \pm 0.00028$ & $-0.04114 \pm 0.00008$ \\
\textsc{scipy.curve\_fit} & $0.46266 \pm 0.00039$ & $-0.04109 \pm 0.00005$ \\
\hline
\end{tabularx}
\end{table}

We have confirmed the robustness of our fits by applying other fitting routines as outlined in Table~\ref{tab:global_fit_results_per_method}. After fitting three different functional forms, we use a combination of parameters to determine which model provides the best fit.

In Table~\ref{tab:global_fit_results_comparison}, the RSS value is the smallest (although only by a small margin) for the quadratic function. When assuming $\sigma^2 = RSS / N$, we can also define a logarithmic likelihood
% in statsmodels:
% https://www.statsmodels.org/dev/_modules/statsmodels/regression/linear_model.html#OLS.loglike
% nobs2 = self.nobs / 2.0
% nobs = float(self.nobs)
% resid = self.endog - np.dot(self.exog, params)
% ssr = np.sum(resid**2)
% llf = -nobs2*np.log(2*np.pi) - nobs2*np.log(ssr / nobs) - nobs2
\begin{equation}
    \ln L = - \frac{N}{2} \ln (2 \pi) - \frac{N}{2} \ln \frac{RSS}{N} - \frac{N}{2}
\end{equation}
for the $N$ data points. For $k$ free parameters, we then calculate the Akaike Information Criterion (AIC) and the Bayesian Information Criterion (BIC) as
\begin{equation} \label{eq:aic_bic}
    AIC = 2 k  - 2 \ln L \qquad BIC = k \ln N - 2 \ln L.
\end{equation}
For these criteria, the quadratic function performs slightly better than the linear or piecewise linear functions (see Table~\ref{tab:global_fit_results_comparison}).

\begin{table*}
\caption{
Fit Evaluation of linear, quadratic, and piecewise linear fits.
Extra parameters are quadratic term and break radius for the quadratic and piecewise fit.
RSS stands for Residual Sum of Squares (Eq.~\ref{eq:rss}).
AIC stands for Akaike Information Criterion and BIC stands for Bayesian Information Criterion (see Eq.~\ref{eq:aic_bic}).
}
\label{tab:global_fit_results_comparison}
\begin{tabularx}{\textwidth}{lCCCccc}
\hline
Function & Intercept ($a_0 \pm \sigma_{a_0}$) & Slope ($a_1 \pm \sigma_{a_1}$) & Extra Parameter & RSS & AIC & BIC \\
\hline
Linear & $0.46266 \pm 0.00039$  & $-0.04109 \pm 0.00005$ & -- & 87.0 & -118270  & -118260 \\ 
Quadratic & $0.48031 \pm 0.00055$  & $-0.04864 \pm 0.00018$ & $0.00045 \pm 0.00001$ & 82.7 & -120150  & -120120 \\ 
Piecewise & $-0.04477 \pm 0.00010$ & $0.47473 \pm 0.00047$ & -- & 82.9 & -120090  & -120050 \\ 
 & -- & $-0.03562 \pm 0.00014$ & $9.3 \pm 0.1$ & & & \\ 
\hline
Linear (bins)& $0.46983 \pm 0.00567$  & $-0.04146 \pm 0.00127$ & -- & 0.0167 & -190  & -190 \\ 
Quadratic (bins) & $0.47457 \pm 0.00674$  & $-0.04585 \pm 0.00359$ & $0.00031 \pm 0.00024$ & 0.0056 & -240  & -230 \\ 
Piecewise (bins) & $-0.04336 \pm 0.00188$ & $0.47274 \pm 0.00606$ & -- & 0.0052 & -240  & -230 \\ 
 & -- & $-0.03360 \pm 0.00589$ & $11.50 \pm 0.25$ & & & \\ 
\hline
\end{tabularx}
\end{table*}

\begin{figure}
    \centering
    \includegraphics[width=\columnwidth]{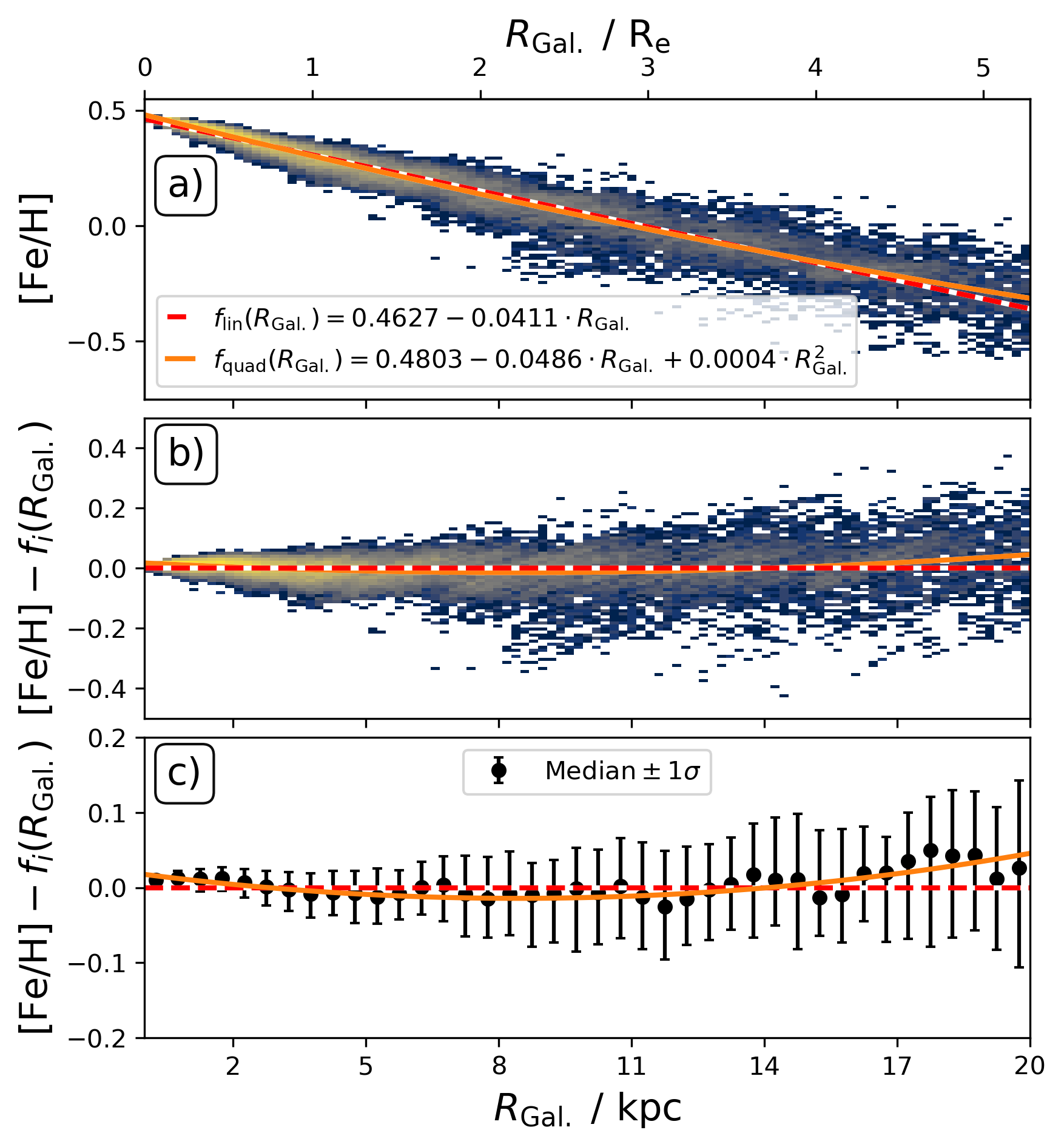}
    \caption{Global fits and deviation to the radial metallicity gradient $R-\mathrm{[Fe/H]}$. Functional forms of the linear (red) and quadratic (orange) lines are shown in the legend. Panel a) shows the underlying data of all data points as logarithmic density and the global fit to them as red dashed line. Panel b) shows the deviation of data from a linear gradient as a logarithmic density plot, whereas panel c) shows the 16th and 84th percentile around the median deviation as error bars in $\Delta R_\mathrm{Gal} = 0.5\,\mathrm{kpc}$ bins.}
    \label{fig:global_r_feh_fit}
\end{figure}

We show the fit residuals in Fig.~\ref{fig:global_r_feh_fit}b as density distribution as well as in Fig.~\ref{fig:global_r_feh_fit}c as percentile distributions in radial bins of $\Delta R_\mathrm{Gal} = 0.5\,\mathrm{kpc}$. While the density distribution shows substructure, which we investigate later in Section~\ref{sec:scatter_radial_metallicity_gradients}, we note an increase in the median residuals of the linear fit in Fig.~\ref{fig:global_r_feh_fit}c towards the inner and outer radii, especially for $R_\mathrm{Gal.} > 17\,\mathrm{kpc}$. A quadratic fit (see orange lines in Fig.~\ref{fig:global_r_feh_fit}) results in a slightly steeper linear component of the gradient (from $-0.0411$ to $-0.0497\,\mathrm{dex\,kpc^{-1}}$), which is counteracted by the quadratic flattening term of $+0.0005\,\mathrm{dex\,kpc^{-2}}$. The latter leads to an effective flattening of $-0.172 + 0.200 = 0.028\,\mathrm{dex}$ (linear vs. quadratic terms) \adjusted{at $R_\mathrm{Gal.} = 20\,\mathrm{kpc}$}. While seemingly only a nuisance correction across the large extent of [Fe/H] and $R_\mathrm{Gal.}$, this quadratic function outperforms the linear fit. This is most apparent in Fig.~\ref{fig:global_r_feh_fit}c, where the orange line better traces the median residuals from the linear function across all radii. This suggests that non-linear functions, such as piecewise linear or quadratic ones describe the gradient better.

\begin{figure}
    \centering
    \includegraphics[width=\columnwidth]{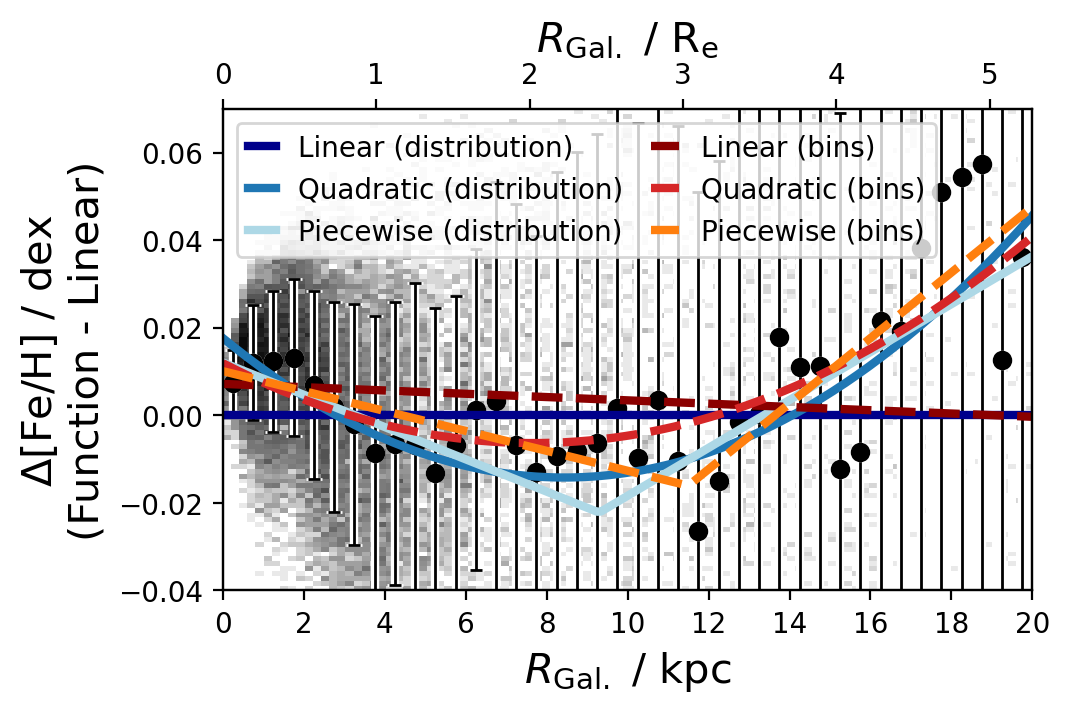}
    \caption{Deviation of different radial metallicity gradient functions from the global linear fit. Shown are the different functions (linear, quadratic, and piecewise linear) estimated from the full distribution (solid lines) or medians and standard deviations (error bars) in $\Delta R_\mathrm{Gal.} = 0.5\,\mathrm{kpc}$ bins (dashed lines).}
    \label{fig:linear_quadratic_piecewise}
\end{figure}

Distinguishing between the two latter functional forms is challenging. The quantitative performance indicators--RSS, AIC, and BIC--show very similar values for both forms, and a closer examination of the residuals in Fig.~\ref{fig:linear_quadratic_piecewise} reveals no clear visual advantage for either the piecewise linear or quadratic model.

\adjusted{When only fitting data up to $R_\mathrm{gal} < 10\,\mathrm{kpc}$, we note that the linear fit performs as well as the other forms, suggesting that the inner disk ($R_\mathrm{gal} < 10\,\mathrm{kpc}$) is indeed showing a linear trend, with non-linear behavior only being imprinted in the outer disk.}

\paragraph*{Take-Away:} Both piecewise linear and quadratic functions provide a better fit to the radial metallicity relation than a simple linear model. However, based on our assessments, there is no clear preference between the piecewise and quadratic functions \adjusted{and a linear function behaves as well as the other forms for the innermost $10\,\mathrm{kpc}$}.

\subsection{The influence of binning}
\label{sec:binning}

In this section, we test the influence of fitting a function to all points of the distribution or binned data in steps of $\Delta R_\mathrm{Gal.} = 0.5\,\mathrm{kpc}$, using median values as data points and standard deviations\footnote{We note that this $\sigma$ is not equivalent to observational uncertainty and can thus not be directly applied onto observational analyses.} as uncertainty \citep[see also][who fitted functions to radially binned IllustrisTNG data]{Hemler2021}. The results are shown in Fig.~\ref{fig:linear_quadratic_piecewise}. Given that more than half of the young star particles of the galaxy are within $R_\mathrm{Gal.} < 4\,\mathrm{kpc}$ \adjusted{as well as a height below $h_{z,\mathrm{gas}} = 1.25\,\mathrm{kpc}$}, this binning -- although counteracted by the smaller spread of [Fe/H] in the inner galaxy -- weighs the distribution of the inner galaxy significantly less than when weighing all particles equally (20 vs. 34\,000 data points). The parameters fitted to the binned data exhibit a larger uncertainty due to our use of the spread of [Fe/H] per bin as absolute uncertainty $\sigma$, but the fitted parameters agree well within the fitting uncertainties.

\paragraph*{Take-Away:} While the specific slopes differ when fitting all points or binned data, they agree within the small fitting uncertainties.

\subsection{The influence of radial coverage on linear fits}
\label{sec:radial_coverage}

Although we have gained useful insight into the global function, observational data will rarely cover the full extent of the stellar disk. Milky Way studies have previously been limited to the range of around $5-15\,\mathrm{kpc}$. There are often similar limitations and even gaps in extragalactic data. Using smaller ranges, observational studies have found hints of piece-wise linear gradients with a break radius in them based on limited radial coverage \citep[e.g.][]{Andrievsky2002, Yong2012, Boeche2013, Hayden2014, Anders2017, Donor2020, Chen2023}. These results are intriguing, since a quadratic function can, to first order, be approximated by two linear functions with a break radius. We therefore want to use our simulation to test if the radial coverage may indeed delude us into identifying broken linear gradients.

\begin{figure}
    \centering
    \includegraphics[width=\columnwidth]{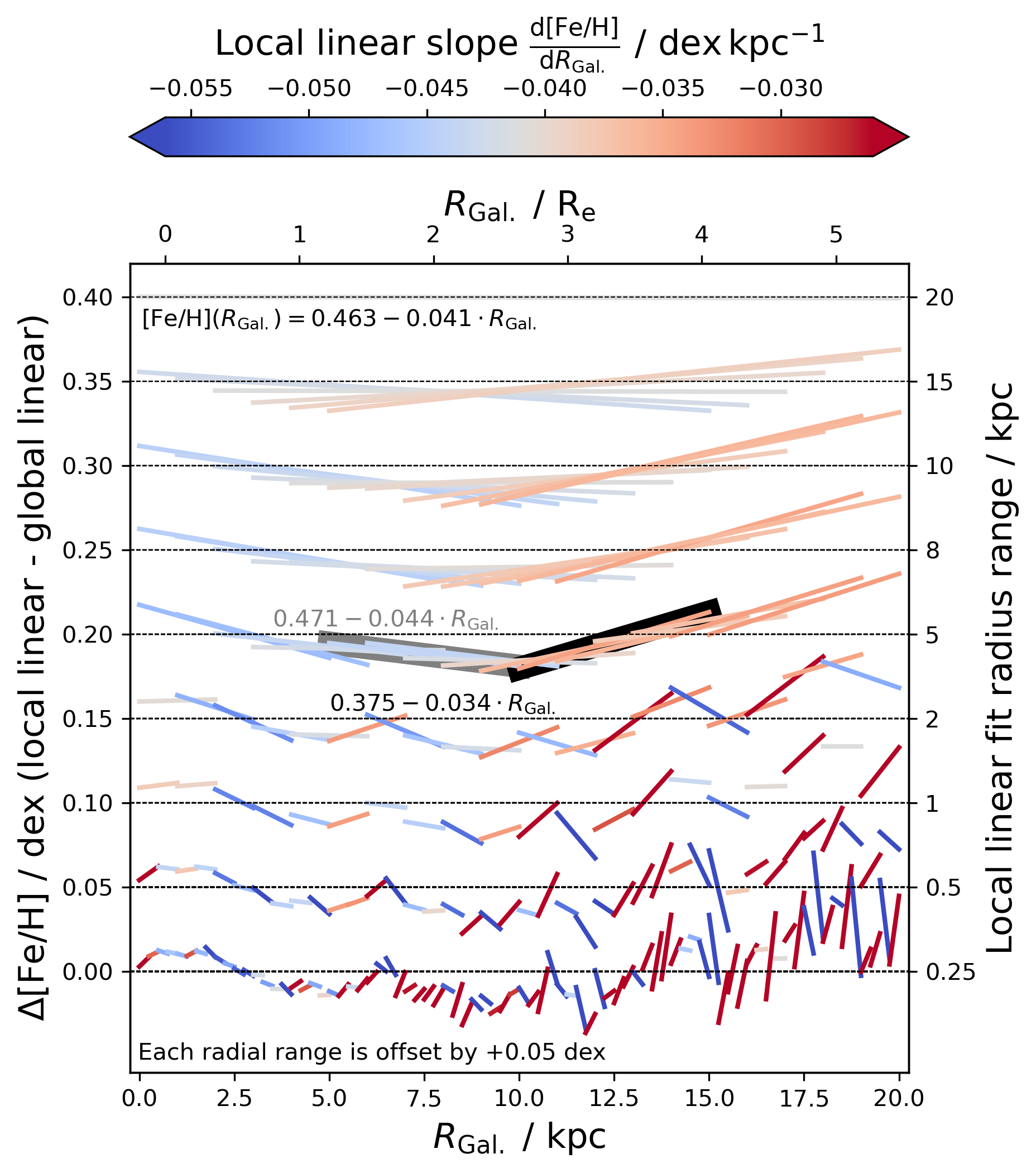}
    \caption{Impact of different coverage in galactocentric radius when fitting a linear radial metallicity gradient to young stars. Each horizontal segment uses a different running radial fitting range between 0.25 and $15\,\mathrm{kpc}$ as outlined on the right. For better contrast, the global linear fit is subtracted from the local gradient estimates and each line is colored by the gradient slope with a color scale centered around the global fit slope. Additionally, the slope of each line segment highlights the difference between global and local slopes. Thus, if the local fit exactly matches the global fit, we display it as a flat line (zero slope difference) on top of the gray dashed line (zero offset deviation) and give it a gray color signaling the same gradient value as the global fit.}
    \label{fig:radial_range_impact}
\end{figure}

We test how smaller coverage and piecewise linear fits could mimic a complex global gradient by fitting in piecewise linear radial ranges of 0.25, 0.5, 1, 2, 5, 8, 10, 15, and $20\,\mathrm{kpc}$. We show their difference with respect to a global linear fit in Fig.~\ref{fig:radial_range_impact}, with color-coding indicating the slope of the local gradient. A horizontal dashed line indicates the same slope as the global fit, whereas the offset of a line from the said horizontal dashed line indicates the local deviation from the global gradient intercept. Differences in line slopes are visualising the difference in gradient slopes between the global and local fits. We see that all ranges suggest more or less significant deviations from a global linear fit. The innermost fit suggests a significantly different gradient than the outermost fit. We also note increasing slope differences towards the smallest scales, hinting at local deviations from a global pattern. We follow these up in Section~\ref{sec:scatter_radial_metallicity_gradients}, but for now, focus on the larger-scale trends.

When directly comparing an inner and outer radius fit, such as between $R_\mathrm{Gal.} = 5-10\,\mathrm{kpc}$ (thick grey line in Fig.~\ref{fig:radial_range_impact}) and $R_\mathrm{Gal.} = 10-15\,\mathrm{kpc}$ (thick black line in Fig.~\ref{fig:radial_range_impact}), we note a significant change, similar to previous estimates of the Milky Way \citep[e.g.][]{Yong2012, Lemasle2008}. In our case, the gradient estimate changes from $\mathrm{[Fe/H]}(R_\mathrm{Gal.}) = 0.471-0.044\cdot R_\mathrm{Gal.}$%
 to $\mathrm{[Fe/H]}(R_\mathrm{Gal.}) = 0.375-0.034\cdot R_\mathrm{Gal.}$%
. When looking at linear gradient fits across the radial coverage of $\Delta R_\mathrm{Gal.} = 5-15\,\mathrm{kpc}$ in Fig.~\ref{fig:radial_range_impact}, the gradient is steeper (bluer color) for smaller radii and flatter (redder) for larger radii. Indeed, a piecewise linear function can well mimic a complex global gradient.

We note that in the simulated data, we see local deviations that become traceable below $\Delta R_\mathrm{Gal.} \leq 2\,\mathrm{kpc} \sim 0.5\,\mathrm{R_e}$ (bottom part of Fig.~\ref{fig:radial_range_impact}). This might indicate the spatial resolution required to see local effects, such as spiral arms, for extragalactic studies \citep[see also][]{Krumholz2018b, Li2024b}. We pursue this observation in the following Section~\ref{sec:scatter_radial_metallicity_gradients}.

\paragraph*{Take-Away:} We find that a piecewise linear function can well mimic a quadratic function across the scales used in Milky Way and extragalactic studies. Local deviations become traceable below \adjusted{a} spatial resolution of $\Delta R_\mathrm{Gal.} \leq 2\,\mathrm{kpc}$ (or $\Delta R_\mathrm{Gal.} \leq 0.5\,\mathrm{R_e}$).

%%%%%%%%%%%%%%%%%%%%%%%%%%%%%%%%%%%%%%%%%%%%%%%%%%
%%%%%%%%%%%%%%%%%%%%%%%%%%%%%%%%%%%%%%%%%%%%%%%%%%
\section{Scatter and local deviations from the gradient}
\label{sec:scatter_radial_metallicity_gradients}
%%%%%%%%%%%%%%%%%%%%%%%%%%%%%%%%%%%%%%%%%%%%%%%%%%
%%%%%%%%%%%%%%%%%%%%%%%%%%%%%%%%%%%%%%%%%%%%%%%%%%

Now that we are sufficiently satisfied that our flattening gradient function reproduces the overall shape of the radial metallicity gradient, we are concerned with both the scatter and local slope deviations across the galactocentric radii in this section. In detail, we analyse the scatter (Section~\ref{sec:coherence_vertical_radial_metallicity_gradients}), vertical variations (Section~\ref{sec:coherence_vertical_radial_metallicity_gradients}), azimuthal variations (Section~\ref{sec:coherence_azimuth_radial_metallicity_gradients}, particularly motivated by the localised, spiral-shaped fit residuals of Figs.~\ref{fig:stars_and_gas_2d_view}c and \ref{fig:stars_and_gas_2d_view}f) and deviations across different ages (Section~\ref{sec:coherence_age_radial_metallicity_gradients}).

\subsection{Scatter}
\label{sec:scatter}

When investigating the change in scatter from the innermost radii to the outermost (see Fig.~\ref{fig:global_r_feh_fit}c), we see a steady increase in $1-\sigma$ spread. This spread increases from \\
$\sigma \mathrm{[Fe/H]} = 0.01\,\mathrm{dex}$%
 at $R_\mathrm{Gal.} = ~~0.25 \pm 0.25\,\mathrm{kpc}$ to \\
$\sigma \mathrm{[Fe/H]} = 0.06\,\mathrm{dex}$%
 at $R_\mathrm{Gal.} = ~~8.25 \pm 0.25\,\mathrm{kpc}$ and reaches \\
$\sigma \mathrm{[Fe/H]} = 0.10\,\mathrm{dex}$%
 at $R_\mathrm{Gal.} = 19.75 \pm 0.25\,\mathrm{kpc}$.

When we recall the observed significant spread in metallicities of young open clusters at the solar radius beyond observational uncertainty \citep[e.g.][]{Donor2020, Spina2021} and our selection of only young ($<$\nihaoAGEmax) stars from the simulation, a strong impact of this scatter by radial migration should be excluded. At this point, we can imagine that this chemical diversity might be caused by less well-mixed gas or non-radial effects (such as vertical or azimuthal ones), which we investigate subsequently.

\subsection{Vertical deviations}
\label{sec:coherence_vertical_radial_metallicity_gradients}

\begin{figure}
    \centering
    \includegraphics[width=0.96\columnwidth]{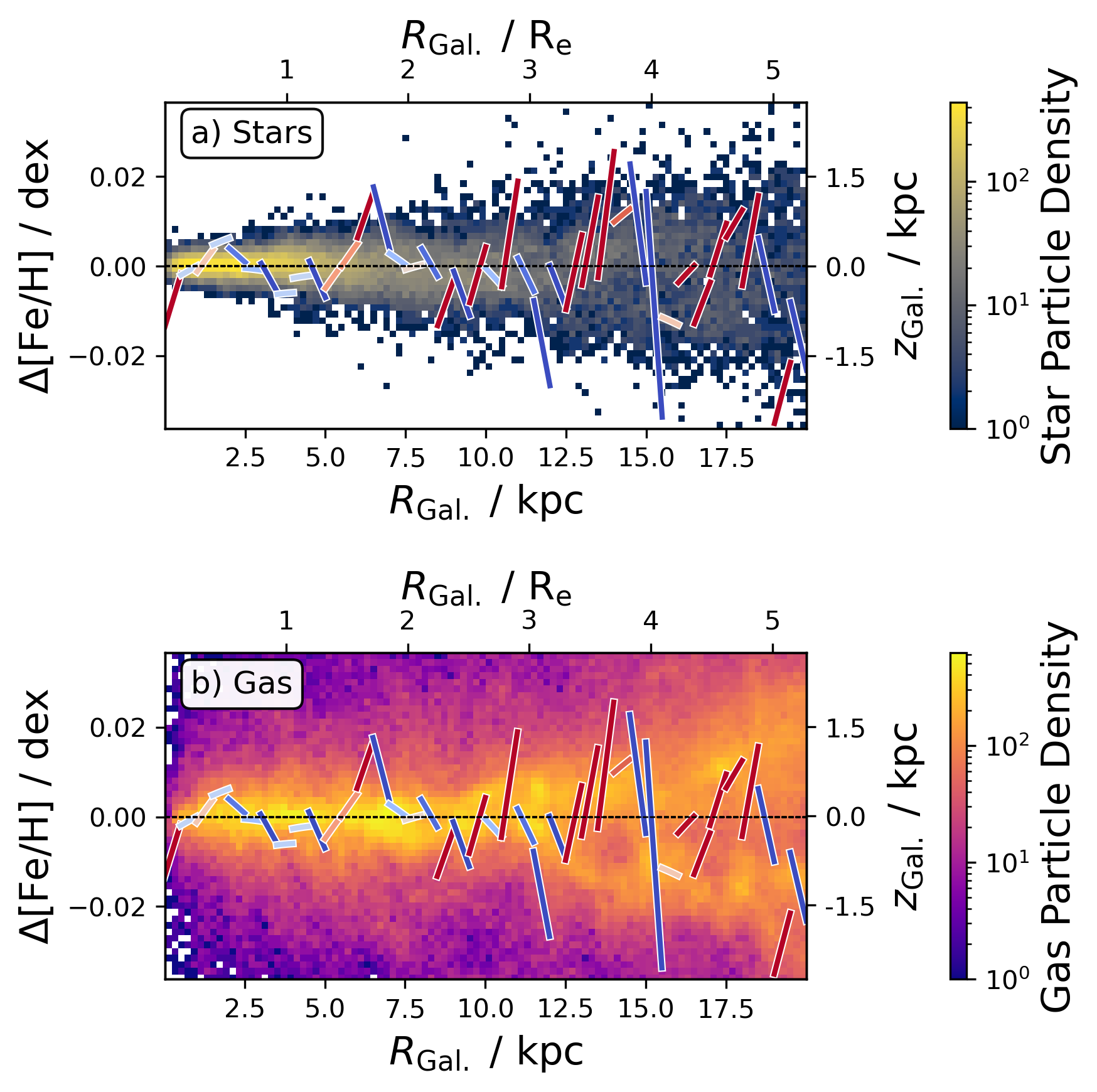}
    \caption{Local gradient deviations (red-blue lines) similar to the second lowest row of Fig.~\ref{fig:radial_range_impact} for radial gradients in $0.5\,\mathrm{kpc}$ steps (but compared to a global quadratic function) overlapped on top of the logarithmic density distribution in $R-z$ for $\vert z \vert < 3\,\mathrm{kpc}$ of gas (panel a) and stars (panel b). We see no strong correlation between local gradient slopes (red-blue lines) and star or gas density in this projection.}
    \label{fig:overlap_local_variation_gas}
\end{figure}

\begin{figure*}
    \centering
    \includegraphics[width=0.9\textwidth]{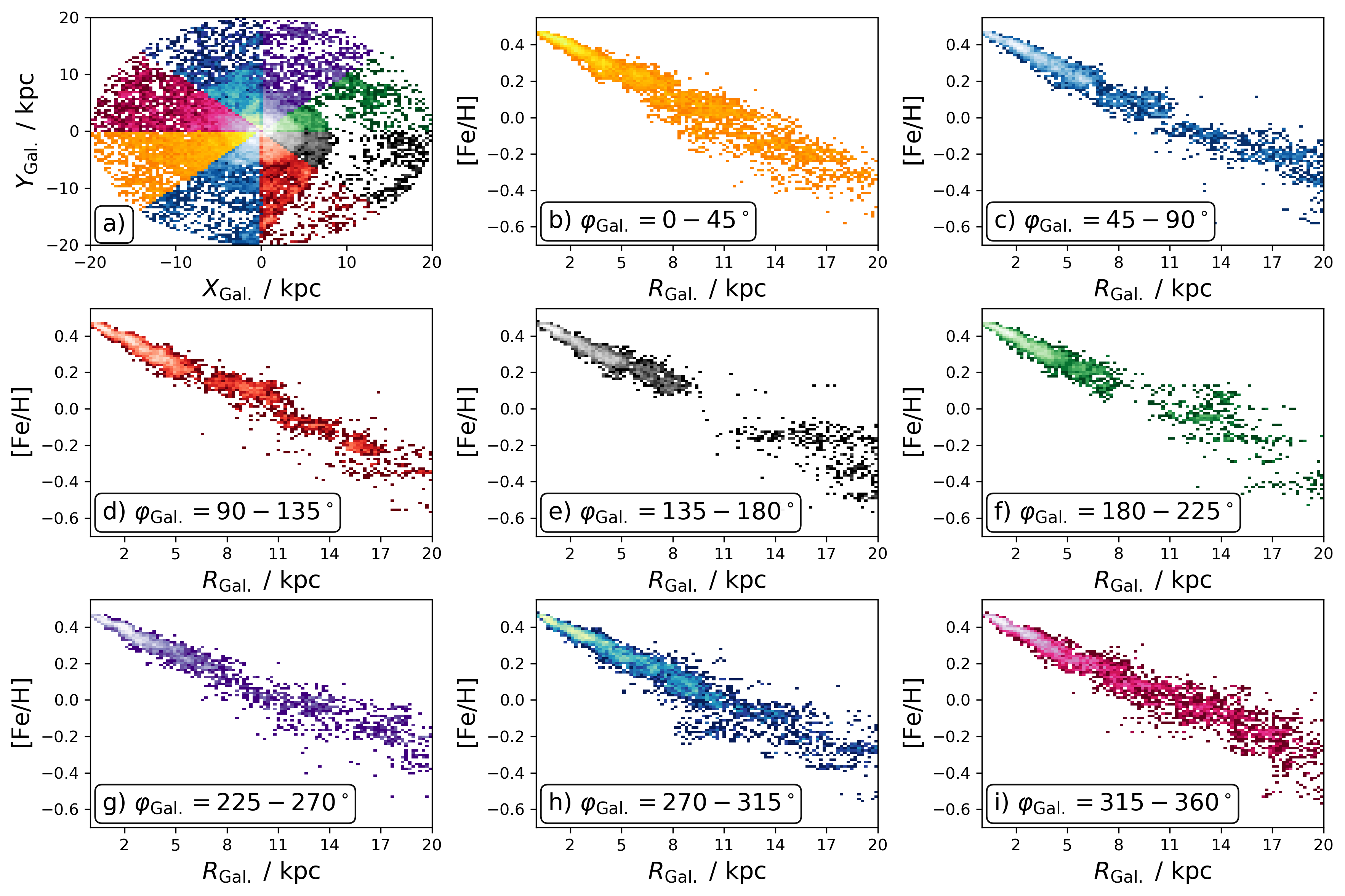}
    \caption{Stellar density variation across 8 different sectors (with color-code visualised in panel a) of the radial metallicity gradient $R-\mathrm{[Fe/H]}$ across 8 different azimuth ranges (panels b-i). A rotating lighthouse-like GIF animation of the median age and median density of the $R-\mathrm{[Fe/H]}$-relation across different azimuths is freely available on a \href{https://github.com/svenbuder/nihao_radial_metallicity_gradients/blob/main/figures/xyz_rfeh.gif}{repository}.}
    \label{fig:radial_metallicity_gradients_mw_in_angles}
\end{figure*}

\begin{figure*}
    \centering
    \includegraphics[width=0.975\textwidth]{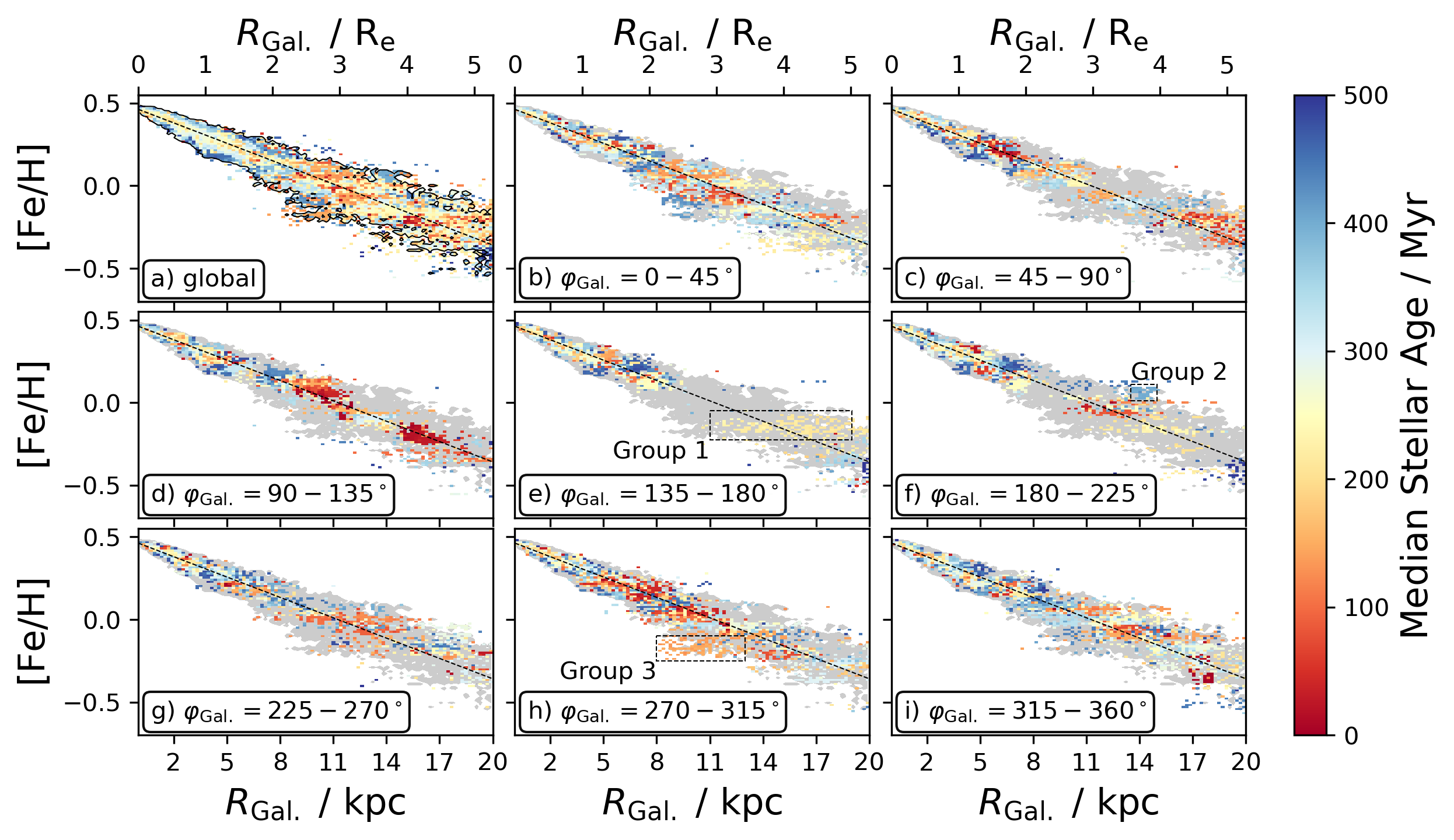}
    \caption{Same as Fig.~\ref{fig:radial_metallicity_gradients_mw_in_angles}, but colored by median age per bin. We identify 3 groups with boxes in panels e, f, and h. A rotating lighthouse-like GIF animation of the median age and median density of the $R-\mathrm{[Fe/H]}$-relation across different azimuths is freely available a \href{https://github.com/svenbuder/nihao_radial_metallicity_gradients/blob/main/figures/xyz_rfeh.gif}{repository}.}
    \label{fig:radial_metallicity_gradients_mw_in_angles_age}
\end{figure*}

In this section, we now look at deviations with respect to the vertical dimension, that is, $R-z$. In Fig.~\ref{fig:overlap_local_variation_gas} we show the previously identified local gradient deviations (lines following the left axis label) on top of the vertical density distribution ($R-z$) of young stars (Fig.~\ref{fig:overlap_local_variation_gas}a) and gas (Fig.~\ref{fig:overlap_local_variation_gas}b) between $-3 < z < 3\,\mathrm{kpc}$. Although the quickly decreasing number of young stars (Fig.~\ref{fig:overlap_local_variation_gas}a) at outer radii does not show substructure in the density plots for reasonable bin sizes, we see more substructure for the gaseous component in Fig.~\ref{fig:overlap_local_variation_gas}b). In particular, we see rather minor deviations at small radii (where most stars and gas are close to the plane). At increasing radii, we notice an increase in both the vertical distribution of stars, and increasing local gradient deviations. In particular, we note a significant deviation of the slope around $R_\mathrm{Gal.} \sim 15\,\mathrm{kpc}$, where the gradient deviation line is steep and blue (indicating a much steeper gradient at this radius), and we notice a significant overdensity of gas around $z \sim 1\,\mathrm{kpc}$. Overall, however, we do not see strong correlations in this particular plane.

This could, however, be caused by a super-position effect of the up- and downturn at larger radii due to the galactic warp (see Figs.~\ref{fig:stars_and_gas_overview}b and \ref{fig:stars_and_gas_overview}e). Although the warp of the stellar disk in Fig.~\ref{fig:stars_and_gas_overview}b is not as clear, we confirm that both the gas disk and the youngest stars below \nihaoAGEmax\ are tracing each other across the simulation in both galactocentric radius $R_\mathrm{Gal.}$ and height $z_\mathrm{Gal.}$ for different sectors in the azimuthal direction. We note that the superposition in Fig.~\ref{fig:overlap_local_variation_gas} could smear out local correlations of slope changes with gas overdensities, for example, by spiral arms. Although such an edge-on view of the galaxy may indeed be the only observable one for extragalactic targets, for example, of the GECKOS survey of edge-on galaxies \citep{GECKOS2023}, we have the luxury of being able to analyse the azimuthal direction of our simulated galaxy, too. 

\paragraph*{Take-Away:} We see no strong correlations of deviations in the vertical direction throughout the simulation. Such correlations could, however, be blurred by azimuthal effects, like the galactic warp, which needs to be disentangled in the azimuthal dimension.

\subsection{Azimuthal deviations}
\label{sec:coherence_azimuth_radial_metallicity_gradients}

To analyse the deviations from a global gradient across different azimuthal viewing angles, we divide the galaxy into 8 sectors with $\Delta \varpi_\mathrm{Gal.} = 45^\circ$ (see Fig.~\ref{fig:radial_metallicity_gradients_mw_in_angles}a). This allows us to study the positions around the upturn and downturn of the galactic warp with the median azimuth of young star particles below and above the plane being $\varphi_\mathrm{Gal.} \sim 183^\circ$ and $\varphi_\mathrm{Gal.} \sim 4^\circ$, respectively (see Fig.~\ref{fig:stars_and_gas_overview}e), while maintaining a reasonable sample size.

At face value, the distribution of $R_\mathrm{Gal.}-\mathrm{[Fe/H]}$ for each sector follows a similar, rather linear shape with most stars being born in the inner $5\,\mathrm{kpc}$ of the galaxy. However, we find significant deviations in different sectors of the galaxy (Fig.~\ref{fig:radial_metallicity_gradients_mw_in_angles}). On the one hand, we find non-linear deviations as bumps with slightly increased or decreased iron abundance - up to $0.1-0.2\,\mathrm{dex}$ - in Figs.~\ref{fig:radial_metallicity_gradients_mw_in_angles}c at $R_\mathrm{Gal} \sim 18\,\mathrm{kpc}$, \ref{fig:radial_metallicity_gradients_mw_in_angles}d at $R_\mathrm{Gal} \sim 10\,\mathrm{kpc}$, \ref{fig:radial_metallicity_gradients_mw_in_angles}f at $R_\mathrm{Gal} \sim 14\,\mathrm{kpc}$, and \ref{fig:radial_metallicity_gradients_mw_in_angles}g at $R_\mathrm{Gal} \sim 17\,\mathrm{kpc}$. On the other hand, we find significant gaps in the distribution at similar [Fe/H], most strikingly at the upturn of the galactic warp in Fig.~\ref{fig:radial_metallicity_gradients_mw_in_angles}e ($\varpi_\mathrm{Gal.} = 135-180^\circ$) at $\mathrm{[Fe/H]} \sim 0\,\mathrm{dex}$ and $R_\mathrm{Gal.} \sim 8-14\,\mathrm{kpc}$. We note that the sector e) with the gap is surrounded by two sectors (d and f) with significant overabundance at the same radius. This could be indicative of stars having formed as a result of gas moving from sector e towards either azimuthal direction, causing a gas overdensity which could in turn lead to higher star formation activity. To establish this observation, we take a closer look at the time-domain, that is, stellar age as well as the spatial domain of $R_\mathrm{Gal.}-\varphi_\mathrm{Gal.}$ in the next section. 

\paragraph*{Take-Away:} We find various deviations from the global trend in the azimuthal direction, including gaps and isolated streaks of stars with similar [Fe/H] throughout $\Delta R_\mathrm{Gal.} = 2-6\,\mathrm{kpc}$. These can introduce local over- and under-enhancement of up to $\pm 0.2\,\mathrm{dex}$ in [Fe/H] at a given radius. In the next section, we analyse whether the stars of these streaks have been born at the same or different time.

\subsection{Deviations with time and age}
\label{sec:coherence_age_radial_metallicity_gradients}

In this section, we examine the radial metallicity gradient in a small age range less than \nihaoAGEmax. To do so, we color Fig.~\ref{fig:radial_metallicity_gradients_mw_in_angles} by the median stellar age rather than the logarithmic density in Fig.~\ref{fig:radial_metallicity_gradients_mw_in_angles_age}. We find an overall significant scatter across time, suggesting a good mix of star formation across all sectors for stars born within less than \nihaoAGEmax. For stars within this restricted age range, we do not see a strong correlation with radius, such as older stars being born further inside, but a larger amount of stars being born closer to the galactic centre. We note that stars with similar [Fe/H] in each sector tend to be formed at similar times (within $50\,\mathrm{Myr}$), that is, as flat lines with the same color (age) in Figs.~\ref{fig:radial_metallicity_gradients_mw_in_angles_age}b-i. To guide the eye, we have identified Group 1 in Fig.~\ref{fig:radial_metallicity_gradients_mw_in_angles_age}e (around $R_\mathrm{Gal.} \sim 14\,\mathrm{kpc}$ at $\varpi_\mathrm{Gal.} = 180-225^\circ$). We further note that the enriched bumps identified earlier are born at similar times, see, for example, Group 2 in Fig.~\ref{fig:radial_metallicity_gradients_mw_in_angles_age}f. The coloring by age also reveals that stars with lower [Fe/H] than expected (see Group 3 in Fig.~\ref{fig:radial_metallicity_gradients_mw_in_angles_age}h) are born at similar times. In some cases, these extend to $\Delta R_\mathrm{Gal.} = 2-6\,\mathrm{kpc}$, see Groups 1, 2, and 3 in Fig.~\ref{fig:radial_metallicity_gradients_mw_in_angles_age}. At a given radius $R_\mathrm{Gal.}$, these streaks cause a significant spread in local [Fe/H] of up to $\pm 0.2\,\mathrm{dex}$ (see Fig.~\ref{fig:radial_metallicity_gradients_mw_in_angles_age}).

\begin{figure}
    \centering
    \includegraphics[width=\columnwidth]{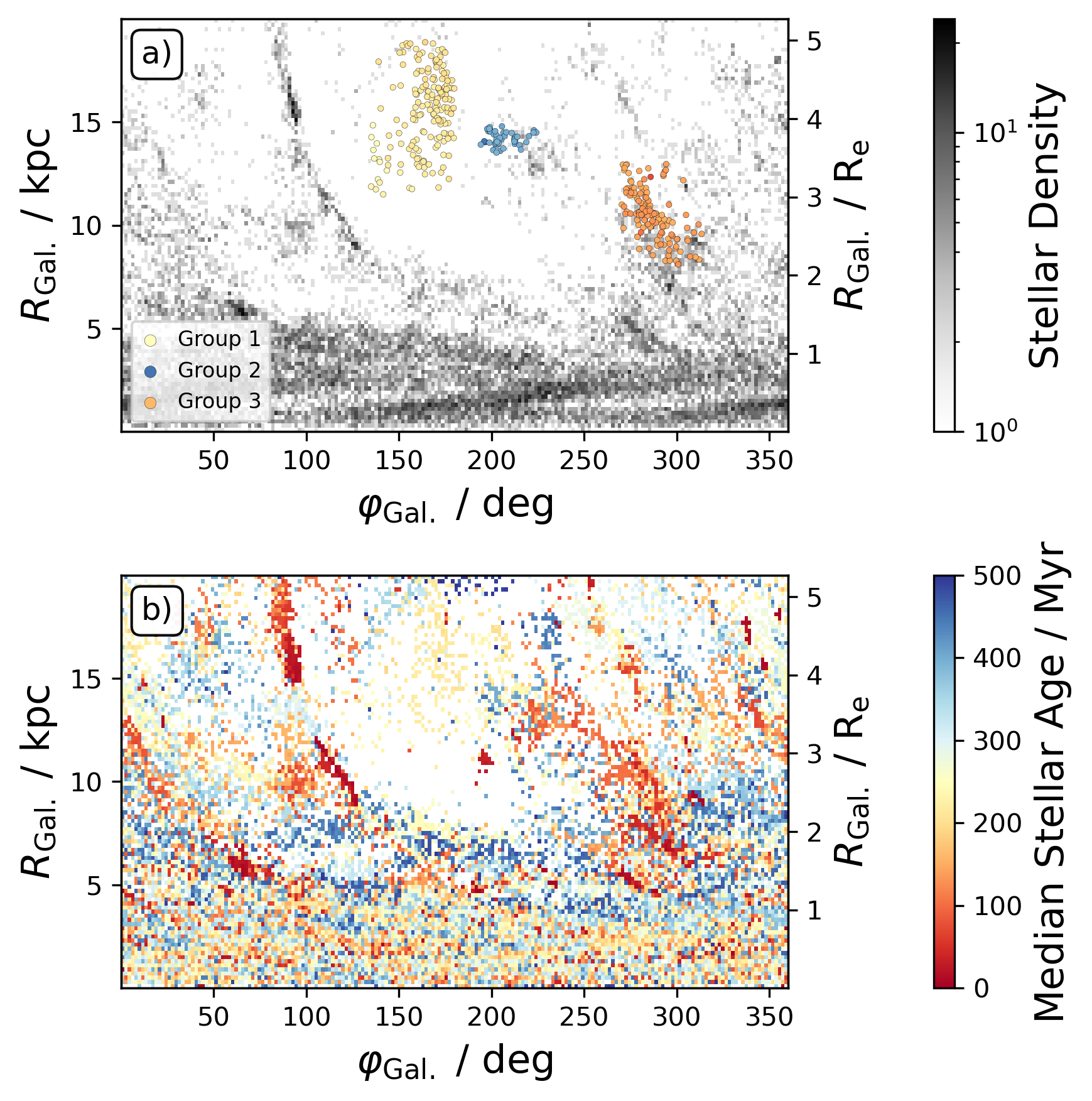}
    \caption{Density distribution (panel a) and age distribution (panel b) of young stars in the azimuthal and radial direction $\varphi_\mathrm{Gal.}-R_\mathrm{Gal.}$. In panel a), we also show the groups previously identified in Fig.~\ref{fig:radial_metallicity_gradients_mw_in_angles_age}.}
    \label{fig:phi_angle_R_follow_up}
\end{figure}

From the analysis of azimuthal sectors, the impression arose that the star formation in this simulated Milky Way analogue is -- as expected for a spiral galaxy -- rather patchy and localised on the smallest timescales. This is confirmed by looking at the spatial distribution of azimuth $\varphi_\mathrm{Gal.}$ and radius $R_\mathrm{Gal.}$ in Fig.~\ref{fig:phi_angle_R_follow_up}. Already when looking at the density distribution of all stars born within less than \nihaoAGEmax\ in Fig.~\ref{fig:phi_angle_R_follow_up}a, multiple streams are visible, stars on spiral patterns \citep[see also][]{Kreckel2019, Chen2024b}. \adjusted{We note for example a narrow sequence around $\varphi_\mathrm{Gal} \sim 100\,\mathrm{deg}$ stretching along $R_\mathrm{Gal} = 8-20\,\mathrm{kpc}$ visible as significant overdensity in Fig.~\ref{fig:phi_angle_R_follow_up}a and distinctly red (young) sequence in Fig.~\ref{fig:phi_angle_R_follow_up}b. This sequence coincides with the red features in Fig.~\ref{fig:radial_metallicity_gradients_mw_in_angles_age}d, which follow the linear metallicity gradient very closely.} When following up the previously identified Groups 1, 2, and 3, we recover them on said spiral patterns (Groups 1 and 3) or a local overdensity (Group 2). Although one could imagine that radial migration might induce such a spiral-like shape for the stars of groups 1 and 3, their low age of less than $250\,\mathrm{Myr}$ would require a significant migration effect of several kpc, while having no influence on the older stars of group 2. When tracing the position of significant overdensities from Fig.~\ref{fig:phi_angle_R_follow_up}a in the same projection colored by age in Fig.~\ref{fig:phi_angle_R_follow_up}b, we note that for radii above $R_\mathrm{Gal.} > 5\,\mathrm{kpc}$ these overdensities are colored in red, that is, containing indeed young stars with ages below $200\,\mathrm{Myr}$ and being consistent with the most recent star formation along these spiral patterns in the outer galaxy.

\paragraph*{Take-Away:} We find significant scatter across the radial metallicity distribution caused by streaks of stars born with similar [Fe/H] at similar times (within $50\,\mathrm{Myr}$) across either very local or radially extended spiral-shaped regions of the galaxy, suggesting local enhancement patterns in small overdensities or along spiral arms.

%%%%%%%%%%%%%%%%%%%%%%%%%%%%%%%%%%%%%%%%%%%%%%%%%%
%%%%%%%%%%%%%%%%%%%%%%%%%%%%%%%%%%%%%%%%%%%%%%%%%%
\section{Discussion} \label{sec:discussion}
%%%%%%%%%%%%%%%%%%%%%%%%%%%%%%%%%%%%%%%%%%%%%%%%%%
%%%%%%%%%%%%%%%%%%%%%%%%%%%%%%%%%%%%%%%%%%%%%%%%%%

Having presented the analysis, we now put our results into the context of other work in terms of our initial aims: to analyse the shape (Section~\ref{sec:discussion_linearity}), scatter (Section~\ref{sec:discussion_scatter}), local deviations (Section~\ref{sec:discussion_coherence_position}), and time-dependence (Section~\ref{sec:discussion_time}) of the radial metallicity gradient. These initial discussions inform our thoughts on the implications of this work for Milky Way studies in Section~\ref{sec:implications_milky_way} and the studies of other galaxies in Section~\ref{sec:implications_extragalactic}.

\subsection{Linearity of the radial metallicity gradient} \label{sec:discussion_linearity}

\begin{figure*}
    \centering
    \includegraphics[width=\textwidth]{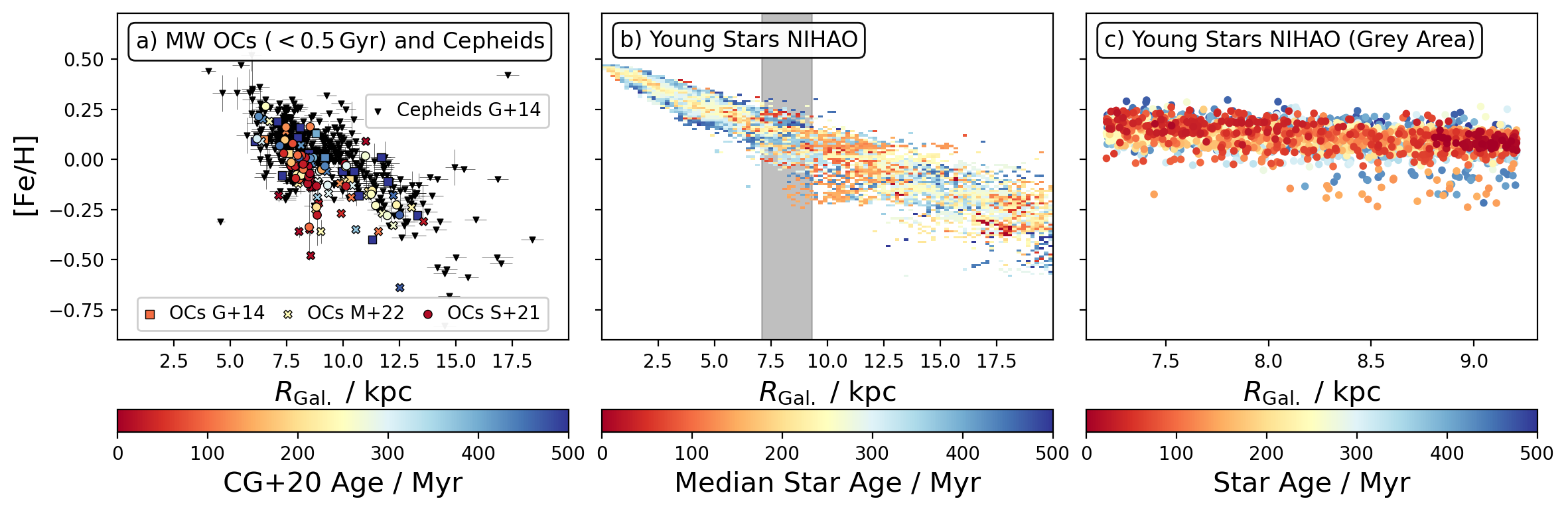}
    \caption{Comparison of the Milky Way's radial metallicity trend as traced by Cepheids \citep[black triangles, compiled from literature by][G+14]{Genovali2014} as well as young ($<$\nihaoAGEmax) open cluster of the Milky Way as traced by the literature compilation from \citet[][G+14 as squares]{Genovali2014}, APOGEE DR17 from \citet[][M+22 as crosses]{Myers2022}, and GALAH DR3 from \citet[][S+21 as circles]{Spina2021}. The latter two are compiled based on the membership and age catalogue by \citet[][CG+20]{CantatGaudin2020}.
    }
    \label{fig:radial_metallicity_gradients_mw_vs_nihao}
\end{figure*}

The radial metallicity gradient of our simulated NIHAO-UHD Milky Way analogue showed an overall decreasing, predominantly linear shape, as established in Section~\ref{sec:linear_radial_metallicity_gradients}. Motivated by previous works by \citet{SanchezMenguiano2016}, among others, we also fitted piecewise linear and quadratic functions to the data in Section~\ref{sec:global_fits}. Both forms perform better than a linear trend. The very similar fitting performances indicate no significant preference between either piecewise linear or quadratic function. Due to both functions' rather good overall fit, we have not tried more exotic non-linear functions as done by \citet{Scarano2013}. Increasing the flexibility of the gradient function could, however, improve the fit at the innermost kpc, where a flattening is predicted by our simulation, but chemical enrichment is also harder to simulate \citep[see also][]{Minchev2013, Sun2024}. We have found no significant influence of binning for our gradient estimates (Section~\ref{sec:binning}), but have found that a limited radial coverage - as is the case for the Milky Way - could mimic a truly quadratic function with two piecewise linear fits (Section~\ref{sec:radial_coverage}). This is important, as it has significant implications for the conclusions we draw from the incomplete data of our Milky Way, as we will discuss in more detail in Section~\ref{sec:implications_milky_way}.

The balance between a quadratic and piecewise linear radial metallicity gradient teeters at the breaking radius. If present, our analysis of the Milky Way analogue would place it \adjusted{at $R_\mathrm{break} \sim 9.3-11.5\,\mathrm{kpc}$}. This radius is strikingly close to the radius of $9\,\mathrm{kpc}$ found by \citet{Hemler2021} for a TNG50 galaxy simulation with a stellar mass of $\log(M_\star/\mathrm{M_\odot}) = 10.72$, that is, close to the Milky Way's (see their Fig.~2). In terms of physical reasons for a breaking radius at this location, a direct and secular influence of a stellar bar with non-symmetric effects around the corotation radius \citep{DiMatteo2013, Scarano2013} should be minor for our specific scenario due to the low ages of the stars considered in our analysis. In particular, our identified break radius is significantly larger than the corotation radius of the Milky Way bar at $4.5-7.0\,\mathrm{kpc}$ \citep[][and references therein]{BlandHawthorn_Gerhard2016} anyway. We are intrigued, however, by the proposition by \citet{Garcia2023} of galactic discs consisting of a star-forming inner disc with a steep gradient and a mixing-dominated outer disc with a flat gradient, with the break radius marking the region of transition between them. In Illustris TNG50-1 data, they found such a transition and break radius to be situated much further out at $30\,\mathrm{kpc}$ for Milky Way mass galaxies ($10.1 \leq \log(M_\star/\mathrm{M_\odot}) \leq 10.6$). While our best-fitting break radius - if present - is inconsistent with theirs, we will follow this up in more detail in Section~\ref{sec:implications_extragalactic}, where we also discuss the implications for extragalactic studies in general.

\begin{figure}
    \centering
    \includegraphics[width=\columnwidth]{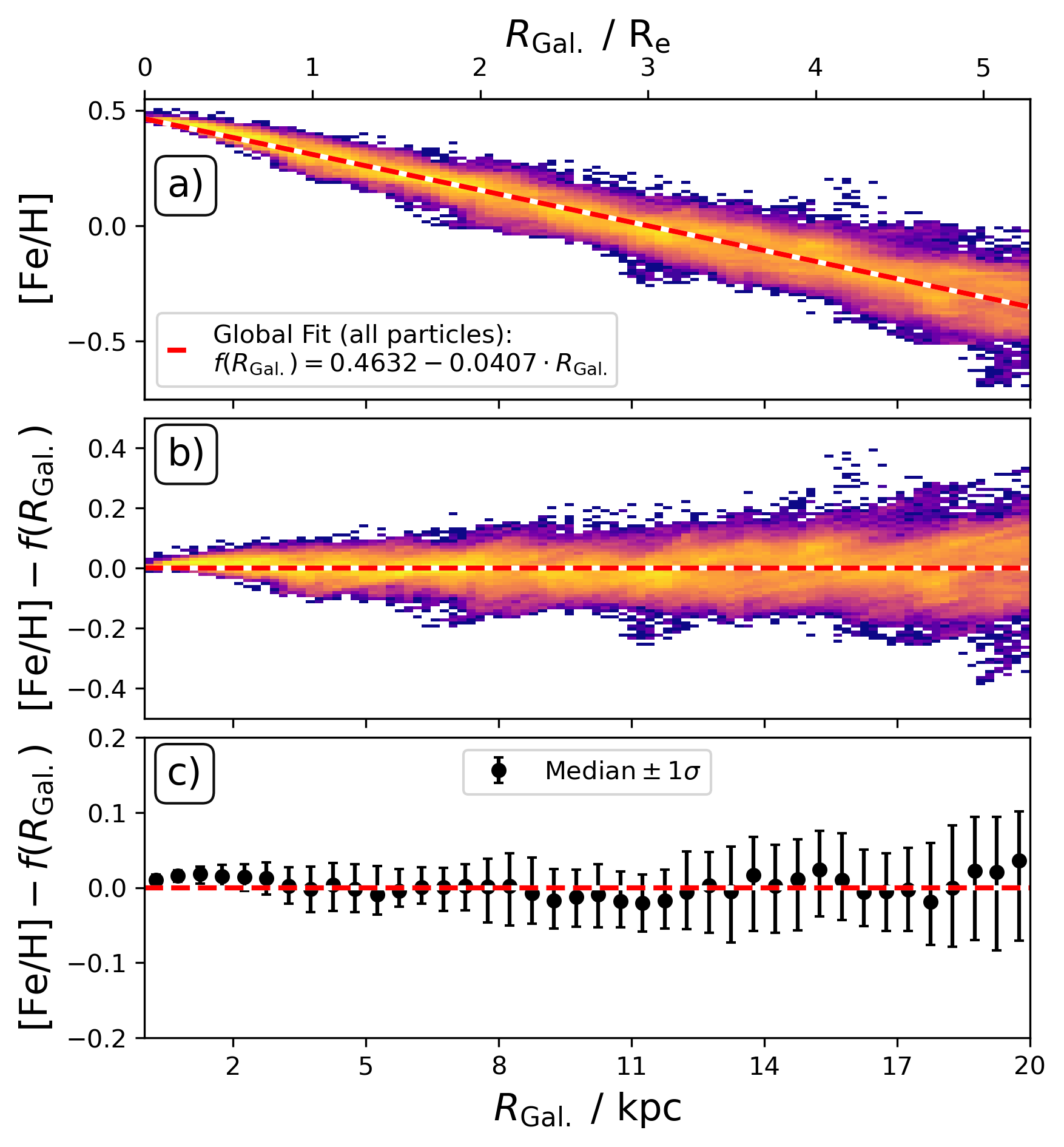}
    \caption{Same as Fig.~\ref{fig:global_r_feh_fit}, but for gas.}
    \label{fig:global_r_feh_fit_gas}
\end{figure}

\subsection{Scatter of the radial metallicity gradient} \label{sec:discussion_scatter}

In Section~\ref{sec:scatter} we found an increasing scatter from $\sigma \mathrm{[Fe/H]} = 0.01\,\mathrm{dex}$%
 in the inner galaxy to $\sigma \mathrm{[Fe/H]} = 0.10\,\mathrm{dex}$%
 around $R_\mathrm{Gal.} \sim 20\,\mathrm{kpc}$. Comparing these values with simulations other than TNG50 with a similar metallicity spread \citep[see Fig.~2 by][]{Hemler2021} is rather difficult, as the literature focuses on the shape and density distribution \citep[see e.g.][their Fig.~10]{Minchev2014b}. When comparing with Milky Way studies \citep[e.g.][]{Anders2017}, the scatter in the simulation is smaller than the observed spread of [Fe/H]. This can be visually appreciated by comparing the combinations of different measurements in the Milky Way \citep{Genovali2014, Spina2021, Myers2022} in Fig.~\ref{fig:radial_metallicity_gradients_mw_vs_nihao}a and our simulation in Fig.~\ref{fig:radial_metallicity_gradients_mw_vs_nihao}b and c. We discuss the implications of this on studies of the Milky Way's gradient in Section~\ref{sec:implications_milky_way}.

\begin{figure*}
    \centering
    \includegraphics[width=0.77\textwidth]{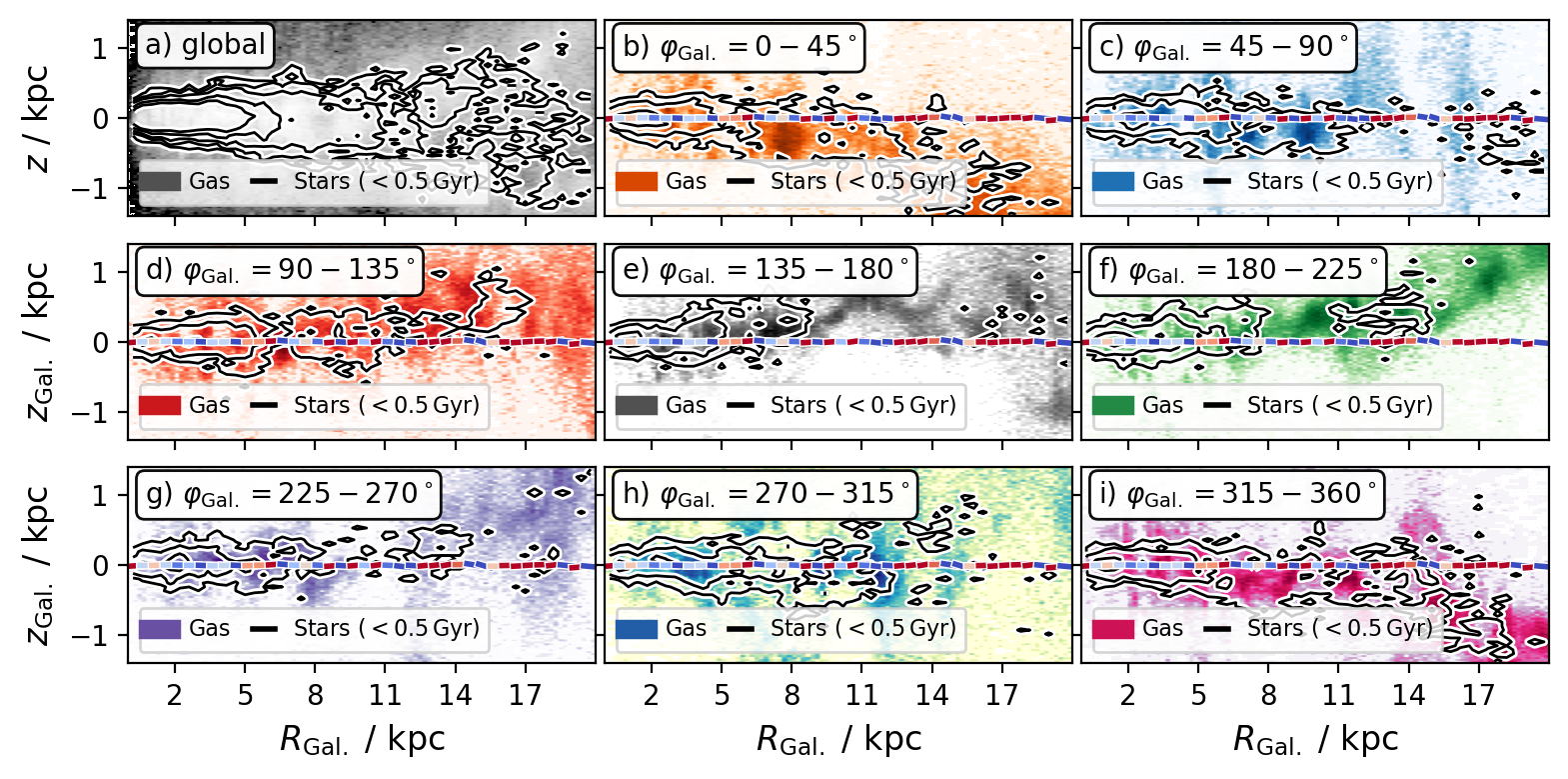}
    \caption{Tracing young stars and gas across galactocentric radii $R_\mathrm{Gal.}$ and height $z_\mathrm{Gal.}$ across the whole galaxy (panel a) and different azimuthal ranges/sectors (panels b-i). Small rectangles with cool-warm colors along the horizontal axis indicate the local gradient slopes as in Fig.~\ref{fig:overlap_local_variation_gas}.}
    \label{fig:tracing_young_stars_and_gas_in_angles}
\end{figure*}

When assuming that young star and gas phase abundances are similar, we find comparable scatter of abundances for example with respect to TYPHOON observations by \citet{Chen2023}. To test this assumption, we also show the gas phase metallicity in Fig.~\ref{fig:global_r_feh_fit_gas}, for which we find a similar shape and scatter of the gradient, but systematically less scatter or spread than observed gas phase abundance, thus urging us to treat the absolute values of abundances and abundance scatters as well as spreads with caution. We furthermore note that the spread of abundance in observations does only increase for some but not all of the observed (and thus observationally limited) galaxies by \citet{Chen2023}. This potentially limits the range of galaxies to which our conclusions may apply. While the simulated abundance scatter is consistent with the predictions by the theoretical forced-diffusion model by \citet{Krumholz2018b}, that is, a scatter of $\sim 0.1\,\mathrm{dex}$ over timescales of $\sim 100-300\,\mathrm{Myr}$, our simulations suggest that the scatter is driven by the radial structure and large-scale spiral arms, which were not included in their model.

\subsection{Localised vertical and azimuthal deviations and their correlation with gas} \label{sec:discussion_coherence_position}

In Sections~\ref{sec:coherence_vertical_radial_metallicity_gradients} and \ref{sec:coherence_azimuth_radial_metallicity_gradients} we established that local deviations contribute significantly to the spread of the global metallicity gradient above $R_\mathrm{Gal.} > 8\,\mathrm{kpc} \sim 2\,\mathrm{R_e}$. We noted in particular a void of stars where we found an upturning warp of the galaxy around $\varphi_\mathrm{Gal.} \sim 180^\circ$ spatially close to regions of the galaxy (Groups 1 and 2) that deviated most significantly from the overall trend in Fig.~\ref{fig:radial_metallicity_gradients_mw_in_angles_age} and Fig.~\ref{fig:phi_angle_R_follow_up}.

\begin{figure*}
    \centering
    \includegraphics[width=0.925\textwidth]{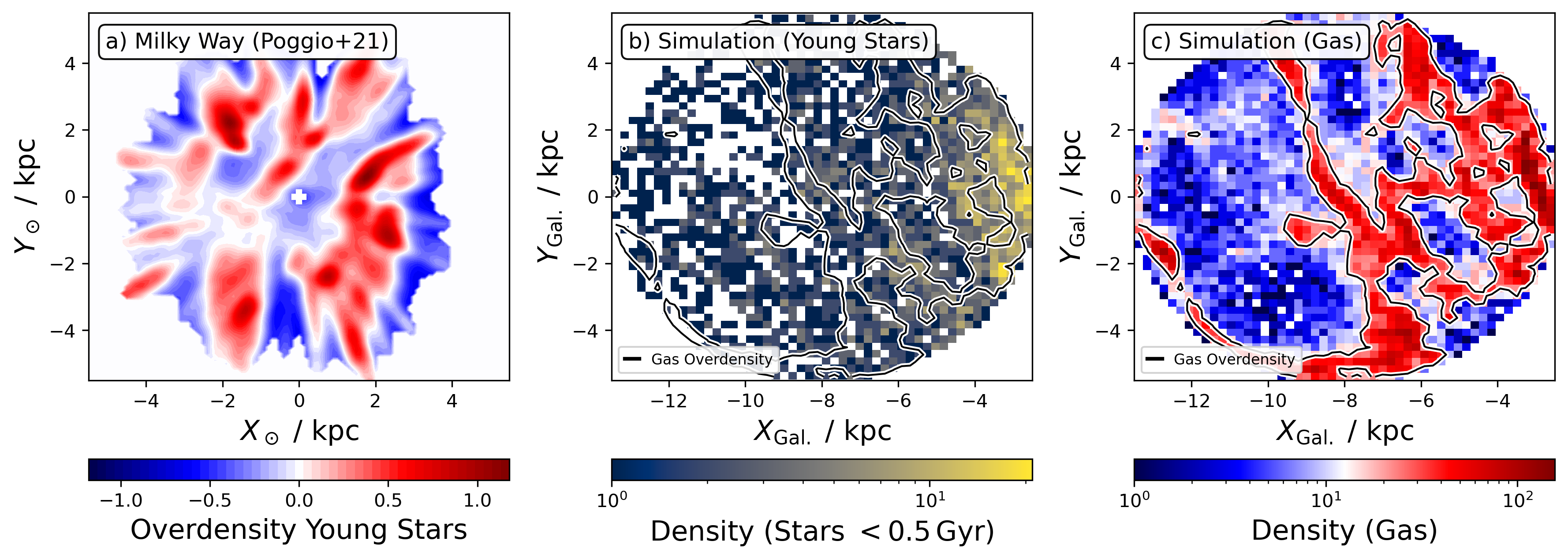}
    \caption{Comparison of the density distribution of young stars and gas in the Milky Way and the NIHAO Milky Way analogue simulation. Panel a) shows the measurements of the Solar vicinity within $5\,\mathrm{kpc}$ by \citet{Poggio2021}. Panels b) and c) show young stars and gas NIHAO, respectively, for a selected region similar to panel a). Black and white contour lines in panel b) trace overdensities in the gas distribution of panel c).}
    \label{fig:overdensities_mw_vs_nihao}
\end{figure*}

\begin{figure*}
    \centering
    \includegraphics[width=\textwidth]{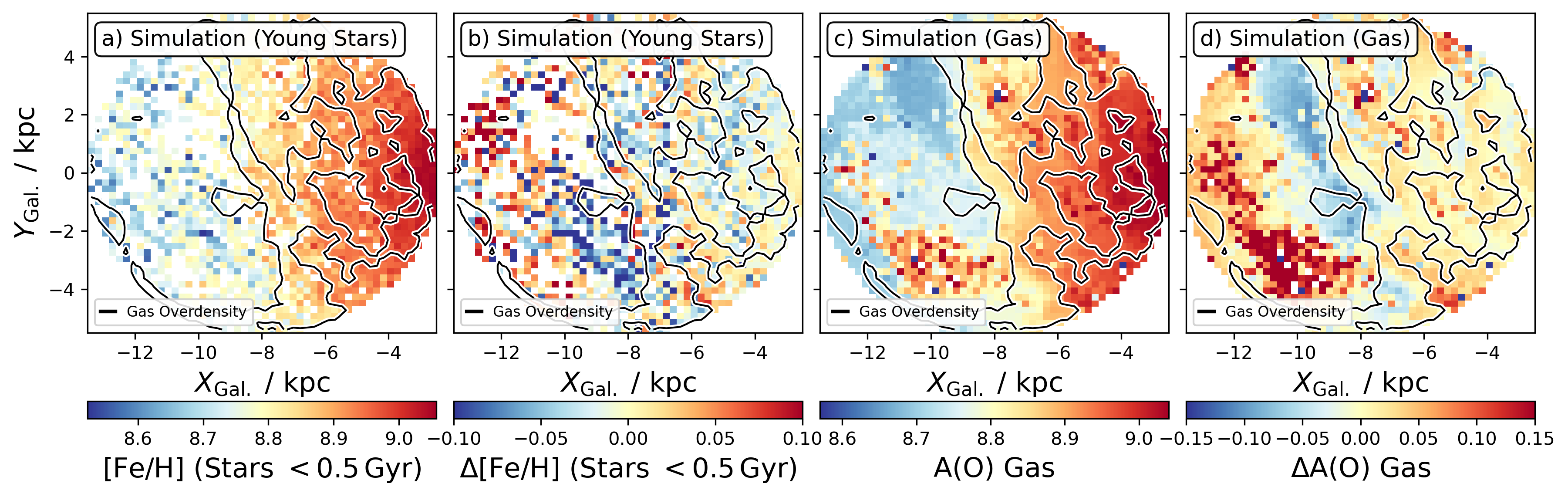}
    \caption{Comparison of density distribution of young stars and gas in the NIHAO-UHD Milky Way analogue simulation for the same regions as Figs.~\ref{fig:overdensities_mw_vs_nihao}b and \ref{fig:overdensities_mw_vs_nihao}c. Panels a) and c) trace median young star Fe and gas O abundances, respectively. Panels b) and d) plot the metallicity residuals of stars and gas, respectively, when correcting with a radial metallicity gradient fit. Black and white contour lines in each panel trace overdensities in the gas distribution of Fig.~\ref{fig:overdensities_mw_vs_nihao}c).}
    \label{fig:nihao_gas_stars_density_overlay_ao}
\end{figure*}

These stellar voids pose the question if they are also void of gas thus suggesting the gas has shifted to the more enhanced regions. In Fig.~\ref{fig:tracing_young_stars_and_gas_in_angles} we are thus tracing both the spatial distribution of gas as colored density distribution and stars as grey contour lines and gas. We find that while stars and gas trace each other in the vertical direction, we do not always see a match between the two tracers in the radial direction. In particular, we do find a significant amount of gas around the stellar void of $\varphi_\mathrm{Gal.} \sim 180^\circ$ and ${R_\mathrm{Gal.} \sim 8-11\,\mathrm{kpc}}$ in Figs.~\ref{fig:tracing_young_stars_and_gas_in_angles}e and \ref{fig:tracing_young_stars_and_gas_in_angles}f. This gas seems to be more tightly concentrated though for example in the tight wave around $R_\mathrm{Gal.} \sim 8-11\,\mathrm{kpc}$ in Fig.~\ref{fig:tracing_young_stars_and_gas_in_angles}e. We also note that significant gas overdensities, for example around $\varphi_\mathrm{Gal.} \sim 0-45^\circ$ and $R_\mathrm{Gal.} \sim 7\,\mathrm{kpc}$ in Fig.~\ref{fig:tracing_young_stars_and_gas_in_angles}a do not seem to correlate with significant overenhancement in iron abundance (compare to Fig.~\ref{fig:radial_metallicity_gradients_mw_in_angles}a). While we see a hint of a coinciding deviation of $\Delta\mathrm{[Fe/H]}$ for larger deviations from the galactic plane $\Delta z$ in the upturning outer region of Fig.~\ref{fig:tracing_young_stars_and_gas_in_angles}f, this does not seem to be the case for the downturning outer region of Figs.~\ref{fig:tracing_young_stars_and_gas_in_angles}b and \ref{fig:tracing_young_stars_and_gas_in_angles}i.

As the edge-on projection is not providing conclusive insights, we are now looking into the phase-on projection in Fig.~\ref{fig:overdensities_mw_vs_nihao}. We have chosen a region of the simulated galaxy whose gas density at solar radius (Fig.~\ref{fig:overdensities_mw_vs_nihao}c) matches with the recently measured distribution of young stars in the Milky Way at face value (Fig.~\ref{fig:overdensities_mw_vs_nihao}a) by \citet{Poggio2021}. Comparing simulated gas and observed young stars is preferable in this case, as the density of simulated stars is too low to easily identify overdensities (Fig.~\ref{fig:overdensities_mw_vs_nihao}b). The region and its gas spiral structures appear to be representative, as these structures exist throughout the whole galaxy (see Fig.~\ref{fig:stars_and_gas_overview}d).

In the different panels of Fig.~\ref{fig:nihao_gas_stars_density_overlay_ao}, we thus show this representative region of the galaxy, but color each spatial bin by stellar metallicity (Fig.~\ref{fig:nihao_gas_stars_density_overlay_ao}a), the deviation from the global linear trend (Fig.~\ref{fig:nihao_gas_stars_density_overlay_ao}b) as well as the gas metallicity (Fig.~\ref{fig:nihao_gas_stars_density_overlay_ao}c) and its deviation from the global linear trend (Fig.~\ref{fig:nihao_gas_stars_density_overlay_ao}d). In all cases, we also overlay the density contours of the significant gas overdensities (red regions in Fig.~\ref{fig:overdensities_mw_vs_nihao}c). As expected, we see that the metallicity color map of the stars in Fig.~\ref{fig:nihao_gas_stars_density_overlay_ao}a shows a decreasing trend from right to left (inner to outer galaxy) and an increasing scatter (more blue and red points towards the left) in the residual plot of Fig.~\ref{fig:nihao_gas_stars_density_overlay_ao}b. We cannot identify a strong correlation between gas overdensities and stellar metallicity or residuals in either plot - possibly caused by the low number density. In Figs.~\ref{fig:nihao_gas_stars_density_overlay_ao}c and \ref{fig:nihao_gas_stars_density_overlay_ao}d, however, the radial gas metallicity gradient shows significant local variations, that is, a trend from left to right that is not very smooth. In particular, we find significant deviations of up to $+0.15\,\mathrm{dex}$ in [Fe/H] behind the outer gas spiral (lower left of Fig.~\ref{fig:nihao_gas_stars_density_overlay_ao}d) and $-0.1\,\mathrm{dex}$ in [Fe/H] in front of the inner gas spiral (upper center of Fig.~\ref{fig:nihao_gas_stars_density_overlay_ao}d) with a steep edge consistent with the gas spiral edge. We have identified the same patterns in both [Fe/H] and A(O) as both elements trace each other rather well in the simulation of \adjusted{young stars and gas (see App.~\ref{sec:app_a} and its Fig.~\ref{fig:fe_h_vs_a_o_gas})}.

\begin{figure}
    \centering
    \includegraphics[width=\columnwidth]{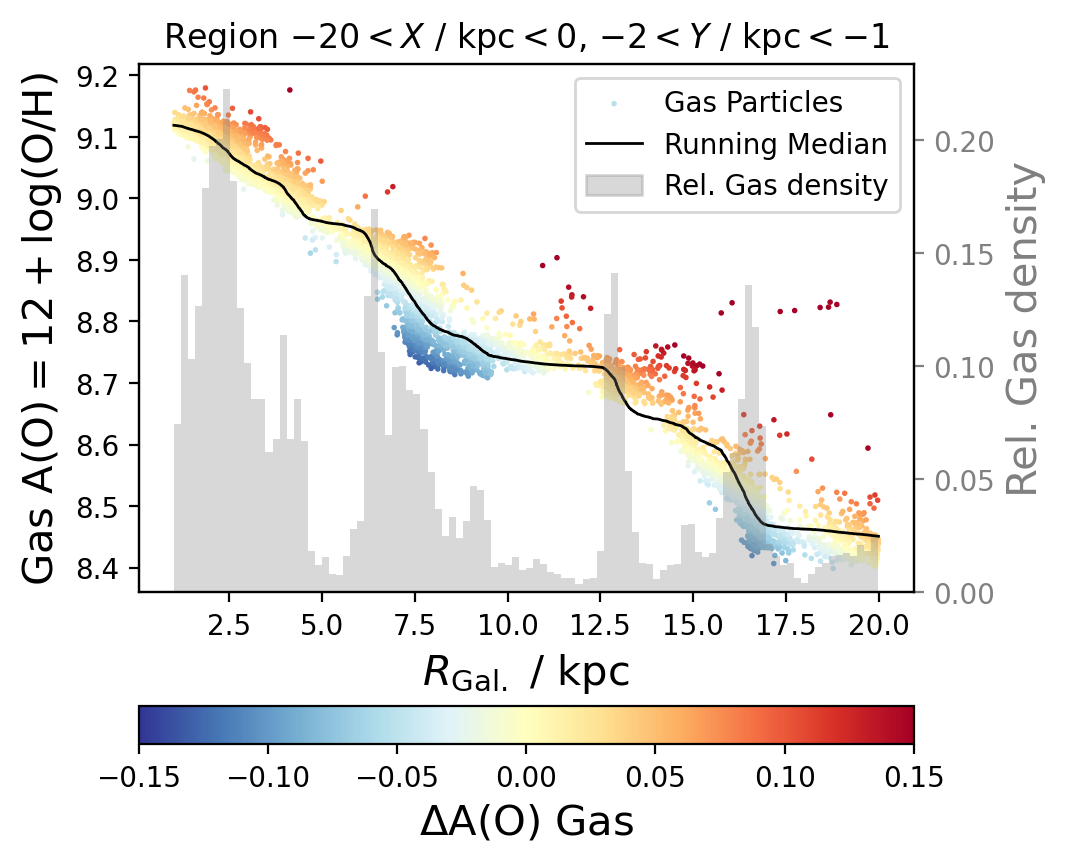}
    \caption{Radial gas metallicity gradient of a slit-like region ($-2 < Y_\mathrm{Gal.}~/~\mathrm{kpc} < -1$) from Fig.~\ref{fig:nihao_gas_stars_density_overlay_ao}. The plot extends towards larger and smaller radii and shows the step-like distribution of individual gas particle metallicities colored by their deviation from a global fit. A running median along 1000 particles is shown as a black line. The gray histogram indicates the gas density along the radius with prominent overdensities coinciding with step edges.}
    \label{fig:region_r_ao_gas_density}
\end{figure}

Tentatively, we even see a slight enhancement of A(O) at the trailing edge of the inner spiral arm (top of Fig.~\ref{fig:nihao_gas_stars_density_overlay_ao}d). We convince ourselves of the step-like behaviour by selecting a small slit-like region of $\varphi_\mathrm{Gal.} \sim 0^\circ$ and $-2 < Y_\mathrm{Gal.}~/~\mathrm{kpc} < -1$ and tracing the gas metallicity and gas density as a function of radius in Fig.~\ref{fig:region_r_ao_gas_density}. We indeed find steps and confirm that they coincide with the location of significant gas overdensities. These step-patterns have also been found by \citet{Grand2015} \adjusted{and \citet{Orr2023} in other simulations} and observationally by \citet{Ho2017c}. \adjusted{While a simplified enrichment model by \citet{Ho2017c} is able to explain oxygen abundance difference on the order of $0.08\,\mathrm{dex}$ as also seen in observations -- similar to those in Fig.~\ref{fig:region_r_ao_gas_density}, we note that the used oxygen yields are notoriously uncertain \citep[see][and references therein]{Vincenzo2016a}, and we thus focus more on the relative changes.} In Fig.~\ref{fig:region_r_ao_gas_density}, we note an extended flat region just beyond $R_\mathrm{Gal.} > 10\,\mathrm{kpc}$, the best fitting $R_\mathrm{break}$ of an assumed piecewise linear fit.

Our analyses suggest that the correlation of void and overdensities with chemical enrichment of gas and young stars is more complicated and should better be followed up by tracing these structures over simulation look-back time in a dedicated follow-up analysis to unravel the physical mechanisms of star formation feedback cycles. This could also involve the tracing of star formation bursts and disk instabilities \citep{Sanchez2014, SanchezBlazquez2014, Ho2015} as well as tracing how much self-enrichment as well as mixing and dilution takes place around the gas spirals \citep{Ho2017c}. \adjusted{While we have confirmed that our results are mainly driven by large amounts of cold gas around $100-1000\,\mathrm{K}$ as well as hot ionized gas below $20\,000\,\mathrm{K}$, we note that smaller amounts of hot gas above $20\,000\,\mathrm{K}$ shows similar trends, but larger abundance spreads that could -- once this gas cools down -- translate into larger abundance spreads of later generations of stars (see App.~\ref{sec:app_b} and its Fig.~\ref{fig:gas_temperature_tracing_r_ao}).} Rather than going back in simulation time, the present simulation data of a single snapshot in time already allows us to look back in terms of stellar lifetime -- similar to Milky Way studies, as we discuss subsequently.

\begin{figure}
    \centering
    \includegraphics[width=\columnwidth]{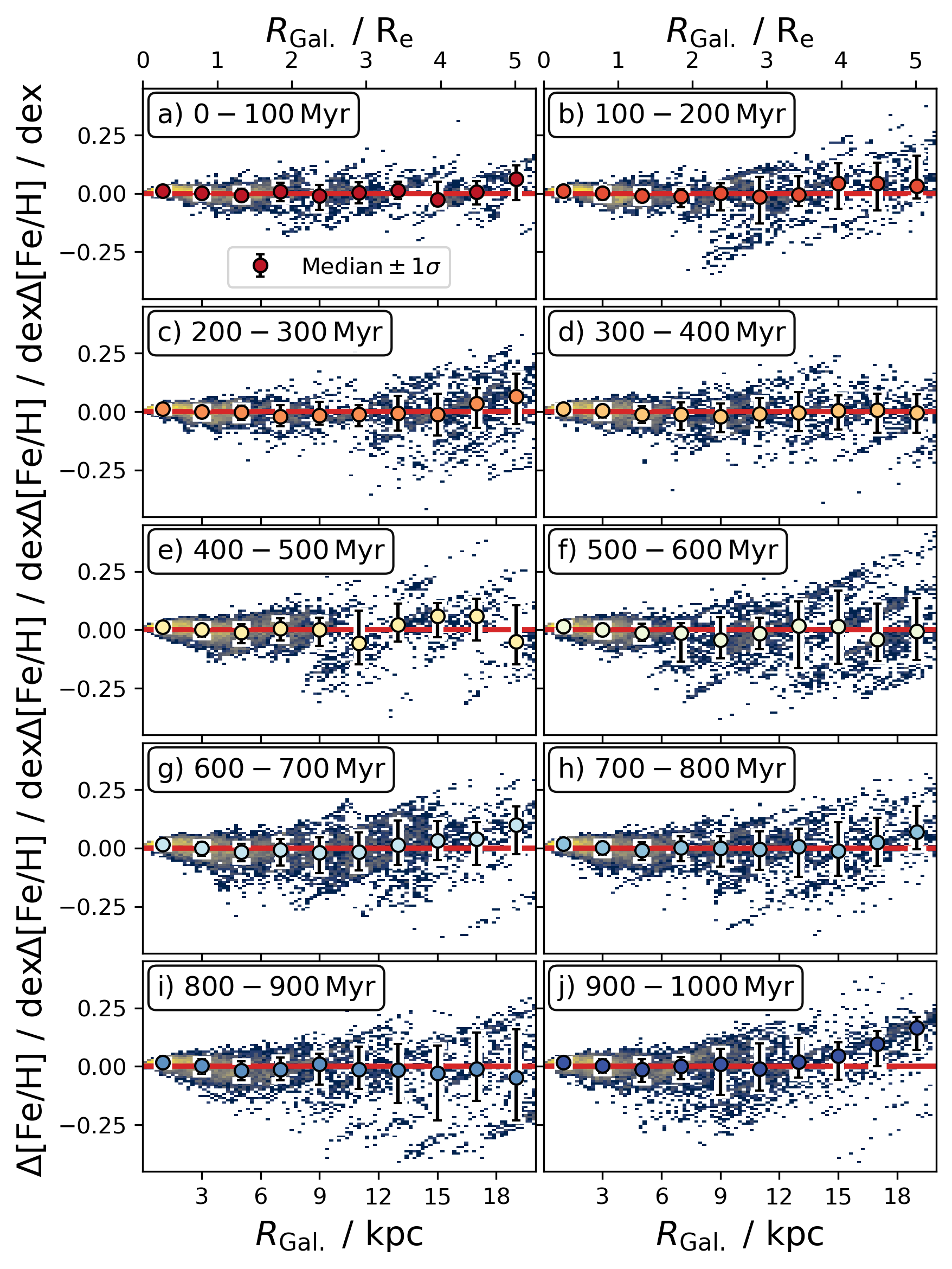}
    \caption{Stellar density distribution and spread of [Fe/H] across different galactocentric radii with respect to a global linear radial metallicity gradient across different age ranges. Panels a-j) show young stars and exhibit a rather similar trend, whereas the scatter increases significantly for stars above \nihaoAGEmax\ in panels k) and l).}
    \label{fig:scatter_with_increasing_age}
\end{figure}

\subsection{The impact of time and age: mixing and migration} \label{sec:discussion_time}

\begin{figure}
    \centering
    \includegraphics[width=\columnwidth]{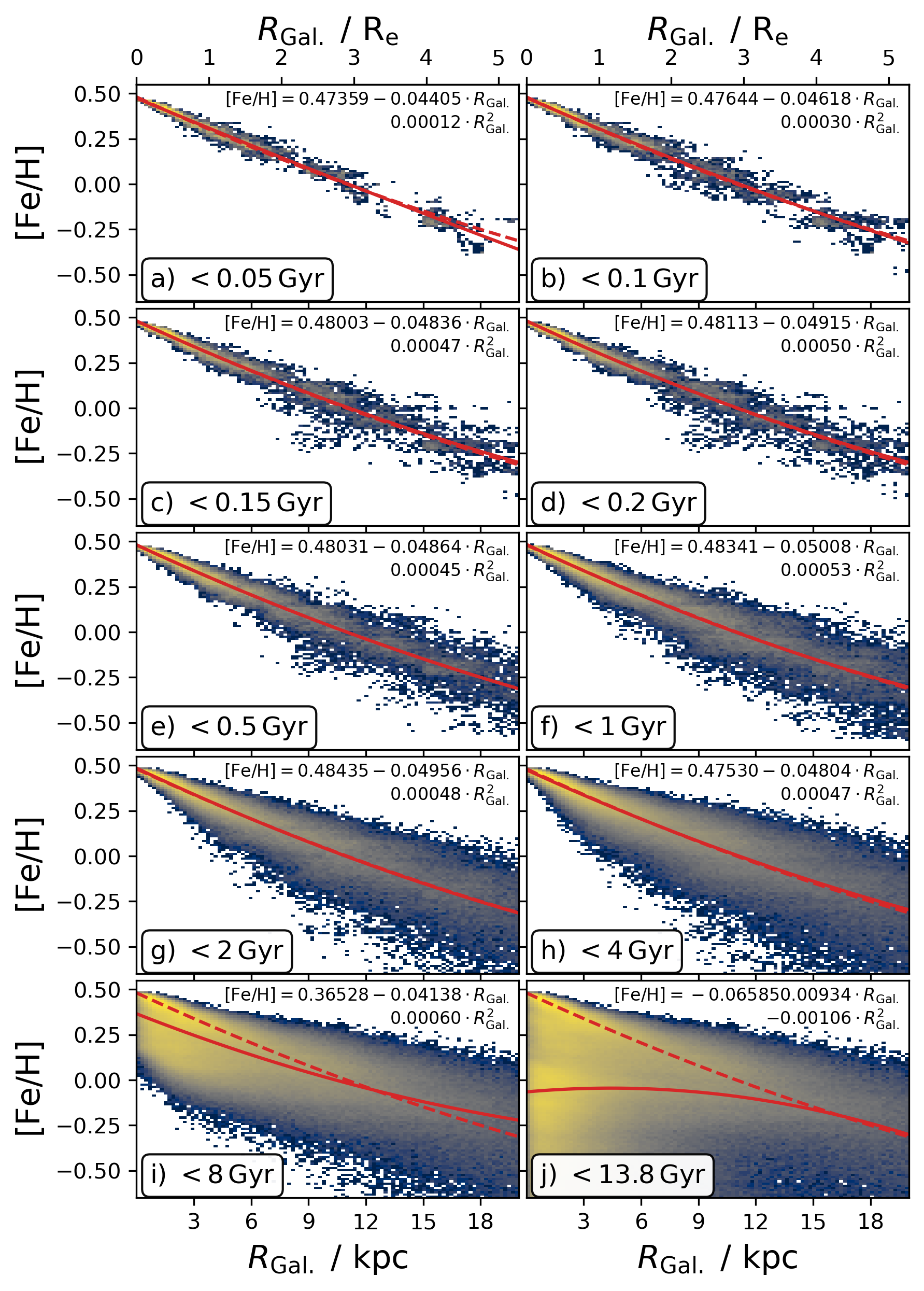}
    \caption{Radial metallicity gradients and quadratic fits for different maximum age ranges. The quadratic fit to stars below $0.5\,\mathrm{Gyr}$ is shown as a dashed red line for reference and the quadratic fit to each shown distribution is overlaid as a solid red line with the functional form given as inset text. At $R_\mathrm{Gal.}=8.21\,\mathrm{kpc}$, spread increases from $\sigma \mathrm{[Fe/H]} = 0.05$ for youngest stars to 0.09 and 0.11 for stars below 4 and $8\,\mathrm{Gyr}$, respectively.}
    \label{fig:quadratic_fit_across_maximum_ages}
\end{figure}

Although we have chosen a rather small stellar age window of \nihaoAGEmax\ to trace the radial metallicity gradient without the expected significant impact of radial mixing and migration, we are testing and discussing this particular choice in this section in two ways. Firstly, we test the deviation of the radial metallicity gradient from the same global shape as well as the abundance spread across smaller age bins of $100\,\mathrm{Myr}$ between $0-1000\,\mathrm{Myr}$ in Fig.~\ref{fig:scatter_with_increasing_age}. Secondly, we trace the distribution of stellar metallicity across galactic radii for increasing age bins from $50\,\mathrm{Myr}$ up to the maximum stellar age of $13.8\,\mathrm{Gyr}$ in Fig.~\ref{fig:quadratic_fit_across_maximum_ages}.

Our first test in Fig.~\ref{fig:scatter_with_increasing_age} shows that the deviation from a global trend remains similar in functional form. We find that the spread of iron abundance does indeed scatter significantly, but the distributions stay within the same overall shape across the ten age bins. We note though, that the smallest age bin of $0-100\,\mathrm{Myr}$ shows the least abundance scatter.

This is consistent with the picture from our second test of increasing age ranges in Fig.~\ref{fig:quadratic_fit_across_maximum_ages}. Here we find the first significant deviation from a tighter and already slightly quadratic relation for an age of $100-150\,\mathrm{Myr}$ in Fig.~\ref{fig:quadratic_fit_across_maximum_ages}c - our previously identified Group 3. As expected from previous simulations and observations, we see an increase in the scatter as we include more and more older stars. We note a still similar albeit more scattered shape for stars below $4\,\mathrm{Gyr}$ in Fig.~\ref{fig:quadratic_fit_across_maximum_ages}h, before we start to see a more metal-poor population of stars in the inner galaxy appear between $4-8\,\mathrm{Gyr}$ in Fig.~\ref{fig:quadratic_fit_across_maximum_ages}i. These also begin to significantly impact the quadratic fit to the radial metallicity distribution, shown as a solid red line, in contrast to our reference fit, represented by a dashed red line. The significant amount of metal-poor stars in the inner galaxy then completely tilts the distribution when also including stars between $8-13.8\,\mathrm{Gyr}$ in Fig.~\ref{fig:quadratic_fit_across_maximum_ages}j \citep[see also][]{Johnson2024}. Similar to the Milky Way \citep{BlandHawthorn_Gerhard2016}, these oldest stars are those of the relatively more metal-poor thick disk that are confined to the inner disk with a shorter scale length. \adjusted{Including such old stars in metallicity gradient studies could significantly bias conclusions.}

While we cannot exclude radial migration playing a role for change of radius for the youngest stars of the simulation, since \citet{Frankel2018} predicted significant shifts even for ages below $0.5\,\mathrm{Gyr}$ (see their Fig.~10), the larger scatter for older stars is certainly suggesting a larger (re-)distribution of stars along the radial axis, as found in previous simulations \citep{Minchev2010, Grand2015}.

\subsection{Implications for Milky Way studies} \label{sec:implications_milky_way}

Our analysis of the radial metallicity gradient in a simulated NIHAO-UHD galaxy offers several insights that are directly applicable to understanding the Milky Way's gradient.

First, the nature of the gradient -- whether it is linear or better described by more complex functional forms -- remains a critical question. Previous studies, such as those by \citet{Lepine2011} and \citet{Donor2020}, have suggested the potential for a break radius, possibly at the corotation radius or further out \citep{Scarano2013}, which could indicate two distinct linear regimes. In our analysis, we find evidence that the gradient is at least is not purely linear, but could also be smoothly flattening. Applying a smooth quadratic function on observational data \citep{Yong2012, Andrievsky2004, Genovali2014}, might provide a better or at least consistent fit for the Milky Way data without the need for a break radius. However, even this may not fully capture the nuances observed in our simulations. Chemical evolution models propose a more sophisticated behaviour \citep[e.g.][]{Chiappini2001, Kubryk2015, Palla2024}, reflecting varying influences of galactic processes at different radii. Understanding this structure in the simulated galaxy provides a framework for interpreting similar complexities in the Milky Way.

Given these complexities, it is also essential to consider how local sampling biases might affect our understanding of the Milky Way's metallicity gradient. For instance, incomplete samples that omit low [Fe/H] clusters or stars could skew gradient estimates, as suggested by our comparisons in Figures~\ref{fig:radial_metallicity_gradients_mw_vs_nihao}a and \ref{fig:radial_metallicity_gradients_mw_vs_nihao}b. Our results indicate that young clusters with lower (or higher) [Fe/H] than expected at a given radius could indicate the previous presence of a spiral arm (see our identified Groups in Figs.~\ref{fig:radial_metallicity_gradients_mw_in_angles_age} and \ref{fig:phi_angle_R_follow_up}). Furthermore, we caution that localised effects -- both intrinsic and in terms of selection function -- could also mimic non-linear shapes and more spatial coverage is needed in the Milky Way. Our results also indicate that older clusters, which have been found more frequently at larger distances than young clusters - are likely influenced by radial migration - and thus complicate the interpretation of these radial metallicity gradients \citep{Magrini2009, Lepine2011}.

Cosmological zoom-in simulations like NIHAO-UHD are approaching the resolution needed to examine regions analogous to the solar vicinity, though the star particle numbers and mass resolution remain a limiting factor. Nonetheless, we observe distinct patterns in the distribution of young stars and gas, including lower [Fe/H] and A(O) in the leading edges of gas overdensities and higher [Fe/H] and A(O) in the trailing edges, consistent with findings by \citet{Grand2016}, \citet{Ho2017c}, and \citet{Kreckel2019}. These trends suggest that local metallicity variations, driven by gas dynamics and bulk motions \citep[see e.g.][]{Orr2023}, may also play a significant role in shaping the observed gradients in the Milky Way.

Additionally, our study hints at the potential for more nuanced variations in [Fe/H] across different regions of the galaxy. In particular, the gas shows a step-like behavior of A(O) and [Fe/H] changes around the edges of gas overdensities (Fig.~\ref{fig:region_r_ao_gas_density}), with significant deviations from the global gradient in specific regions. We have also found a larger stellar void around $-12 < R_\mathrm{Gal} < -10\mathrm{kpc}$. Although further investigation is needed, these findings could have important implications for understanding localized star formation events and their impact on the overall metallicity distribution in the Milky Way \citep{Sanchez2014, SanchezBlazquez2014, Ho2015}. It will certainly be exciting to see how much more insights \citep{Poggio2021, Hackshaw2024} we will get from the more extended data of future data releases of \textit{Gaia} and spectroscopic surveys.

We cannot directly link spiral arms to bar resonances or bar-driven mixing in our simulation, because of a negligible bar strength in our galaxy\footnote{The second Fourier component of the density distribution has an amplitude of only 0.02.} \citep[but see][]{Minchev2010, DiMatteo2013}. However, the influence of a galactic bar on the spiral arms and, by extension, on the radial metallicity gradient, remains a possibility \citep[see again][]{Chen2023}. Disk instabilities and warps might further complicate the interpretation of these gradients and progress will likely rely on the detailed disentangling of these effects from both cosmological simulations as well as idealised simulations and models \citep{Minchev2013, Grand2015, Grand2016, Krumholz2018, Sharda2021, BlandHawthorn2024, TepperGarcia2024}.

\subsection{Implications for extragalactic studies} \label{sec:implications_extragalactic}

The insights gained from our analysis of the radial metallicity gradient in a simulated NIHAO-UHD galaxy extend beyond the Milky Way, offering valuable implications for the study of extragalactic systems.

One key observation is that deviations from a purely linear metallicity gradient, as seen in our Milky Way analogue, are common in other galaxies as well. When fitting a piecewise linear fit to our data, we found a break radius \adjusted{at $R_\mathrm{Gal.} = 9.3-11.5\,\mathrm{kpc}$}. Converted to effective radii $\mathrm{R_e}$ or radii $\mathrm{R_{25}}$ covering the $25\,\mathrm{mag\,arcsec^{-2}}$ isophote\footnote{We assume $\mathrm{R_{25}} = 3.6\,\mathrm{R_e}$ \citep{Williams2009, Chen2023}.}, this corresponds \adjusted{to $R_\mathrm{break} \sim 2.4-3.0\,\mathrm{R_e} \equiv 0.7-0.8\,\mathrm{R_{25}}$ for our simulation}. This would be consistent with the observational results by \citet{Sanchez2014} who found that breaks in metallicity gradients are common in both spiral and barred galaxies, with flattening of the abundance being evident beyond $\sim 2\,\mathrm{R_e}$ \citep[compare also to][]{Belfiore2017}. Similar to our suggestion for Milky Way studies, we suggest to also test a smooth function, such as a quadratic one, on extragalactic observational data \citep[e.g.][]{Bresolin2012, Chen2023} to test the preference of a distinct break radius.

\begin{figure*}
    \centering
    \includegraphics[width=\textwidth]{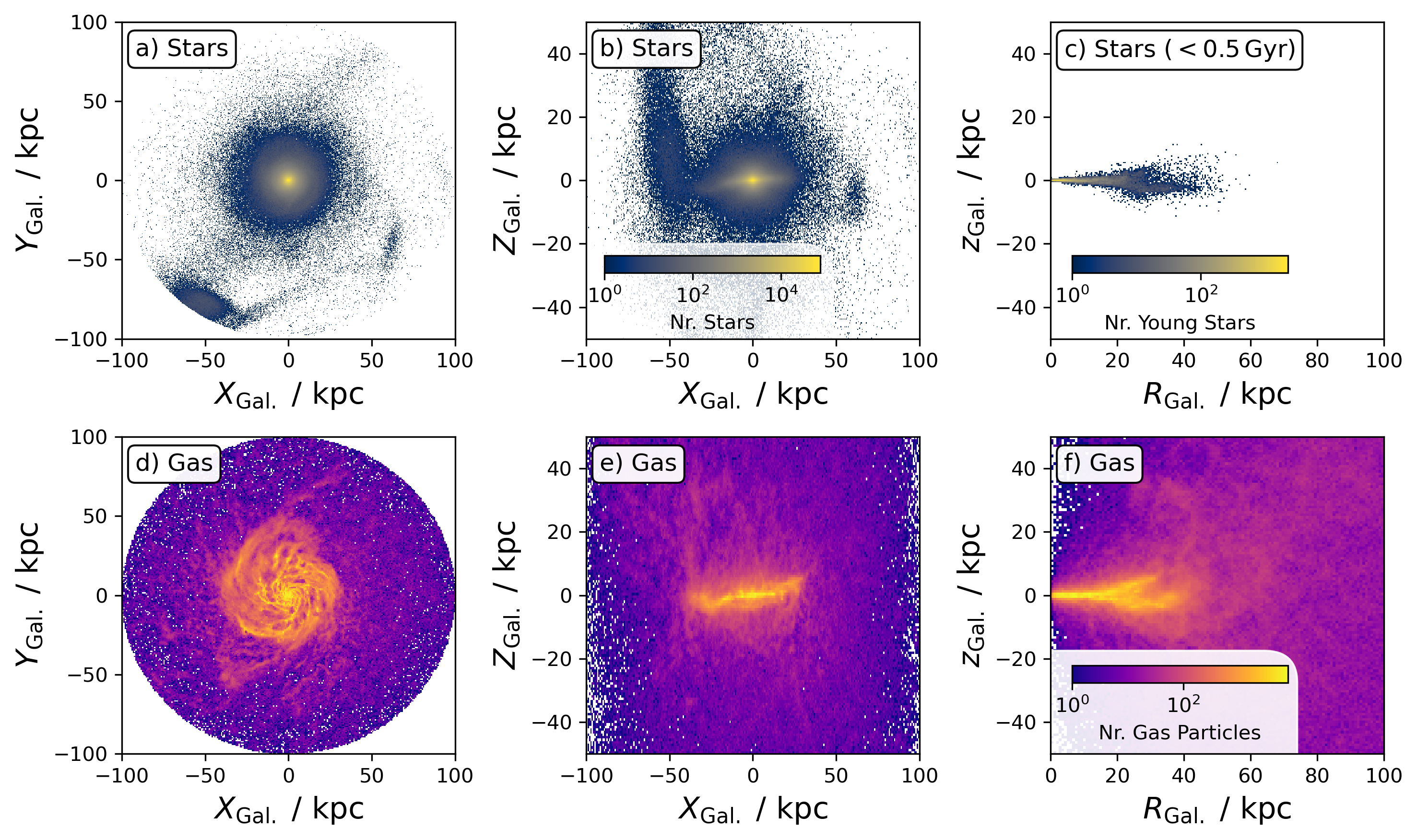}
    \caption{Same as Fig.~\ref{fig:stars_and_gas_overview}, but for an extended $R_\mathrm{Gal.} \leq 100\,\mathrm{kpc}$ and $\vert z_\mathrm{Gal.} \vert \leq 50\,\mathrm{kpc}$.}
    \label{fig:stars_and_gas_overview_100kpc}
\end{figure*}

\begin{figure}
    \centering
    \includegraphics[width=0.95\columnwidth]{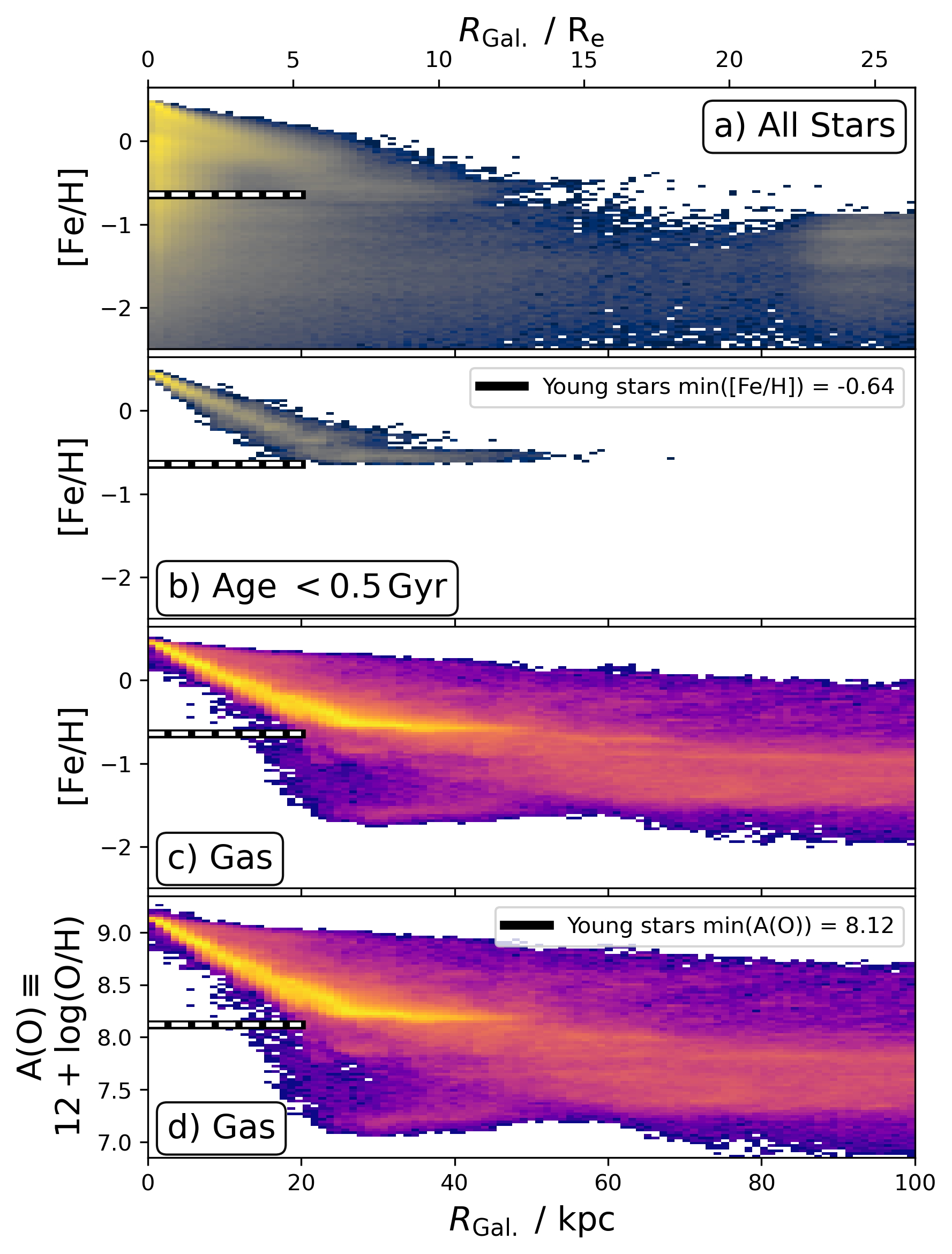}
    \caption{Radial metallicity functions for all stars (panel a), young stars (panel b), and gas (panels c and d for iron and oxygen as metallicity tracers) out to $R_\mathrm{Gal.} \leq 100\,\mathrm{kpc}$. Panels b and c are comparable to Figs.~\ref{fig:global_r_feh_fit}a and \ref{fig:global_r_feh_fit_gas}a for a smaller radial coverage.}
    \label{fig:trace_stars_and_gas_100kpc}
\end{figure}

Although the focus of this research lies on the observable region of the Milky Way ($R_\mathrm{Gal.} < 20\,\mathrm{kpc}$) and most other galaxies ($R_\mathrm{Gal.} < 2.5\,\mathrm{R_e}$), the finding of significant gradient changes in the outskirts of galaxies by \citet{Garcia2023}, suggests to also test this region of our Milky Way analogue. \citet[][see their Fig.~4]{Garcia2023} found a metallicity floor in IllustrisTNG galaxies. When using their sample to identify a metallicity floor radius for a Milky Way mass galaxy with $\log(M_\star/\mathrm{M_\odot}) = 10.7$ \citep{BlandHawthorn_Gerhard2016} at redshift $z \sim 0$, we would expect to find it around $25-30\,\mathrm{kpc}$. We therefore extend the analysed radius to $R_\mathrm{Gal.} \leq 100\,\mathrm{kpc}$ (see Fig.~\ref{fig:stars_and_gas_overview_100kpc}) and indeed find a similar abundance floor of $\mathrm{[Fe/H]} \geq -0.64$ for young stars and $\mathrm{A(O)} \geq 8.12$ for the majority of gas (see Fig.~\ref{fig:trace_stars_and_gas_100kpc} at a similar radius. We note that another galaxy without gas in this figure is a sufficiently large distance of $92\,\mathrm{kpc}$, that is, $(Y,Y,Z) = (-50,-75,20)\,\mathrm{kpc}$.

These lowest abundances remind us of two observational results. Firstly, the iron abundance floor is consistent with the lower end of the Milky Way thin - and coincidentally outer - disk of ${\mathrm{[Fe/H]} \sim -0.7}$ \citep{Bensby2014, Buder2019}. Secondly this oxygen abundance floor is consistent with the results by \citet{Grasha2022} from TYPHOON galaxy observations. \citet{Grasha2022} suggested this could be caused by changes in the ratio of supernovae II and AGB reflected by a changing ratio of nitrogen to oxygen abundance N/O which also flattens towards a lower plateau below metallicities of ${\mathrm{A(O)} \sim 8.0}$ \citep{Nicholls2017}. While we cannot follow this observation up with the present simulation, a similar simulation used by \citet{Buder2024} has traced the relative contribution of both supernovae II and AGB and should be used to test this hypothesis in the future.

It is important to note that the chemical evolution model in the NIHAO-UHD simulations is constrained by the current, incomplete understanding of evolutionary pathways and yields \citep{Buck2021}, as well as by limitations in resolution and the imperfect physics inherent to cosmological zoom-in simulations \citep{Buck2020}.
Both could contribute to the identified differences in absolute and relative abundances across different scales - including a different scatter of abundances for example of the gas phase metallicity between NIHAO-UHD of up to $0.1\,\mathrm{dex}$ and the low scatter of $0.03-0.05\,\mathrm{dex}$ (and even lower on local scales) found by PHANGS-MUSE face-on observations \citep{Kreckel2020}. Extending our analysis to other simulations and further improving the resolution and physics of the simulations will be key in uniting the observational and theoretical insights into galactic chemical evolution on small and large scales. \adjusted{Furthermore, it would be worth to further investigate the highly enriched gas (top-right of Fig.~\ref{fig:trace_stars_and_gas_100kpc}d). An initial investigation confirms that this gas is not only hotter, but also less dense than the typical gas in the plane and shows typically larger vertical velocities -- consistent with outflows. Most excitingly, a significant fraction of this enhanced gas can be traced back to confined regions around $x_\mathrm{Gal.}\sim30\,\mathrm{kpc}$, $-30 < y_\mathrm{Gal.}~/~\mathrm{kpc} < 15$ and $z_\mathrm{Gal.}\sim-5\,\mathrm{kpc}$, where several supernovae went off recently.}

Similar to more resolved and higher quality observations in the Milky Way, we also expect more, better, and diverse face-on and edge-on observations and analyses across a range of wavelengths by the PHANGS and GECKOS teams \citep{Kreckel2019, Kreckel2020, GECKOS2023} as well as the SDSS-V and MAGPI collaborations \citep{Kollmeier2017, MAGPI2021, Mun2024, Chen2024}, among many other ongoing efforts.

%%%%%%%%%%%%%%%%%%%%%%%%%%%%%%%%%%%%%%%%%%%%%%%%%%
%%%%%%%%%%%%%%%%%%%%%%%%%%%%%%%%%%%%%%%%%%%%%%%%%%
\section{Conclusions}
\label{sec:conc}
%%%%%%%%%%%%%%%%%%%%%%%%%%%%%%%%%%%%%%%%%%%%%%%%%%
%%%%%%%%%%%%%%%%%%%%%%%%%%%%%%%%%%%%%%%%%%%%%%%%%%

To conclude our study, we first iterate the main take-away of our research in Section~\ref{sec:take_away} before giving suggestions for future research in Section~\ref{sec:future_research}.

% \newpage
\subsection{Take-Away} \label{sec:take_away}

We have analysed the radial metallicity distribution of young stars and gas in the inner \nihaoRmax\ of a NIHAO-UHD Milky Way analogue (Fig.~\ref{fig:stars_and_gas_overview}), finding a predominantly linear decrease (Figs.~\ref{fig:stars_and_gas_2d_view} and \ref{fig:global_r_feh_fit}). Although our analysis of a single spiral galaxy simulation has limited applicability to the entire population of diverse spiral galaxies, it reveals several intriguing findings about the shape and local metallicity variations. The results we find in this work hold relevance for both the Milky Way and extragalactic research communities:

\begin{itemize}[left=0.3cm]
    \item Looking into the shape in detail, we find that \adjusted{a linear function can reproduce the trend of the innermost $R_\mathrm{gal} < 10\,\mathrm{kpc}$}, but piecewise linear and quadratic functions both perform better than a linear fit to the radial metallicity relation \adjusted{out to $R_\mathrm{gal} < 20\,\mathrm{kpc}$}. However, we see no significant preference between piecewise and quadratic functions based on our assessments (Fig.~\ref{fig:linear_quadratic_piecewise}). While the specific slopes differ when fitting all points or binned data, they agree within the still rather small fitting uncertainties.
    \item We find that a piecewise linear function can effectively approximate a quadratic function across scales commonly applied in Milky Way and extragalactic studies. Local deviations become traceable below a spatial resolution of $\Delta R_\mathrm{Gal.} \leq 2\,\mathrm{kpc}$ (Fig.~\ref{fig:radial_range_impact}).
    \item We see no strong correlations of deviations in the vertical direction across the whole simulation (Fig.~\ref{fig:overlap_local_variation_gas}). Such correlations could, however, be blurred by azimuthal effects, like the galactic warp, which needs to be disentangled in the azimuthal dimension. 
    \item We find various deviations from the global trend in azimuthal direction, including gaps as well as isolated streaks of stars with similar [Fe/H] across $\Delta R_\mathrm{Gal.} = 2-6\,\mathrm{kpc}$ (Fig.~\ref{fig:radial_metallicity_gradients_mw_in_angles}). These can introduce significant local over-/under-enhancement of up to $\pm 0.2\,\mathrm{dex}$ in [Fe/H] at a given radius.
    \item We find significant scatter across the radial metallicity distribution caused by streaks of stars born with similar [Fe/H] at similar times and similar but slightly extended regions of the galaxy (Figs.~\ref{fig:radial_metallicity_gradients_mw_in_angles_age} and \ref{fig:phi_angle_R_follow_up}).
    \item Our results imply the need for more careful consideration of local intrinsic effects and selection effects on radial metallicity gradient and scatter studies in the Milky Way (Fig.~\ref{fig:radial_metallicity_gradients_mw_vs_nihao}).
    \item Expanding our work to the gas phase metallicity gradient (Fig.~\ref{fig:global_r_feh_fit_gas}), we perform a preliminary comparison of observed and simulated young stars as well as simulated gas distribution and chemistry (Figs.~\ref{fig:tracing_young_stars_and_gas_in_angles} and \ref{fig:overdensities_mw_vs_nihao}), finding significant step-like changes in the gas chemistry at the leading and trailing edges of gas spirals, with lower and higher enhancement respectively (Figs.~\ref{fig:nihao_gas_stars_density_overlay_ao} and \ref{fig:region_r_ao_gas_density}).
    \item We have further identified that the abundance scatter, which increases towards larger radii, is as large as $0.1\,\mathrm{dex}$ and already present at the youngest ages of $100\,\mathrm{Myr}$ (Fig.~\ref{fig:scatter_with_increasing_age}). While not the focus of our analysis, we have also confirmed that the scatter significantly increases towards larger ages (Fig.~\ref{fig:quadratic_fit_across_maximum_ages}).
    \item We have discussed the implications of our findings for studies of the Milky Way (Section~\ref{sec:implications_milky_way}) as well as external galaxies (Section.~\ref{sec:implications_extragalactic}). Here, we firstly suggest to explore the spread of abundances across different radii in more detail. Secondly, we suggest approaching the fitting of gradients in external galaxies in a more agnostic way to the shape. This will be particularly interesting when we can observe the outermost regions of galaxies, where simulations predict an abundance floor (Figs.~\ref{fig:stars_and_gas_overview_100kpc} and \ref{fig:trace_stars_and_gas_100kpc}).
\end{itemize}

\subsection{Future Research} \label{sec:future_research}

\adjusted{This study focused on a present-day snapshot of the NIHAO-UHD Milky Way analogue, akin to observations possible in our local Universe. Since the simulation traces particles and gas over time, a natural next step is to examine the temporal evolution and coherence of spatial and chemical over- and underdensities across different elements \citep[see also][]{Zhang2025}. This would enable tracking abundance changes in the leading and trailing edges of spiral arms and follow their mixing and blurring over time. Establishing links to physical mechanisms and quantifying their roles requires further study. Upcoming datasets with broader coverage and improved stellar measurements \citep[e.g.][]{Barbillon2025} will also allow us to move beyond one-dimensional gradient analyses and better model local variations.}

\adjusted{While some elemental abundance trends are broadly consistent with observations \citep{Buck2021, Buder2024}, uncertainties in enrichment sites, yields, and environments limit the predictive power of absolute abundances. Because of these limitations, and the simulation's mass resolution constraints, we refrain from detailed comparisons to the actual Milky Way. Variations may also arise from different formation pathways, such as merger histories \citep{Buck2023, Buder2024}. For example, our simulated galaxy exhibits a weak bar and a strong bulge ($B/T = 0.48$ when selecting $j_z/j_c < 0.5$ as bulge stars versus $j_z/j_c > 0.7$ as disk stars via actions $j$), similar to values found by \citet{Obreja2019} in a lower-resolution run of this simulation. These structural features align with results from \citet{Chen2023}, who show that bar strength affects whether radial metallicity gradients change smoothly or feature distinct break radii. Motivated by the bar analysis of \citet{Tuntipong2024}, future studies should investigate a larger sample of simulated galaxies to assess the role of bars in shaping gradient profiles. Expanding our work to other simulations, such as the \textsc{VINTERGATAN} suite \citep{Renaud2025}, and comparing to galaxies with varying mass, bar strength, formation history, or environment will help quantify their impact on metallicity gradients and disentangle the influence of different enrichment processes on galactic chemical evolution.}

% \newpage
%%%%%%%%%%%%%%%%%%%%%%%%%%%%%%%%%%%%%%%%%%%%%%%%%%
%%%%%%%%%%%%%%%%%%%%%%%%%%%%%%%%%%%%%%%%%%%%%%%%%%
\section*{Software}

The research for this publication was coded in \textsc{python} (version 3.7.4) and included its packages
\textsc{astropy} \citep[v. 3.2.2;][]{Robitaille2013,PriceWhelan2018},
\textsc{IPython} \citep[v. 7.8.0;][]{ipython},
\textsc{matplotlib} \citep[v. 3.1.3;][]{matplotlib},
\textsc{NumPy} \citep[v. 1.17.2;][]{numpy},
\textsc{pynbody} \citep[v. 1.1.0;][]{pynbody},
\textsc{scipy} \citep[v. 1.3.1;][]{Scipy},
\textsc{sklearn} \citep[v. 1.5.1][]{scikit-learn}
\textsc{statsmodels} \citep[v. 0.14.2][]{statsmodels}
We further made use of \textsc{topcat} \citep[version 4.7;][]{Taylor2005};

%%%%%%%%%%%%%%%%%%%%%%%%%%%%%%%%%%%%%%%%%%%%%%%%%
\section*{Data Availability}

All code to reproduce the analysis and figures can be publicly accessed via \url{https://github.com/svenbuder/nihao_radial_metallicity_gradients}. The used simulationsnapshot can be accessed as FITS file via \url{https://github.com/svenbuder/preparing_NIHAO}. Original data, more snapshots and other galaxies can be found at \url{https://tobias-buck.de/#sim_data}. We encourage interested readers to get in contact with the authors for full data access and advice for use and cite \citet{Buck2020b, Buck2021}.

\section*{Acknowledgments}

We acknowledge the traditional owners of the land on which the ANU stands, the Ngunnawal and Ngambri people. We pay our respects to elders past, and present and are proud to continue their tradition of surveying the night sky and its mysteries to better understand our Universe.

This work was supported by the Australian Research Council Centre of Excellence for All Sky Astrophysics in 3 Dimensions (ASTRO 3D), through project number CE170100013. SB and KG acknowledge support from the Australian Research Council under grant numbers DE240100150 and DE220100766, respectively. TB acknowledges funding from the Carl Zeiss Stiftung and support from the European Research Council under ERC-CoG grant CRAGSMAN-646955. We gratefully acknowledge the Gauss Centre for Supercomputing e.V. (\url{www.gaus s-centre.eu}) for funding this project by providing computing time on the GCS Supercomputer SuperMUC at Leibniz Supercomputing Centre (\url{www.lrz.de}). Simulations were partially computed with High Performance Computing resources at New York University, Abu Dhabi.

\end{CJK*}

\bibliographystyle{mnras}
\bibliography{bib} % if your bibtex file is called example.bib

\begin{thebibliography}{}
\makeatletter
\relax
\def\mn@urlcharsother{\let\do\@makeother \do\$\do\&\do\#\do\^\do\_\do\%\do\~}
\def\mn@doi{\begingroup\mn@urlcharsother \@ifnextchar [ {\mn@doi@}
  {\mn@doi@[]}}
\def\mn@doi@[#1]#2{\def\@tempa{#1}\ifx\@tempa\@empty \href
  {http://dx.doi.org/#2} {doi:#2}\else \href {http://dx.doi.org/#2} {#1}\fi
  \endgroup}
\def\mn@eprint#1#2{\mn@eprint@#1:#2::\@nil}
\def\mn@eprint@arXiv#1{\href {http://arxiv.org/abs/#1} {{\tt arXiv:#1}}}
\def\mn@eprint@dblp#1{\href {http://dblp.uni-trier.de/rec/bibtex/#1.xml}
  {dblp:#1}}
\def\mn@eprint@#1:#2:#3:#4\@nil{\def\@tempa {#1}\def\@tempb {#2}\def\@tempc
  {#3}\ifx \@tempc \@empty \let \@tempc \@tempb \let \@tempb \@tempa \fi \ifx
  \@tempb \@empty \def\@tempb {arXiv}\fi \@ifundefined
  {mn@eprint@\@tempb}{\@tempb:\@tempc}{\expandafter \expandafter \csname
  mn@eprint@\@tempb\endcsname \expandafter{\@tempc}}}

\bibitem[\protect\citeauthoryear{{Agertz} et~al.,}{{Agertz}
  et~al.}{2021}]{Agertz2021}
{Agertz} O.,  et~al., 2021, \mn@doi [\mnras] {10.1093/mnras/stab322}, \href
  {https://ui.adsabs.harvard.edu/abs/2021MNRAS.503.5826A} {503, 5826}

\bibitem[\protect\citeauthoryear{{Allende Prieto}, {Beers}, {Wilhelm},
  {Newberg}, {Rockosi}, {Yanny}  \& {Lee}}{{Allende Prieto}
  et~al.}{2006}]{AllendePrieto2006}
{Allende Prieto} C.,  {Beers} T.~C.,  {Wilhelm} R.,  {Newberg} H.~J.,
  {Rockosi} C.~M.,  {Yanny} B.,   {Lee} Y.~S.,  2006, \mn@doi [\apj]
  {10.1086/498131}, \href
  {https://ui.adsabs.harvard.edu/abs/2006ApJ...636..804A} {636, 804}

\bibitem[\protect\citeauthoryear{{Anders} et~al.,}{{Anders}
  et~al.}{2014}]{Anders2014}
{Anders} F.,  et~al., 2014, \mn@doi [\aap] {10.1051/0004-6361/201323038}, \href
  {https://ui.adsabs.harvard.edu/abs/2014A&A...564A.115A} {564, A115}

\bibitem[\protect\citeauthoryear{{Anders} et~al.,}{{Anders}
  et~al.}{2017}]{Anders2017}
{Anders} F.,  et~al., 2017, \mn@doi [\aap] {10.1051/0004-6361/201629363}, \href
  {https://ui.adsabs.harvard.edu/abs/2017A&A...600A..70A} {600, A70}

\bibitem[\protect\citeauthoryear{{Andrievsky} et~al.,}{{Andrievsky}
  et~al.}{2002a}]{Andrievsky2002}
{Andrievsky} S.~M.,  et~al., 2002a, \mn@doi [\aap]
  {10.1051/0004-6361:20011488}, \href
  {https://ui.adsabs.harvard.edu/abs/2002A&A...381...32A} {381, 32}

\bibitem[\protect\citeauthoryear{{Andrievsky}, {Bersier}, {Kovtyukh}, {Luck},
  {Maciel}, {L{\'e}pine}  \& {Beletsky}}{{Andrievsky}
  et~al.}{2002b}]{Andrievsky2002b}
{Andrievsky} S.~M.,  {Bersier} D.,  {Kovtyukh} V.~V.,  {Luck} R.~E.,  {Maciel}
  W.~J.,  {L{\'e}pine} J.~R.~D.,   {Beletsky} Y.~V.,  2002b, \mn@doi [\aap]
  {10.1051/0004-6361:20020016}, \href
  {https://ui.adsabs.harvard.edu/abs/2002A&A...384..140A} {384, 140}

\bibitem[\protect\citeauthoryear{{Andrievsky}, {Luck}, {Martin}  \&
  {L{\'e}pine}}{{Andrievsky} et~al.}{2004}]{Andrievsky2004}
{Andrievsky} S.~M.,  {Luck} R.~E.,  {Martin} P.,   {L{\'e}pine} J.~R.~D.,
  2004, \mn@doi [\aap] {10.1051/0004-6361:20031528}, \href
  {https://ui.adsabs.harvard.edu/abs/2004A&A...413..159A} {413, 159}

\bibitem[\protect\citeauthoryear{{Astropy Collaboration} et~al.,}{{Astropy
  Collaboration} et~al.}{2013}]{Robitaille2013}
{Astropy Collaboration} et~al., 2013, \mn@doi [\aap]
  {10.1051/0004-6361/201322068}, \href
  {http://adsabs.harvard.edu/abs/2013A%26A...558A..33A} {558, A33}

\bibitem[\protect\citeauthoryear{{Astropy Collaboration} et~al.,}{{Astropy
  Collaboration} et~al.}{2018}]{PriceWhelan2018}
{Astropy Collaboration} et~al., 2018, \mn@doi [\aj] {10.3847/1538-3881/aabc4f},
  \href {https://ui.adsabs.harvard.edu/abs/2018AJ....156..123A} {156, 123}

\bibitem[\protect\citeauthoryear{{Barbillon}, {Recio-Blanco}, {Poggio},
  {Palicio}, {Spitoni}, {de Laverny}  \& {Cescutti}}{{Barbillon}
  et~al.}{2025}]{Barbillon2025}
{Barbillon} M.,  {Recio-Blanco} A.,  {Poggio} E.,  {Palicio} P.~A.,  {Spitoni}
  E.,  {de Laverny} P.,   {Cescutti} G.,  2025, \mn@doi [\aap]
  {10.1051/0004-6361/202450868}, \href
  {https://ui.adsabs.harvard.edu/abs/2025A&A...693A...3B} {693, A3}

\bibitem[\protect\citeauthoryear{{Belfiore} et~al.,}{{Belfiore}
  et~al.}{2016}]{Belfiore2016}
{Belfiore} F.,  et~al., 2016, \mn@doi [\mnras] {10.1093/mnras/stw1234}, \href
  {https://ui.adsabs.harvard.edu/abs/2016MNRAS.461.3111B} {461, 3111}

\bibitem[\protect\citeauthoryear{{Belfiore} et~al.,}{{Belfiore}
  et~al.}{2017}]{Belfiore2017}
{Belfiore} F.,  et~al., 2017, \mn@doi [\mnras] {10.1093/mnras/stx789}, \href
  {https://ui.adsabs.harvard.edu/abs/2017MNRAS.469..151B} {469, 151}

\bibitem[\protect\citeauthoryear{{Bellardini}, {Wetzel}, {Loebman},
  {Faucher-Gigu{\`e}re}, {Ma}  \& {Feldmann}}{{Bellardini}
  et~al.}{2021}]{Bellardini2021}
{Bellardini} M.~A.,  {Wetzel} A.,  {Loebman} S.~R.,  {Faucher-Gigu{\`e}re}
  C.-A.,  {Ma} X.,   {Feldmann} R.,  2021, \mn@doi [\mnras]
  {10.1093/mnras/stab1606}, \href
  {https://ui.adsabs.harvard.edu/abs/2021MNRAS.505.4586B} {505, 4586}

\bibitem[\protect\citeauthoryear{{Bellardini}, {Wetzel}, {Loebman}  \&
  {Bailin}}{{Bellardini} et~al.}{2022}]{Bellardini2022}
{Bellardini} M.~A.,  {Wetzel} A.,  {Loebman} S.~R.,   {Bailin} J.,  2022,
  \mn@doi [\mnras] {10.1093/mnras/stac1637}, \href
  {https://ui.adsabs.harvard.edu/abs/2022MNRAS.514.4270B} {514, 4270}

\bibitem[\protect\citeauthoryear{{Bensby}, {Feltzing}  \& {Oey}}{{Bensby}
  et~al.}{2014}]{Bensby2014}
{Bensby} T.,  {Feltzing} S.,   {Oey} M.~S.,  2014, \mn@doi [\aap]
  {10.1051/0004-6361/201322631}, \href
  {http://adsabs.harvard.edu/abs/2014A%26A...562A..71B} {562, A71}

\bibitem[\protect\citeauthoryear{{Bergemann} et~al.,}{{Bergemann}
  et~al.}{2014}]{Bergemann2014}
{Bergemann} M.,  et~al., 2014, \mn@doi [\aap] {10.1051/0004-6361/201423456},
  \href {http://adsabs.harvard.edu/abs/2014A%26A...565A..89B} {565, A89}

\bibitem[\protect\citeauthoryear{{Binney} \& {Tremaine}}{{Binney} \&
  {Tremaine}}{2008}]{Binney2008}
{Binney} J.,  {Tremaine} S.,  2008, {Galactic Dynamics: Second Edition}.
Princeton University Press

\bibitem[\protect\citeauthoryear{{Bird}, {Kazantzidis}, {Weinberg}, {Guedes},
  {Callegari}, {Mayer}  \& {Madau}}{{Bird} et~al.}{2013}]{Bird2013}
{Bird} J.~C.,  {Kazantzidis} S.,  {Weinberg} D.~H.,  {Guedes} J.,  {Callegari}
  S.,  {Mayer} L.,   {Madau} P.,  2013, \mn@doi [\apj]
  {10.1088/0004-637X/773/1/43}, \href
  {http://adsabs.harvard.edu/abs/2013ApJ...773...43B} {773, 43}

\bibitem[\protect\citeauthoryear{{Bland-Hawthorn} \&
  {Gerhard}}{{Bland-Hawthorn} \& {Gerhard}}{2016}]{BlandHawthorn_Gerhard2016}
{Bland-Hawthorn} J.,  {Gerhard} O.,  2016, \mn@doi [\araa]
  {10.1146/annurev-astro-081915-023441}, \href
  {http://adsabs.harvard.edu/abs/2016ARA%26A..54..529B} {54, 529}

\bibitem[\protect\citeauthoryear{{Bland-Hawthorn}, {Tepper-Garcia}, {Agertz}
  \& {Federrath}}{{Bland-Hawthorn} et~al.}{2024}]{BlandHawthorn2024}
{Bland-Hawthorn} J.,  {Tepper-Garcia} T.,  {Agertz} O.,   {Federrath} C.,
  2024, \mn@doi [\apj] {10.3847/1538-4357/ad4118}, \href
  {https://ui.adsabs.harvard.edu/abs/2024ApJ...968...86B} {968, 86}

\bibitem[\protect\citeauthoryear{{Boeche} et~al.,}{{Boeche}
  et~al.}{2013}]{Boeche2013}
{Boeche} C.,  et~al., 2013, \mn@doi [\aap] {10.1051/0004-6361/201322085}, \href
  {https://ui.adsabs.harvard.edu/abs/2013A&A...559A..59B} {559, A59}

\bibitem[\protect\citeauthoryear{{Bresolin}, {Kennicutt}  \&
  {Ryan-Weber}}{{Bresolin} et~al.}{2012}]{Bresolin2012}
{Bresolin} F.,  {Kennicutt} R.~C.,   {Ryan-Weber} E.,  2012, \mn@doi [\apj]
  {10.1088/0004-637X/750/2/122}, \href
  {https://ui.adsabs.harvard.edu/abs/2012ApJ...750..122B} {750, 122}

\bibitem[\protect\citeauthoryear{{Buck}}{{Buck}}{2020}]{Buck2020}
{Buck} T.,  2020, \mn@doi [\mnras] {10.1093/mnras/stz3289}, \href
  {https://ui.adsabs.harvard.edu/abs/2020MNRAS.491.5435B} {491, 5435}

\bibitem[\protect\citeauthoryear{{Buck}, {Macci{\`o}}, {Dutton}, {Obreja}  \&
  {Frings}}{{Buck} et~al.}{2019}]{Buck2019b}
{Buck} T.,  {Macci{\`o}} A.~V.,  {Dutton} A.~A.,  {Obreja} A.,   {Frings} J.,
  2019, \mn@doi [\mnras] {10.1093/mnras/sty2913}, \href
  {https://ui.adsabs.harvard.edu/abs/2019MNRAS.483.1314B} {483, 1314}

\bibitem[\protect\citeauthoryear{{Buck}, {Obreja}, {Macci{\`o}}, {Minchev},
  {Dutton}  \& {Ostriker}}{{Buck} et~al.}{2020}]{Buck2020b}
{Buck} T.,  {Obreja} A.,  {Macci{\`o}} A.~V.,  {Minchev} I.,  {Dutton} A.~A.,
  {Ostriker} J.~P.,  2020, \mn@doi [\mnras] {10.1093/mnras/stz3241}, \href
  {https://ui.adsabs.harvard.edu/abs/2020MNRAS.491.3461B} {491, 3461}

\bibitem[\protect\citeauthoryear{{Buck}, {Rybizki}, {Buder}, {Obreja},
  {Macci{\`o}}, {Pfrommer}, {Steinmetz}  \& {Ness}}{{Buck}
  et~al.}{2021}]{Buck2021}
{Buck} T.,  {Rybizki} J.,  {Buder} S.,  {Obreja} A.,  {Macci{\`o}} A.~V.,
  {Pfrommer} C.,  {Steinmetz} M.,   {Ness} M.,  2021, \mn@doi [\mnras]
  {10.1093/mnras/stab2736}, \href
  {https://ui.adsabs.harvard.edu/abs/2021MNRAS.508.3365B} {508, 3365}

\bibitem[\protect\citeauthoryear{{Buck}, {Obreja}, {Ratcliffe}, {Lu}, {Minchev}
   \& {Macci{\`o}}}{{Buck} et~al.}{2023}]{Buck2023}
{Buck} T.,  {Obreja} A.,  {Ratcliffe} B.,  {Lu} Y.,  {Minchev} I.,
  {Macci{\`o}} A.~V.,  2023, \mn@doi [\mnras] {10.1093/mnras/stad1503}, \href
  {https://ui.adsabs.harvard.edu/abs/2023MNRAS.523.1565B} {523, 1565}

\bibitem[\protect\citeauthoryear{{Buder} et~al.,}{{Buder}
  et~al.}{2019}]{Buder2019}
{Buder} S.,  et~al., 2019, \mn@doi [\aap] {10.1051/0004-6361/201833218}, \href
  {http://adsabs.harvard.edu/abs/2019A%26A...624A..19B} {624, A19}

\bibitem[\protect\citeauthoryear{{Buder} et~al.,}{{Buder}
  et~al.}{2024a}]{Buder2024b}
{Buder} S.,  et~al., 2024a, \mn@doi [arXiv e-prints]
  {10.48550/arXiv.2409.19858}, \href
  {https://ui.adsabs.harvard.edu/abs/2024arXiv240919858B} {p. arXiv:2409.19858}

\bibitem[\protect\citeauthoryear{{Buder}, {Mijnarends}  \& {Buck}}{{Buder}
  et~al.}{2024b}]{Buder2024}
{Buder} S.,  {Mijnarends} L.,   {Buck} T.,  2024b, \mn@doi [\mnras]
  {10.1093/mnras/stae1552}, \href
  {https://ui.adsabs.harvard.edu/abs/2024MNRAS.532.1010B} {532, 1010}

\bibitem[\protect\citeauthoryear{{Cantat-Gaudin} \& {Anders}}{{Cantat-Gaudin}
  \& {Anders}}{2020}]{CantatGaudin2020}
{Cantat-Gaudin} T.,  {Anders} F.,  2020, \mn@doi [\aap]
  {10.1051/0004-6361/201936691}, \href
  {https://ui.adsabs.harvard.edu/abs/2020A&A...633A..99C} {633, A99}

\bibitem[\protect\citeauthoryear{{Casamiquela} et~al.,}{{Casamiquela}
  et~al.}{2019}]{Casamiquela2019}
{Casamiquela} L.,  et~al., 2019, \mn@doi [\mnras] {10.1093/mnras/stz2595},
  \href {https://ui.adsabs.harvard.edu/abs/2019MNRAS.490.1821C} {490, 1821}

\bibitem[\protect\citeauthoryear{{Chabrier}}{{Chabrier}}{2003}]{Chabrier2003}
{Chabrier} G.,  2003, \mn@doi [\pasp] {10.1086/376392}, \href
  {https://ui.adsabs.harvard.edu/abs/2003PASP..115..763C} {115, 763}

\bibitem[\protect\citeauthoryear{{Chen}, {Grasha}, {Battisti}, {Kewley},
  {Madore}, {Seibert}, {Rich}  \& {Beaton}}{{Chen} et~al.}{2023}]{Chen2023}
{Chen} Q.-H.,  {Grasha} K.,  {Battisti} A.~J.,  {Kewley} L.~J.,  {Madore}
  B.~F.,  {Seibert} M.,  {Rich} J.~A.,   {Beaton} R.~L.,  2023, \mn@doi
  [\mnras] {10.1093/mnras/stac3790}, \href
  {https://ui.adsabs.harvard.edu/abs/2023MNRAS.519.4801C} {519, 4801}

\bibitem[\protect\citeauthoryear{{Chen} et~al.,}{{Chen}
  et~al.}{2024a}]{Chen2024}
{Chen} Q.-H.,  et~al., 2024a, \mn@doi [\mnras] {10.1093/mnras/stad3348}, \href
  {https://ui.adsabs.harvard.edu/abs/2024MNRAS.527.2991C} {527, 2991}

\bibitem[\protect\citeauthoryear{{Chen} et~al.,}{{Chen}
  et~al.}{2024b}]{Chen2024b}
{Chen} Q.-H.,  et~al., 2024b, \mn@doi [\mnras] {10.1093/mnras/stae2119}, \href
  {https://ui.adsabs.harvard.edu/abs/2024MNRAS.534..883C} {534, 883}

\bibitem[\protect\citeauthoryear{{Chiappini}}{{Chiappini}}{2002}]{Chiappini2002}
{Chiappini} C.,  2002, \mn@doi [\apss] {10.1023/A:1019575123561}, \href
  {https://ui.adsabs.harvard.edu/abs/2002Ap&SS.281..253C} {281, 253}

\bibitem[\protect\citeauthoryear{{Chiappini}, {Matteucci}  \&
  {Gratton}}{{Chiappini} et~al.}{1997}]{Chiappini1997}
{Chiappini} C.,  {Matteucci} F.,   {Gratton} R.,  1997, \mn@doi [\apj]
  {10.1086/303726}, \href {http://adsabs.harvard.edu/abs/1997ApJ...477..765C}
  {477, 765}

\bibitem[\protect\citeauthoryear{{Chiappini}, {Matteucci}  \&
  {Romano}}{{Chiappini} et~al.}{2001}]{Chiappini2001}
{Chiappini} C.,  {Matteucci} F.,   {Romano} D.,  2001, \mn@doi [\apj]
  {10.1086/321427}, \href
  {https://ui.adsabs.harvard.edu/#abs/2001ApJ...554.1044C} {554, 1044}

\bibitem[\protect\citeauthoryear{Chieffi \& Limongi}{Chieffi \&
  Limongi}{2004}]{Chieffi2004}
Chieffi A.,  Limongi M.,  2004, \mn@doi [\apj] {10.1086/392523}, 608, 405

\bibitem[\protect\citeauthoryear{{Chiosi}}{{Chiosi}}{1980}]{Chiosi1980}
{Chiosi} C.,  1980, \aap, \href
  {https://ui.adsabs.harvard.edu/abs/1980A&A....83..206C} {83, 206}

\bibitem[\protect\citeauthoryear{{Cunha} et~al.,}{{Cunha}
  et~al.}{2016}]{Cunha2016}
{Cunha} K.,  et~al., 2016, \mn@doi [Astronomische Nachrichten]
  {10.1002/asna.201612398}, \href
  {http://adsabs.harvard.edu/abs/2016AN....337..922C} {337, 922}

\bibitem[\protect\citeauthoryear{{Di Matteo}, {Haywood}, {Combes}, {Semelin}
  \& {Snaith}}{{Di Matteo} et~al.}{2013}]{DiMatteo2013}
{Di Matteo} P.,  {Haywood} M.,  {Combes} F.,  {Semelin} B.,   {Snaith} O.~N.,
  2013, \mn@doi [\aap] {10.1051/0004-6361/201220539}, \href
  {https://ui.adsabs.harvard.edu/abs/2013A&A...553A.102D} {553, A102}

\bibitem[\protect\citeauthoryear{{Donor} et~al.,}{{Donor}
  et~al.}{2020}]{Donor2020}
{Donor} J.,  et~al., 2020, \mn@doi [\aj] {10.3847/1538-3881/ab77bc}, \href
  {https://ui.adsabs.harvard.edu/abs/2020AJ....159..199D} {159, 199}

\bibitem[\protect\citeauthoryear{{Foster} et~al.,}{{Foster}
  et~al.}{2021}]{MAGPI2021}
{Foster} C.,  et~al., 2021, \mn@doi [\pasa] {10.1017/pasa.2021.25}, \href
  {https://ui.adsabs.harvard.edu/abs/2021PASA...38...31F} {38, e031}

\bibitem[\protect\citeauthoryear{{Frankel}, {Rix}, {Ting}, {Ness}  \&
  {Hogg}}{{Frankel} et~al.}{2018}]{Frankel2018}
{Frankel} N.,  {Rix} H.-W.,  {Ting} Y.-S.,  {Ness} M.,   {Hogg} D.~W.,  2018,
  \mn@doi [\apj] {10.3847/1538-4357/aadba5}, \href
  {http://adsabs.harvard.edu/abs/2018ApJ...865...96F} {865, 96}

\bibitem[\protect\citeauthoryear{{Frankel}, {Sanders}, {Ting}  \&
  {Rix}}{{Frankel} et~al.}{2020}]{Frankel2020}
{Frankel} N.,  {Sanders} J.,  {Ting} Y.-S.,   {Rix} H.-W.,  2020, \mn@doi
  [\apj] {10.3847/1538-4357/ab910c}, \href
  {https://ui.adsabs.harvard.edu/abs/2020ApJ...896...15F} {896, 15}

\bibitem[\protect\citeauthoryear{{Fraser-McKelvie} et~al.,}{{Fraser-McKelvie}
  et~al.}{2022}]{FraserMcKelvie2022}
{Fraser-McKelvie} A.,  et~al., 2022, \mn@doi [\mnras] {10.1093/mnras/stab3430},
  \href {https://ui.adsabs.harvard.edu/abs/2022MNRAS.510..320F} {510, 320}

\bibitem[\protect\citeauthoryear{{Gaia Collaboration} et~al.,}{{Gaia
  Collaboration} et~al.}{2016}]{Gaia-Collaboration2016}
{Gaia Collaboration} et~al., 2016, \mn@doi [\aap]
  {10.1051/0004-6361/201629272}, \href
  {http://adsabs.harvard.edu/abs/2016A%26A...595A...1G} {595, A1}

\bibitem[\protect\citeauthoryear{{Garcia} et~al.,}{{Garcia}
  et~al.}{2023}]{Garcia2023}
{Garcia} A.~M.,  et~al., 2023, \mn@doi [\mnras] {10.1093/mnras/stac3749}, \href
  {https://ui.adsabs.harvard.edu/abs/2023MNRAS.519.4716G} {519, 4716}

\bibitem[\protect\citeauthoryear{{Genovali} et~al.,}{{Genovali}
  et~al.}{2014}]{Genovali2014}
{Genovali} K.,  et~al., 2014, \mn@doi [\aap] {10.1051/0004-6361/201323198},
  \href {https://ui.adsabs.harvard.edu/abs/2014A&A...566A..37G} {566, A37}

\bibitem[\protect\citeauthoryear{{Graf}, {Wetzel}, {Bellardini}  \&
  {Bailin}}{{Graf} et~al.}{2025}]{Graf2025}
{Graf} R.~L.,  {Wetzel} A.,  {Bellardini} M.~A.,   {Bailin} J.,  2025, \mn@doi
  [\apj] {10.3847/1538-4357/adacd7}, \href
  {https://ui.adsabs.harvard.edu/abs/2025ApJ...981...47G} {981, 47}

\bibitem[\protect\citeauthoryear{{Grand}, {Kawata}  \& {Cropper}}{{Grand}
  et~al.}{2015}]{Grand2015}
{Grand} R. J.~J.,  {Kawata} D.,   {Cropper} M.,  2015, \mn@doi [\mnras]
  {10.1093/mnras/stv016}, \href
  {https://ui.adsabs.harvard.edu/abs/2015MNRAS.447.4018G} {447, 4018}

\bibitem[\protect\citeauthoryear{{Grand} et~al.,}{{Grand}
  et~al.}{2016}]{Grand2016}
{Grand} R. J.~J.,  et~al., 2016, \mn@doi [\mnras] {10.1093/mnrasl/slw086},
  \href {https://ui.adsabs.harvard.edu/abs/2016MNRAS.460L..94G} {460, L94}

\bibitem[\protect\citeauthoryear{{Grasha} et~al.,}{{Grasha}
  et~al.}{2022}]{Grasha2022}
{Grasha} K.,  et~al., 2022, \mn@doi [\apj] {10.3847/1538-4357/ac5ab2}, \href
  {https://ui.adsabs.harvard.edu/abs/2022ApJ...929..118G} {929, 118}

\bibitem[\protect\citeauthoryear{{Hackshaw}, {Hawkins}, {Filion}, {Horta},
  {Laporte}, {Carr}  \& {Price-Whelan}}{{Hackshaw} et~al.}{2024}]{Hackshaw2024}
{Hackshaw} Z.,  {Hawkins} K.,  {Filion} C.,  {Horta} D.,  {Laporte} C. F.~P.,
  {Carr} C.,   {Price-Whelan} A.~M.,  2024, \mn@doi [\apj]
  {10.3847/1538-4357/ad900e}, \href
  {https://ui.adsabs.harvard.edu/abs/2024ApJ...977..143H} {977, 143}

\bibitem[\protect\citeauthoryear{{Hayden} et~al.,}{{Hayden}
  et~al.}{2014}]{Hayden2014}
{Hayden} M.~R.,  et~al., 2014, \mn@doi [\aj] {10.1088/0004-6256/147/5/116},
  \href {https://ui.adsabs.harvard.edu/abs/2014AJ....147..116H} {147, 116}

\bibitem[\protect\citeauthoryear{{Hayden} et~al.,}{{Hayden}
  et~al.}{2015}]{Hayden2015}
{Hayden} M.~R.,  et~al., 2015, \mn@doi [\apj] {10.1088/0004-637X/808/2/132},
  \href {http://adsabs.harvard.edu/abs/2015ApJ...808..132H} {808, 132}

\bibitem[\protect\citeauthoryear{{Hemler} et~al.,}{{Hemler}
  et~al.}{2021}]{Hemler2021}
{Hemler} Z.~S.,  et~al., 2021, \mn@doi [\mnras] {10.1093/mnras/stab1803}, \href
  {https://ui.adsabs.harvard.edu/abs/2021MNRAS.506.3024H} {506, 3024}

\bibitem[\protect\citeauthoryear{{Ho}, {Kudritzki}, {Kewley}, {Zahid},
  {Dopita}, {Bresolin}  \& {Rupke}}{{Ho} et~al.}{2015}]{Ho2015}
{Ho} I.~T.,  {Kudritzki} R.-P.,  {Kewley} L.~J.,  {Zahid} H.~J.,  {Dopita}
  M.~A.,  {Bresolin} F.,   {Rupke} D. S.~N.,  2015, \mn@doi [\mnras]
  {10.1093/mnras/stv067}, \href
  {https://ui.adsabs.harvard.edu/abs/2015MNRAS.448.2030H} {448, 2030}

\bibitem[\protect\citeauthoryear{{Ho} et~al.,}{{Ho} et~al.}{2017}]{Ho2017c}
{Ho} I.~T.,  et~al., 2017, \mn@doi [\apj] {10.3847/1538-4357/aa8460}, \href
  {https://ui.adsabs.harvard.edu/abs/2017ApJ...846...39H} {846, 39}

\bibitem[\protect\citeauthoryear{{Hogg}, {Eilers}  \& {Rix}}{{Hogg}
  et~al.}{2019}]{Hogg2019}
{Hogg} D.~W.,  {Eilers} A.-C.,   {Rix} H.-W.,  2019, \mn@doi [\aj]
  {10.3847/1538-3881/ab398c}, \href
  {https://ui.adsabs.harvard.edu/abs/2019AJ....158..147H} {158, 147}

\bibitem[\protect\citeauthoryear{Hunter}{Hunter}{2007}]{matplotlib}
Hunter J.~D.,  2007, \mn@doi [Comput Sci Eng] {10.1109/MCSE.2007.55}, 9, 90

\bibitem[\protect\citeauthoryear{{Imig} et~al.,}{{Imig}
  et~al.}{2023}]{Imig2023}
{Imig} J.,  et~al., 2023, \mn@doi [\apj] {10.3847/1538-4357/ace9b8}, \href
  {https://ui.adsabs.harvard.edu/abs/2023ApJ...954..124I} {954, 124}

\bibitem[\protect\citeauthoryear{{Janes}}{{Janes}}{1979}]{Janes1979}
{Janes} K.~A.,  1979, \mn@doi [\apjs] {10.1086/190568}, \href
  {https://ui.adsabs.harvard.edu/abs/1979ApJS...39..135J} {39, 135}

\bibitem[\protect\citeauthoryear{{Johnson} et~al.,}{{Johnson}
  et~al.}{2024}]{Johnson2024}
{Johnson} J.~W.,  et~al., 2024, \mn@doi [arXiv e-prints]
  {10.48550/arXiv.2410.13256}, \href
  {https://ui.adsabs.harvard.edu/abs/2024arXiv241013256J} {p. arXiv:2410.13256}

\bibitem[\protect\citeauthoryear{Karakas \& Lugaro}{Karakas \&
  Lugaro}{2016}]{Karakas2016}
Karakas A.~I.,  Lugaro M.,  2016, \mn@doi [ApJ] {10.3847/0004-637X/825/1/26},
  825, 26

\bibitem[\protect\citeauthoryear{{Kauffmann}}{{Kauffmann}}{1996}]{Kauffman1996}
{Kauffmann} G.,  1996, \mn@doi [\mnras] {10.1093/mnras/281.2.475}, \href
  {https://ui.adsabs.harvard.edu/abs/1996MNRAS.281..475K} {281, 475}

\bibitem[\protect\citeauthoryear{{Khoperskov}, {Sivkova}, {Saburova},
  {Vasiliev}, {Shustov}, {Minchev}  \& {Walcher}}{{Khoperskov}
  et~al.}{2023}]{Khoperskov2023e}
{Khoperskov} S.,  {Sivkova} E.,  {Saburova} A.,  {Vasiliev} E.,  {Shustov} B.,
  {Minchev} I.,   {Walcher} C.~J.,  2023, \mn@doi [\aap]
  {10.1051/0004-6361/202142581}, \href
  {https://ui.adsabs.harvard.edu/abs/2023A&A...671A..56K} {671, A56}

\bibitem[\protect\citeauthoryear{{Knollmann} \& {Knebe}}{{Knollmann} \&
  {Knebe}}{2009}]{Knollman2009}
{Knollmann} S.~R.,  {Knebe} A.,  2009, \mn@doi [\apjs]
  {10.1088/0067-0049/182/2/608}, \href
  {https://ui.adsabs.harvard.edu/abs/2009ApJS..182..608K} {182, 608}

\bibitem[\protect\citeauthoryear{{Kollmeier} et~al.,}{{Kollmeier}
  et~al.}{2017}]{Kollmeier2017}
{Kollmeier} J.~A.,  et~al., 2017, arXiv e-prints, \href
  {https://ui.adsabs.harvard.edu/abs/2017arXiv171103234K} {p. arXiv:1711.03234}

\bibitem[\protect\citeauthoryear{{Kreckel} et~al.,}{{Kreckel}
  et~al.}{2019}]{Kreckel2019}
{Kreckel} K.,  et~al., 2019, \mn@doi [\apj] {10.3847/1538-4357/ab5115}, \href
  {https://ui.adsabs.harvard.edu/abs/2019ApJ...887...80K} {887, 80}

\bibitem[\protect\citeauthoryear{{Kreckel} et~al.,}{{Kreckel}
  et~al.}{2020}]{Kreckel2020}
{Kreckel} K.,  et~al., 2020, \mn@doi [\mnras] {10.1093/mnras/staa2743}, \href
  {https://ui.adsabs.harvard.edu/abs/2020MNRAS.499..193K} {499, 193}

\bibitem[\protect\citeauthoryear{{Krumholz} \& {Ting}}{{Krumholz} \&
  {Ting}}{2018}]{Krumholz2018b}
{Krumholz} M.~R.,  {Ting} Y.-S.,  2018, \mn@doi [\mnras]
  {10.1093/mnras/stx3286}, \href
  {https://ui.adsabs.harvard.edu/abs/2018MNRAS.475.2236K} {475, 2236}

\bibitem[\protect\citeauthoryear{{Krumholz}, {Burkhart}, {Forbes}  \&
  {Crocker}}{{Krumholz} et~al.}{2018}]{Krumholz2018}
{Krumholz} M.~R.,  {Burkhart} B.,  {Forbes} J.~C.,   {Crocker} R.~M.,  2018,
  \mn@doi [\mnras] {10.1093/mnras/sty852}, \href
  {https://ui.adsabs.harvard.edu/abs/2018MNRAS.477.2716K} {477, 2716}

\bibitem[\protect\citeauthoryear{{Kubryk}, {Prantzos}  \&
  {Athanassoula}}{{Kubryk} et~al.}{2015}]{Kubryk2015}
{Kubryk} M.,  {Prantzos} N.,   {Athanassoula} E.,  2015, \mn@doi [\aap]
  {10.1051/0004-6361/201424171}, \href
  {https://ui.adsabs.harvard.edu/abs/2015A&A...580A.126K} {580, A126}

\bibitem[\protect\citeauthoryear{{Lacey} \& {Fall}}{{Lacey} \&
  {Fall}}{1985}]{Lacey1985}
{Lacey} C.~G.,  {Fall} S.~M.,  1985, \mn@doi [\apj] {10.1086/162970}, \href
  {https://ui.adsabs.harvard.edu/abs/1985ApJ...290..154L} {290, 154}

\bibitem[\protect\citeauthoryear{{Larson}}{{Larson}}{1976}]{Larson1976}
{Larson} R.~B.,  1976, \mn@doi [\mnras] {10.1093/mnras/176.1.31}, \href
  {https://ui.adsabs.harvard.edu/abs/1976MNRAS.176...31L} {176, 31}

\bibitem[\protect\citeauthoryear{{Lemasle}, {Fran{\c{c}}ois}, {Bono},
  {Mottini}, {Primas}  \& {Romaniello}}{{Lemasle} et~al.}{2007}]{Lemasle2007}
{Lemasle} B.,  {Fran{\c{c}}ois} P.,  {Bono} G.,  {Mottini} M.,  {Primas} F.,
  {Romaniello} M.,  2007, \mn@doi [\aap] {10.1051/0004-6361:20066375}, \href
  {https://ui.adsabs.harvard.edu/abs/2007A&A...467..283L} {467, 283}

\bibitem[\protect\citeauthoryear{{Lemasle}, {Fran{\c{c}}ois}, {Piersimoni},
  {Pedicelli}, {Bono}, {Laney}, {Primas}  \& {Romaniello}}{{Lemasle}
  et~al.}{2008}]{Lemasle2008}
{Lemasle} B.,  {Fran{\c{c}}ois} P.,  {Piersimoni} A.,  {Pedicelli} S.,  {Bono}
  G.,  {Laney} C.~D.,  {Primas} F.,   {Romaniello} M.,  2008, \mn@doi [\aap]
  {10.1051/0004-6361:200810192}, \href
  {https://ui.adsabs.harvard.edu/abs/2008A&A...490..613L} {490, 613}

\bibitem[\protect\citeauthoryear{{Lemasle} et~al.,}{{Lemasle}
  et~al.}{2013}]{Lemasle2013}
{Lemasle} B.,  et~al., 2013, \mn@doi [\aap] {10.1051/0004-6361/201322115},
  \href {https://ui.adsabs.harvard.edu/abs/2013A&A...558A..31L} {558, A31}

\bibitem[\protect\citeauthoryear{{Lemasle} et~al.,}{{Lemasle}
  et~al.}{2022}]{Lemasle2022}
{Lemasle} B.,  et~al., 2022, \mn@doi [\aap] {10.1051/0004-6361/202243273},
  \href {https://ui.adsabs.harvard.edu/abs/2022A&A...668A..40L} {668, A40}

\bibitem[\protect\citeauthoryear{{L{\'e}pine} et~al.,}{{L{\'e}pine}
  et~al.}{2011}]{Lepine2011}
{L{\'e}pine} J.~R.~D.,  et~al., 2011, \mn@doi [\mnras]
  {10.1111/j.1365-2966.2011.19314.x}, \href
  {https://ui.adsabs.harvard.edu/abs/2011MNRAS.417..698L} {417, 698}

\bibitem[\protect\citeauthoryear{{Li} et~al.,}{{Li} et~al.}{2024}]{Li2024b}
{Li} Z.,  et~al., 2024, \mn@doi [\mnras] {10.1093/mnras/stae480}, \href
  {https://ui.adsabs.harvard.edu/abs/2024MNRAS.528.7103L} {528, 7103}

\bibitem[\protect\citeauthoryear{{Lilly}, {Carollo}, {Pipino}, {Renzini}  \&
  {Peng}}{{Lilly} et~al.}{2013}]{Lilly2013}
{Lilly} S.~J.,  {Carollo} C.~M.,  {Pipino} A.,  {Renzini} A.,   {Peng} Y.,
  2013, \mn@doi [\apj] {10.1088/0004-637X/772/2/119}, \href
  {https://ui.adsabs.harvard.edu/abs/2013ApJ...772..119L} {772, 119}

\bibitem[\protect\citeauthoryear{{Lu}, {Buck}, {Minchev}  \& {Ness}}{{Lu}
  et~al.}{2022}]{Lu2022}
{Lu} Y.,  {Buck} T.,  {Minchev} I.,   {Ness} M.~K.,  2022, \mn@doi [\mnras]
  {10.1093/mnrasl/slac065}, \href
  {https://ui.adsabs.harvard.edu/abs/2022MNRAS.515L..34L} {515, L34}

\bibitem[\protect\citeauthoryear{{Ma}, {Hopkins}, {Feldmann}, {Torrey},
  {Faucher-Gigu{\`e}re}  \& {Kere{\v{s}}}}{{Ma} et~al.}{2017a}]{Ma2017b}
{Ma} X.,  {Hopkins} P.~F.,  {Feldmann} R.,  {Torrey} P.,  {Faucher-Gigu{\`e}re}
  C.-A.,   {Kere{\v{s}}} D.,  2017a, \mn@doi [\mnras] {10.1093/mnras/stx034},
  \href {https://ui.adsabs.harvard.edu/abs/2017MNRAS.466.4780M} {466, 4780}

\bibitem[\protect\citeauthoryear{{Ma}, {Hopkins}, {Wetzel}, {Kirby},
  {Angl{\'e}s-Alc{\'a}zar}, {Faucher-Gigu{\`e}re}, {Kere{\v{s}}}  \&
  {Quataert}}{{Ma} et~al.}{2017b}]{Ma2017}
{Ma} X.,  {Hopkins} P.~F.,  {Wetzel} A.~R.,  {Kirby} E.~N.,
  {Angl{\'e}s-Alc{\'a}zar} D.,  {Faucher-Gigu{\`e}re} C.-A.,  {Kere{\v{s}}} D.,
    {Quataert} E.,  2017b, \mn@doi [\mnras] {10.1093/mnras/stx273}, \href
  {https://ui.adsabs.harvard.edu/abs/2017MNRAS.467.2430M} {467, 2430}

\bibitem[\protect\citeauthoryear{{Magrini}, {Sestito}, {Randich}  \&
  {Galli}}{{Magrini} et~al.}{2009}]{Magrini2009}
{Magrini} L.,  {Sestito} P.,  {Randich} S.,   {Galli} D.,  2009, \mn@doi [\aap]
  {10.1051/0004-6361:200810634}, \href
  {https://ui.adsabs.harvard.edu/abs/2009A&A...494...95M} {494, 95}

\bibitem[\protect\citeauthoryear{{Magrini} et~al.,}{{Magrini}
  et~al.}{2017}]{Magrini2017}
{Magrini} L.,  et~al., 2017, \mn@doi [\aap] {10.1051/0004-6361/201630294},
  \href {https://ui.adsabs.harvard.edu/abs/2017A&A...603A...2M} {603, A2}

\bibitem[\protect\citeauthoryear{{Matteucci} \& {Recchi}}{{Matteucci} \&
  {Recchi}}{2001}]{Matteucci2001}
{Matteucci} F.,  {Recchi} S.,  2001, \mn@doi [\apj] {10.1086/322472}, \href
  {http://adsabs.harvard.edu/abs/2001ApJ...558..351M} {558, 351}

\bibitem[\protect\citeauthoryear{{Metha}, {Trenti}  \& {Chu}}{{Metha}
  et~al.}{2021}]{Metha2021}
{Metha} B.,  {Trenti} M.,   {Chu} T.,  2021, \mn@doi [\mnras]
  {10.1093/mnras/stab2554}, \href
  {https://ui.adsabs.harvard.edu/abs/2021MNRAS.508..489M} {508, 489}

\bibitem[\protect\citeauthoryear{{Minchev} \& {Famaey}}{{Minchev} \&
  {Famaey}}{2010}]{Minchev2010}
{Minchev} I.,  {Famaey} B.,  2010, \mn@doi [\apj]
  {10.1088/0004-637X/722/1/112}, \href
  {http://adsabs.harvard.edu/abs/2010ApJ...722..112M} {722, 112}

\bibitem[\protect\citeauthoryear{{Minchev}, {Chiappini}  \& {Martig}}{{Minchev}
  et~al.}{2013}]{Minchev2013}
{Minchev} I.,  {Chiappini} C.,   {Martig} M.,  2013, \mn@doi [\aap]
  {10.1051/0004-6361/201220189}, \href
  {https://ui.adsabs.harvard.edu/abs/2013A&A...558A...9M} {558, A9}

\bibitem[\protect\citeauthoryear{{Minchev}, {Chiappini}  \& {Martig}}{{Minchev}
  et~al.}{2014}]{Minchev2014b}
{Minchev} I.,  {Chiappini} C.,   {Martig} M.,  2014, \mn@doi [\aap]
  {10.1051/0004-6361/201423487}, \href
  {https://ui.adsabs.harvard.edu/abs/2014A&A...572A..92M} {572, A92}

\bibitem[\protect\citeauthoryear{{Minchev} et~al.,}{{Minchev}
  et~al.}{2018}]{Minchev2018}
{Minchev} I.,  et~al., 2018, \mn@doi [\mnras] {10.1093/mnras/sty2033}, \href
  {https://ui.adsabs.harvard.edu/abs/2018MNRAS.481.1645M} {481, 1645}

\bibitem[\protect\citeauthoryear{{Moran} et~al.,}{{Moran}
  et~al.}{2012}]{Moran2012}
{Moran} S.~M.,  et~al., 2012, \mn@doi [\apj] {10.1088/0004-637X/745/1/66},
  \href {https://ui.adsabs.harvard.edu/abs/2012ApJ...745...66M} {745, 66}

\bibitem[\protect\citeauthoryear{{Mun} et~al.,}{{Mun} et~al.}{2024}]{Mun2024}
{Mun} M.,  et~al., 2024, \mn@doi [\mnras] {10.1093/mnras/stae1132}, \href
  {https://ui.adsabs.harvard.edu/abs/2024MNRAS.530.5072M} {530, 5072}

\bibitem[\protect\citeauthoryear{{Myers} et~al.,}{{Myers}
  et~al.}{2022}]{Myers2022}
{Myers} N.,  et~al., 2022, \mn@doi [\aj] {10.3847/1538-3881/ac7ce5}, \href
  {https://ui.adsabs.harvard.edu/abs/2022AJ....164...85M} {164, 85}

\bibitem[\protect\citeauthoryear{{Nicholls}, {Sutherland}, {Dopita}, {Kewley}
  \& {Groves}}{{Nicholls} et~al.}{2017}]{Nicholls2017}
{Nicholls} D.~C.,  {Sutherland} R.~S.,  {Dopita} M.~A.,  {Kewley} L.~J.,
  {Groves} B.~A.,  2017, \mn@doi [\mnras] {10.1093/mnras/stw3235}, \href
  {http://adsabs.harvard.edu/abs/2017MNRAS.466.4403N} {466, 4403}

\bibitem[\protect\citeauthoryear{{Obreja} et~al.,}{{Obreja}
  et~al.}{2019}]{Obreja2019}
{Obreja} A.,  et~al., 2019, \mn@doi [\mnras] {10.1093/mnras/stz1563}, \href
  {https://ui.adsabs.harvard.edu/abs/2019MNRAS.487.4424O} {487, 4424}

\bibitem[\protect\citeauthoryear{{Obreja}, {Buck}  \& {Macci{\`o}}}{{Obreja}
  et~al.}{2022}]{Obreja2022}
{Obreja} A.,  {Buck} T.,   {Macci{\`o}} A.~V.,  2022, \mn@doi [\aap]
  {10.1051/0004-6361/202140983}, \href
  {https://ui.adsabs.harvard.edu/abs/2022A&A...657A..15O} {657, A15}

\bibitem[\protect\citeauthoryear{{Orr} et~al.,}{{Orr} et~al.}{2023}]{Orr2023}
{Orr} M.~E.,  et~al., 2023, \mn@doi [\mnras] {10.1093/mnras/stad676}, \href
  {https://ui.adsabs.harvard.edu/abs/2023MNRAS.521.3708O} {521, 3708}

\bibitem[\protect\citeauthoryear{{Osterbrock}}{{Osterbrock}}{1989}]{Osterbrock1989}
{Osterbrock} D.~E.,  1989, {Astrophysics of gaseous nebulae and active galactic
  nuclei}.
University Science Books

\bibitem[\protect\citeauthoryear{{Palla}, {Magrini}, {Spitoni}, {Matteucci},
  {Viscasillas V{\'a}zquez}, {Franchini}, {Molero}  \& {Randich}}{{Palla}
  et~al.}{2024}]{Palla2024}
{Palla} M.,  {Magrini} L.,  {Spitoni} E.,  {Matteucci} F.,  {Viscasillas
  V{\'a}zquez} C.,  {Franchini} M.,  {Molero} M.,   {Randich} S.,  2024,
  \mn@doi [\aap] {10.1051/0004-6361/202451395}, \href
  {https://ui.adsabs.harvard.edu/abs/2024A&A...690A.334P} {690, A334}

\bibitem[\protect\citeauthoryear{Pedregosa et~al.,}{Pedregosa
  et~al.}{2011}]{scikit-learn}
Pedregosa F.,  et~al., 2011, J Mach Learn Res, 12, 2825

\bibitem[\protect\citeauthoryear{P\'erez \& Granger}{P\'erez \&
  Granger}{2007}]{ipython}
P\'erez F.,  Granger B.~E.,  2007, \mn@doi [Comput Sci Eng]
  {10.1109/MCSE.2007.53}, 9, 21

\bibitem[\protect\citeauthoryear{{Perktold} et~al.,}{{Perktold}
  et~al.}{2024}]{statsmodels}
{Perktold} J.,  et~al., 2024, {statsmodels/statsmodels: Release 0.14.2},
  \mn@doi{10.5281/zenodo.593847}

\bibitem[\protect\citeauthoryear{{Pilkington} et~al.,}{{Pilkington}
  et~al.}{2012}]{Pilkington2012}
{Pilkington} K.,  et~al., 2012, \mn@doi [\aap] {10.1051/0004-6361/201117466},
  \href {https://ui.adsabs.harvard.edu/abs/2012A&A...540A..56P} {540, A56}

\bibitem[\protect\citeauthoryear{{Pilyugin}}{{Pilyugin}}{2003}]{Pilyugin2003}
{Pilyugin} L.~S.,  2003, \mn@doi [\aap] {10.1051/0004-6361:20021505}, \href
  {https://ui.adsabs.harvard.edu/abs/2003A&A...397..109P} {397, 109}

\bibitem[\protect\citeauthoryear{{Pilyugin} \&
  {Tautvai{\v{s}}ien{\.{e}}}}{{Pilyugin} \&
  {Tautvai{\v{s}}ien{\.{e}}}}{2024}]{Pilyugin2024}
{Pilyugin} L.~S.,  {Tautvai{\v{s}}ien{\.{e}}} G.,  2024, \mn@doi [\aap]
  {10.1051/0004-6361/202347032}, \href
  {https://ui.adsabs.harvard.edu/abs/2024A&A...682A..41P} {682, A41}

\bibitem[\protect\citeauthoryear{{Pilyugin}, {Grebel}, {Zinchenko}, {Nefedyev}
  \& {V{\'\i}lchez}}{{Pilyugin} et~al.}{2017}]{Pilyugin2017}
{Pilyugin} L.~S.,  {Grebel} E.~K.,  {Zinchenko} I.~A.,  {Nefedyev} Y.~A.,
  {V{\'\i}lchez} J.~M.,  2017, \mn@doi [\aap] {10.1051/0004-6361/201731256},
  \href {https://ui.adsabs.harvard.edu/abs/2017A&A...608A.127P} {608, A127}

\bibitem[\protect\citeauthoryear{{Planck Collaboration} et~al.,}{{Planck
  Collaboration} et~al.}{2014}]{Planck2014}
{Planck Collaboration} et~al., 2014, \mn@doi [\aap]
  {10.1051/0004-6361/201321591}, \href
  {https://ui.adsabs.harvard.edu/abs/2014A&A...571A..16P} {571, A16}

\bibitem[\protect\citeauthoryear{{Poggio} et~al.,}{{Poggio}
  et~al.}{2018}]{Poggio2018}
{Poggio} E.,  et~al., 2018, \mn@doi [\mnras] {10.1093/mnrasl/sly148}, \href
  {https://ui.adsabs.harvard.edu/abs/2018MNRAS.481L..21P} {481, L21}

\bibitem[\protect\citeauthoryear{{Poggio} et~al.,}{{Poggio}
  et~al.}{2021}]{Poggio2021}
{Poggio} E.,  et~al., 2021, \mn@doi [\aap] {10.1051/0004-6361/202140687}, \href
  {https://ui.adsabs.harvard.edu/abs/2021A&A...651A.104P} {651, A104}

\bibitem[\protect\citeauthoryear{{Poggio} et~al.,}{{Poggio}
  et~al.}{2022}]{Poggio2022}
{Poggio} E.,  et~al., 2022, \mn@doi [\aap] {10.1051/0004-6361/202244361}, \href
  {https://ui.adsabs.harvard.edu/abs/2022A&A...666L...4P} {666, L4}

\bibitem[\protect\citeauthoryear{{Pontzen}, {Ro{\v s}kar}, {Stinson}  \&
  {Woods}}{{Pontzen} et~al.}{2013}]{pynbody}
{Pontzen} A.,  {Ro{\v s}kar} R.,  {Stinson} G.~S.,   {Woods} R.,  2013,
  {pynbody: Astrophysics Simulation Analysis for Python}, Astrophysics Source
  Code Library, record ascl:1305.002

\bibitem[\protect\citeauthoryear{{Quirk} \& {Tinsley}}{{Quirk} \&
  {Tinsley}}{1973}]{Quirk1973}
{Quirk} W.~J.,  {Tinsley} B.~M.,  1973, \mn@doi [\apj] {10.1086/151847}, \href
  {https://ui.adsabs.harvard.edu/abs/1973ApJ...179...69Q} {179, 69}

\bibitem[\protect\citeauthoryear{{Ratcliffe}, {Ness}, {Buck}, {Johnston},
  {Sen}, {Beraldo e Silva}  \& {Debattista}}{{Ratcliffe}
  et~al.}{2022}]{Ratcliffe2022}
{Ratcliffe} B.~L.,  {Ness} M.~K.,  {Buck} T.,  {Johnston} K.~V.,  {Sen} B.,
  {Beraldo e Silva} L.,   {Debattista} V.~P.,  2022, \mn@doi [\apj]
  {10.3847/1538-4357/ac3481}, \href
  {https://ui.adsabs.harvard.edu/abs/2022ApJ...924...60R} {924, 60}

\bibitem[\protect\citeauthoryear{{Ratcliffe} et~al.,}{{Ratcliffe}
  et~al.}{2023}]{Ratcliffe2023}
{Ratcliffe} B.,  et~al., 2023, \mn@doi [\mnras] {10.1093/mnras/stad1573}, \href
  {https://ui.adsabs.harvard.edu/abs/2023MNRAS.525.2208R} {525, 2208}

\bibitem[\protect\citeauthoryear{{Ratcliffe}, {Khoperskov}, {Minchev}, {Lee},
  {Buck}, {Marques}, {Lu}  \& {Steinmetz}}{{Ratcliffe}
  et~al.}{2024}]{Ratcliffe2024b}
{Ratcliffe} B.,  {Khoperskov} S.,  {Minchev} I.,  {Lee} N.~D.,  {Buck} T.,
  {Marques} L.,  {Lu} L.,   {Steinmetz} M.,  2024, \mn@doi [arXiv e-prints]
  {10.48550/arXiv.2410.17326}, \href
  {https://ui.adsabs.harvard.edu/abs/2024arXiv241017326R} {p. arXiv:2410.17326}

\bibitem[\protect\citeauthoryear{{Renaud}, {Ratcliffe}, {Minchev}, {Haywood},
  {Di Matteo}, {Agertz}  \& {Romeo}}{{Renaud} et~al.}{2025}]{Renaud2025}
{Renaud} F.,  {Ratcliffe} B.,  {Minchev} I.,  {Haywood} M.,  {Di Matteo} P.,
  {Agertz} O.,   {Romeo} A.~B.,  2025, \mn@doi [\aap]
  {10.1051/0004-6361/202452219}, \href
  {https://ui.adsabs.harvard.edu/abs/2025A&A...694A..56R} {694, A56}

\bibitem[\protect\citeauthoryear{{Rolleston}, {Smartt}, {Dufton}  \&
  {Ryans}}{{Rolleston} et~al.}{2000}]{Rolleston2000}
{Rolleston} W.~R.~J.,  {Smartt} S.~J.,  {Dufton} P.~L.,   {Ryans} R.~S.~I.,
  2000, \aap, \href {https://ui.adsabs.harvard.edu/abs/2000A&A...363..537R}
  {363, 537}

\bibitem[\protect\citeauthoryear{{Rosales-Ortega}, {D{\'\i}az}, {Kennicutt}  \&
  {S{\'a}nchez}}{{Rosales-Ortega} et~al.}{2011}]{RosalesOrtega2011}
{Rosales-Ortega} F.~F.,  {D{\'\i}az} A.~I.,  {Kennicutt} R.~C.,   {S{\'a}nchez}
  S.~F.,  2011, \mn@doi [\mnras] {10.1111/j.1365-2966.2011.18870.x}, \href
  {https://ui.adsabs.harvard.edu/abs/2011MNRAS.415.2439R} {415, 2439}

\bibitem[\protect\citeauthoryear{{Rybizki}, {Just}  \& {Rix}}{{Rybizki}
  et~al.}{2017}]{Rybizki2017}
{Rybizki} J.,  {Just} A.,   {Rix} H.-W.,  2017, \mn@doi [\aap]
  {10.1051/0004-6361/201730522}, \href
  {http://adsabs.harvard.edu/abs/2017A%26A...605A..59R} {605, A59}

\bibitem[\protect\citeauthoryear{{S{\'a}nchez-Bl{\'a}zquez}
  et~al.,}{{S{\'a}nchez-Bl{\'a}zquez} et~al.}{2014}]{SanchezBlazquez2014}
{S{\'a}nchez-Bl{\'a}zquez} P.,  et~al., 2014, \mn@doi [\aap]
  {10.1051/0004-6361/201423635}, \href
  {https://ui.adsabs.harvard.edu/abs/2014A&A...570A...6S} {570, A6}

\bibitem[\protect\citeauthoryear{{S{\'a}nchez-Menguiano}
  et~al.,}{{S{\'a}nchez-Menguiano} et~al.}{2016}]{SanchezMenguiano2016}
{S{\'a}nchez-Menguiano} L.,  et~al., 2016, \mn@doi [\aap]
  {10.1051/0004-6361/201527450}, \href
  {https://ui.adsabs.harvard.edu/abs/2016A&A...587A..70S} {587, A70}

\bibitem[\protect\citeauthoryear{{S{\'a}nchez} et~al.,}{{S{\'a}nchez}
  et~al.}{2013}]{Sanchez2013}
{S{\'a}nchez} S.~F.,  et~al., 2013, \mn@doi [\aap]
  {10.1051/0004-6361/201220669}, \href
  {https://ui.adsabs.harvard.edu/abs/2013A&A...554A..58S} {554, A58}

\bibitem[\protect\citeauthoryear{{S{\'a}nchez} et~al.,}{{S{\'a}nchez}
  et~al.}{2014}]{Sanchez2014}
{S{\'a}nchez} S.~F.,  et~al., 2014, \mn@doi [\aap]
  {10.1051/0004-6361/201322343}, \href
  {https://ui.adsabs.harvard.edu/abs/2014A&A...563A..49S} {563, A49}

\bibitem[\protect\citeauthoryear{{Scarano} \& {L{\'e}pine}}{{Scarano} \&
  {L{\'e}pine}}{2013}]{Scarano2013}
{Scarano} S.,  {L{\'e}pine} J.~R.~D.,  2013, \mn@doi [\mnras]
  {10.1093/mnras/sts048}, \href
  {https://ui.adsabs.harvard.edu/abs/2013MNRAS.428..625S} {428, 625}

\bibitem[\protect\citeauthoryear{{Sch{\"o}nrich} \& {Binney}}{{Sch{\"o}nrich}
  \& {Binney}}{2009}]{Schoenrich2009b}
{Sch{\"o}nrich} R.,  {Binney} J.,  2009, \mn@doi [\mnras]
  {10.1111/j.1365-2966.2009.14750.x}, \href
  {https://ui.adsabs.harvard.edu/abs/2009MNRAS.396..203S} {396, 203}

\bibitem[\protect\citeauthoryear{{Searle}}{{Searle}}{1971}]{Searle1971}
{Searle} L.,  1971, \mn@doi [\apj] {10.1086/151090}, \href
  {https://ui.adsabs.harvard.edu/abs/1971ApJ...168..327S} {168, 327}

\bibitem[\protect\citeauthoryear{{Seitenzahl} et~al.,}{{Seitenzahl}
  et~al.}{2013}]{Seitenzahl2013}
{Seitenzahl} I.~R.,  et~al., 2013, \mn@doi [\mnras] {10.1093/mnras/sts402},
  \href {http://adsabs.harvard.edu/abs/2013MNRAS.429.1156S} {429, 1156}

\bibitem[\protect\citeauthoryear{{Sharda}, {Krumholz}, {Wisnioski}, {Forbes},
  {Federrath}  \& {Acharyya}}{{Sharda} et~al.}{2021}]{Sharda2021}
{Sharda} P.,  {Krumholz} M.~R.,  {Wisnioski} E.,  {Forbes} J.~C.,  {Federrath}
  C.,   {Acharyya} A.,  2021, \mn@doi [\mnras] {10.1093/mnras/stab252}, \href
  {https://ui.adsabs.harvard.edu/abs/2021MNRAS.502.5935S} {502, 5935}

\bibitem[\protect\citeauthoryear{{Shaver}, {McGee}, {Newton}, {Danks}  \&
  {Pottasch}}{{Shaver} et~al.}{1983}]{Shaver1983}
{Shaver} P.~A.,  {McGee} R.~X.,  {Newton} L.~M.,  {Danks} A.~C.,   {Pottasch}
  S.~R.,  1983, \mn@doi [\mnras] {10.1093/mnras/204.1.53}, \href
  {https://ui.adsabs.harvard.edu/abs/1983MNRAS.204...53S} {204, 53}

\bibitem[\protect\citeauthoryear{{Spina} et~al.,}{{Spina}
  et~al.}{2021}]{Spina2021}
{Spina} L.,  et~al., 2021, \mn@doi [\mnras] {10.1093/mnras/stab471}, \href
  {https://ui.adsabs.harvard.edu/abs/2021MNRAS.503.3279S} {503, 3279}

\bibitem[\protect\citeauthoryear{{Spitoni} et~al.,}{{Spitoni}
  et~al.}{2023}]{Spitoni2023}
{Spitoni} E.,  et~al., 2023, \mn@doi [\aap] {10.1051/0004-6361/202347325},
  \href {https://ui.adsabs.harvard.edu/abs/2023A&A...680A..85S} {680, A85}

\bibitem[\protect\citeauthoryear{{Stanghellini}, {Magrini}  \&
  {Casasola}}{{Stanghellini} et~al.}{2015}]{Stanghellini2015}
{Stanghellini} L.,  {Magrini} L.,   {Casasola} V.,  2015, \mn@doi [\apj]
  {10.1088/0004-637X/812/1/39}, \href
  {https://ui.adsabs.harvard.edu/abs/2015ApJ...812...39S} {812, 39}

\bibitem[\protect\citeauthoryear{{Stinson}, {Seth}, {Katz}, {Wadsley},
  {Governato}  \& {Quinn}}{{Stinson} et~al.}{2006}]{Stinson2006}
{Stinson} G.,  {Seth} A.,  {Katz} N.,  {Wadsley} J.,  {Governato} F.,   {Quinn}
  T.,  2006, \mn@doi [\mnras] {10.1111/j.1365-2966.2006.11097.x}, \href
  {https://ui.adsabs.harvard.edu/abs/2006MNRAS.373.1074S} {373, 1074}

\bibitem[\protect\citeauthoryear{{Stinson}, {Brook}, {Macci{\`o}}, {Wadsley},
  {Quinn}  \& {Couchman}}{{Stinson} et~al.}{2013}]{Stinson2013}
{Stinson} G.~S.,  {Brook} C.,  {Macci{\`o}} A.~V.,  {Wadsley} J.,  {Quinn}
  T.~R.,   {Couchman} H.~M.~P.,  2013, \mn@doi [\mnras] {10.1093/mnras/sts028},
  \href {https://ui.adsabs.harvard.edu/abs/2013MNRAS.428..129S} {428, 129}

\bibitem[\protect\citeauthoryear{{Sun} et~al.,}{{Sun} et~al.}{2024}]{Sun2024}
{Sun} X.,  et~al., 2024, \mn@doi [arXiv e-prints] {10.48550/arXiv.2409.09290},
  \href {https://ui.adsabs.harvard.edu/abs/2024arXiv240909290S} {p.
  arXiv:2409.09290}

\bibitem[\protect\citeauthoryear{{Taylor}}{{Taylor}}{2005}]{Taylor2005}
{Taylor} M.~B.,  2005, ASPC, \href
  {http://adsabs.harvard.edu/abs/2005ASPC..347...29T} {347, 29}

\bibitem[\protect\citeauthoryear{{Tepper-Garc{\'\i}a}, {Bland-Hawthorn},
  {Vasiliev}, {Agertz}, {Teyssier}  \& {Federrath}}{{Tepper-Garc{\'\i}a}
  et~al.}{2024}]{TepperGarcia2024}
{Tepper-Garc{\'\i}a} T.,  {Bland-Hawthorn} J.,  {Vasiliev} E.,  {Agertz} O.,
  {Teyssier} R.,   {Federrath} C.,  2024, \mn@doi [\mnras]
  {10.1093/mnras/stae2372}, \href
  {https://ui.adsabs.harvard.edu/abs/2024MNRAS.535..187T} {535, 187}

\bibitem[\protect\citeauthoryear{{Tinsley}}{{Tinsley}}{1980}]{Tinsley1980}
{Tinsley} B.~M.,  1980, \fcp, \href
  {https://ui.adsabs.harvard.edu/abs/1980FCPh....5..287T} {5, 287}

\bibitem[\protect\citeauthoryear{{Tissera}, {Rosas-Guevara}, {Bower}, {Crain},
  {del P Lagos}, {Schaller}, {Schaye}  \& {Theuns}}{{Tissera}
  et~al.}{2019}]{Tissera2019}
{Tissera} P.~B.,  {Rosas-Guevara} Y.,  {Bower} R.~G.,  {Crain} R.~A.,  {del P
  Lagos} C.,  {Schaller} M.,  {Schaye} J.,   {Theuns} T.,  2019, \mn@doi
  [\mnras] {10.1093/mnras/sty2817}, \href
  {https://ui.adsabs.harvard.edu/abs/2019MNRAS.482.2208T} {482, 2208}

\bibitem[\protect\citeauthoryear{{Tuntipong} et~al.,}{{Tuntipong}
  et~al.}{2024}]{Tuntipong2024}
{Tuntipong} S.,  et~al., 2024, \mn@doi [\mnras] {10.1093/mnras/stae2042}, \href
  {https://ui.adsabs.harvard.edu/abs/2024MNRAS.533.4334T} {533, 4334}

\bibitem[\protect\citeauthoryear{{Twarog}}{{Twarog}}{1980}]{Twarog1980}
{Twarog} B.~A.,  1980, \mn@doi [\apj] {10.1086/158460}, \href
  {http://adsabs.harvard.edu/abs/1980ApJ...242..242T} {242, 242}

\bibitem[\protect\citeauthoryear{{Twarog}, {Ashman}  \&
  {Anthony-Twarog}}{{Twarog} et~al.}{1997}]{Twarog1997}
{Twarog} B.~A.,  {Ashman} K.~M.,   {Anthony-Twarog} B.~J.,  1997, \mn@doi [\aj]
  {10.1086/118667}, \href
  {https://ui.adsabs.harvard.edu/abs/1997AJ....114.2556T} {114, 2556}

\bibitem[\protect\citeauthoryear{{Vickers}, {Shen}  \& {Li}}{{Vickers}
  et~al.}{2021}]{Vickers2021}
{Vickers} J.~J.,  {Shen} J.,   {Li} Z.-Y.,  2021, \mn@doi [\apj]
  {10.3847/1538-4357/ac27a9}, \href
  {https://ui.adsabs.harvard.edu/abs/2021ApJ...922..189V} {922, 189}

\bibitem[\protect\citeauthoryear{{Vilchez} \& {Esteban}}{{Vilchez} \&
  {Esteban}}{1996}]{Vilchez1996}
{Vilchez} J.~M.,  {Esteban} C.,  1996, \mn@doi [\mnras]
  {10.1093/mnras/280.3.720}, \href
  {https://ui.adsabs.harvard.edu/abs/1996MNRAS.280..720V} {280, 720}

\bibitem[\protect\citeauthoryear{{Vincenzo}, {Matteucci}, {Belfiore}  \&
  {Maiolino}}{{Vincenzo} et~al.}{2016}]{Vincenzo2016a}
{Vincenzo} F.,  {Matteucci} F.,  {Belfiore} F.,   {Maiolino} R.,  2016, \mn@doi
  [\mnras] {10.1093/mnras/stv2598}, \href
  {https://ui.adsabs.harvard.edu/abs/2016MNRAS.455.4183V} {455, 4183}

\bibitem[\protect\citeauthoryear{Virtanen et~al.,}{Virtanen
  et~al.}{2020}]{Scipy}
Virtanen P.,  et~al., 2020, \mn@doi [Nature Methods]
  {10.1038/s41592-019-0686-2}, \href {https://rdcu.be/b08Wh} {17, 261}

\bibitem[\protect\citeauthoryear{{Wadsley}, {Keller}  \& {Quinn}}{{Wadsley}
  et~al.}{2017}]{Wadsley2017}
{Wadsley} J.~W.,  {Keller} B.~W.,   {Quinn} T.~R.,  2017, \mn@doi [\mnras]
  {10.1093/mnras/stx1643}, \href
  {https://ui.adsabs.harvard.edu/abs/2017MNRAS.471.2357W} {471, 2357}

\bibitem[\protect\citeauthoryear{Walt, Colbert  \& Varoquaux}{Walt
  et~al.}{2011}]{numpy}
Walt S. v.~d.,  Colbert S.~C.,   Varoquaux G.,  2011, \mn@doi [Comput Sci Eng]
  {10.1109/MCSE.2011.37}, 13, 22

\bibitem[\protect\citeauthoryear{{Wang}, {Dutton}, {Stinson}, {Macci{\`o}},
  {Penzo}, {Kang}, {Keller}  \& {Wadsley}}{{Wang} et~al.}{2015}]{Wang2015}
{Wang} L.,  {Dutton} A.~A.,  {Stinson} G.~S.,  {Macci{\`o}} A.~V.,  {Penzo} C.,
   {Kang} X.,  {Keller} B.~W.,   {Wadsley} J.,  2015, \mn@doi [\mnras]
  {10.1093/mnras/stv1937}, \href
  {http://adsabs.harvard.edu/abs/2015MNRAS.454...83W} {454, 83}

\bibitem[\protect\citeauthoryear{{Willett} et~al.,}{{Willett}
  et~al.}{2023}]{Willett2023}
{Willett} E.,  et~al., 2023, \mn@doi [\mnras] {10.1093/mnras/stad2374}, \href
  {https://ui.adsabs.harvard.edu/abs/2023MNRAS.526.2141W} {526, 2141}

\bibitem[\protect\citeauthoryear{{Williams}, {Bureau}  \&
  {Cappellari}}{{Williams} et~al.}{2009}]{Williams2009}
{Williams} M.~J.,  {Bureau} M.,   {Cappellari} M.,  2009, \mn@doi [\mnras]
  {10.1111/j.1365-2966.2009.15582.x}, \href
  {https://ui.adsabs.harvard.edu/abs/2009MNRAS.400.1665W} {400, 1665}

\bibitem[\protect\citeauthoryear{{Wyse} \& {Silk}}{{Wyse} \&
  {Silk}}{1989}]{Wyse1989}
{Wyse} R. F.~G.,  {Silk} J.,  1989, \mn@doi [\apj] {10.1086/167329}, \href
  {https://ui.adsabs.harvard.edu/abs/1989ApJ...339..700W} {339, 700}

\bibitem[\protect\citeauthoryear{{Yong}, {Carney}  \& {Friel}}{{Yong}
  et~al.}{2012}]{Yong2012}
{Yong} D.,  {Carney} B.~W.,   {Friel} E.~D.,  2012, \mn@doi [\aj]
  {10.1088/0004-6256/144/4/95}, \href
  {https://ui.adsabs.harvard.edu/abs/2012AJ....144...95Y} {144, 95}

\bibitem[\protect\citeauthoryear{{Zari}, {Hashemi}, {Brown}, {Jardine}  \& {de
  Zeeuw}}{{Zari} et~al.}{2018}]{Zari2018}
{Zari} E.,  {Hashemi} H.,  {Brown} A.~G.~A.,  {Jardine} K.,   {de Zeeuw} P.~T.,
   2018, \mn@doi [\aap] {10.1051/0004-6361/201834150}, \href
  {https://ui.adsabs.harvard.edu/abs/2018A&A...620A.172Z} {620, A172}

\bibitem[\protect\citeauthoryear{{Zari}, {Rix}, {Frankel}, {Xiang}, {Poggio},
  {Drimmel}  \& {Tkachenko}}{{Zari} et~al.}{2021}]{Zari2021}
{Zari} E.,  {Rix} H.~W.,  {Frankel} N.,  {Xiang} M.,  {Poggio} E.,  {Drimmel}
  R.,   {Tkachenko} A.,  2021, \mn@doi [\aap] {10.1051/0004-6361/202039726},
  \href {https://ui.adsabs.harvard.edu/abs/2021A&A...650A.112Z} {650, A112}

\bibitem[\protect\citeauthoryear{{Zaritsky}, {Kennicutt}  \&
  {Huchra}}{{Zaritsky} et~al.}{1994}]{Zaritsky1994}
{Zaritsky} D.,  {Kennicutt} Robert~C. J.,   {Huchra} J.~P.,  1994, \mn@doi
  [\apj] {10.1086/173544}, \href
  {https://ui.adsabs.harvard.edu/abs/1994ApJ...420...87Z} {420, 87}

\bibitem[\protect\citeauthoryear{{Zhang}, {Li}, {Hu}  \& {Krumholz}}{{Zhang}
  et~al.}{2025}]{Zhang2025}
{Zhang} C.,  {Li} Z.,  {Hu} Z.,   {Krumholz} M.~R.,  2025, \mn@doi [\mnras]
  {10.1093/mnras/staf194}, \href
  {https://ui.adsabs.harvard.edu/abs/2025MNRAS.tmp..198Z} {}

\bibitem[\protect\citeauthoryear{{van de Sande}, {Fraser-McKelvie}, {Fisher},
  {Martig}, {Hayden}  \& {the GECKOS Survey collaboration}}{{van de Sande}
  et~al.}{2023}]{GECKOS2023}
{van de Sande} J.,  {Fraser-McKelvie} A.,  {Fisher} D.~B.,  {Martig} M.,
  {Hayden} M.~R.,   {the GECKOS Survey collaboration} 2023, \mn@doi [arXiv
  e-prints] {10.48550/arXiv.2306.00059}, \href
  {https://ui.adsabs.harvard.edu/abs/2023arXiv230600059V} {p. arXiv:2306.00059}

\makeatother
\end{thebibliography}

$\,$
\clearpage$\,$

\appendix

\section{Gas phase oxygen and iron abundances} \label{sec:app_a}

\begin{figure}[!ht]
    \centering
    \includegraphics[width=0.9\columnwidth]{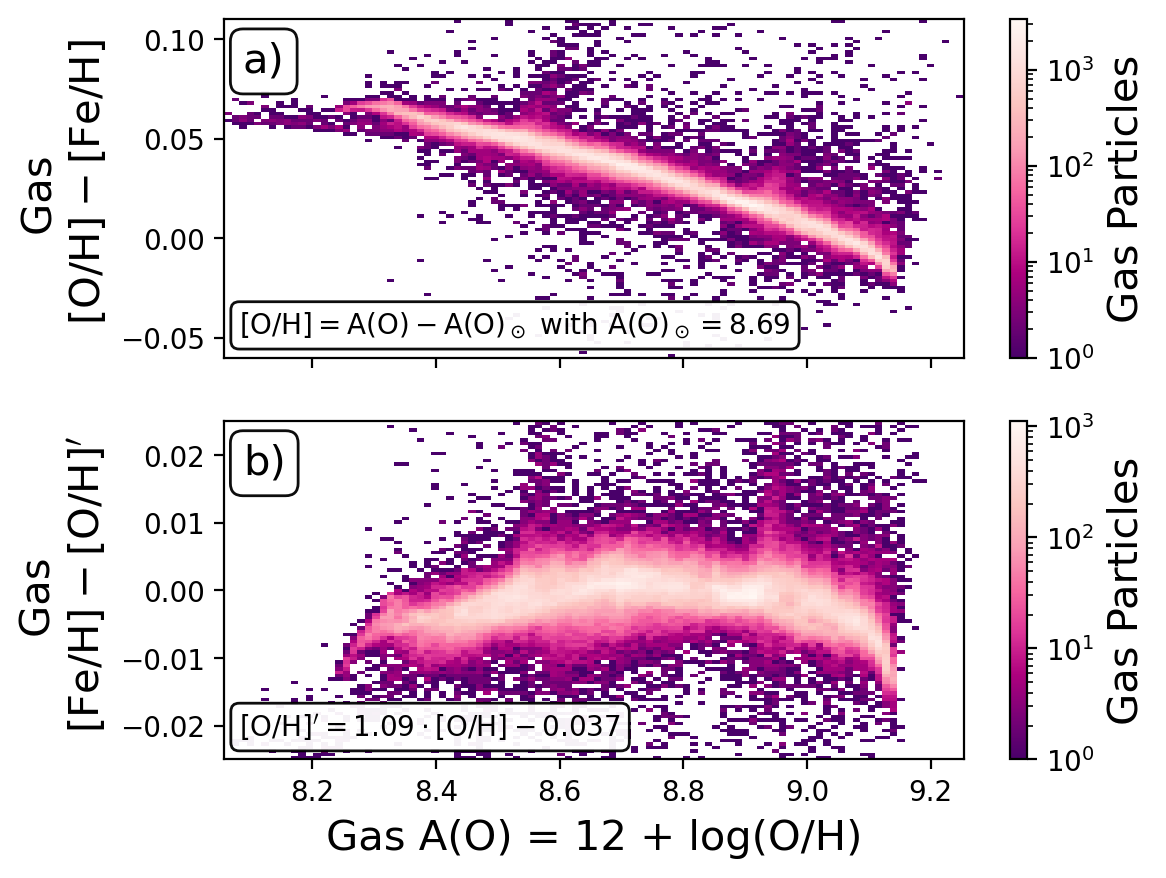}
    \caption{Comparison of gas abundances for oxygen and iron. Shown are absolute oxygen abundances A(O) and the comparison of relative iron and oxygen abundances [Fe/H] - [O/H]. The top panel shows values at face values, whereas the bottom panel shows the comparison for a linear approximation of [Fe/H] from [O/H].}    \label{fig:fe_h_vs_a_o_gas}
\end{figure}

Fig.~\ref{fig:fe_h_vs_a_o_gas} demonstrates the tight correlation of gas phase iron and oxygen abundance and how to approximate them linearly. \adjusted{While observations typically show higher [O/H] at lower [Fe/H] \citep[e.g.][]{Buder2024b}, we attribute the rather flat relation to the dominating contribution from core-collapse supernovae of massive stars to both O and Fe, a similar level of contribution of asymptotic giant branch stars to O and Fe as well as a lower amount of delayed contributions from supernovae type Ia across time in the simulation \citep[see also][]{Buck2021}.}

\section{Spatial distribution and oxygen abundances for different gas temperatures} \label{sec:app_b}

\begin{figure*}
    \centering
    \includegraphics[width=0.98\textwidth]{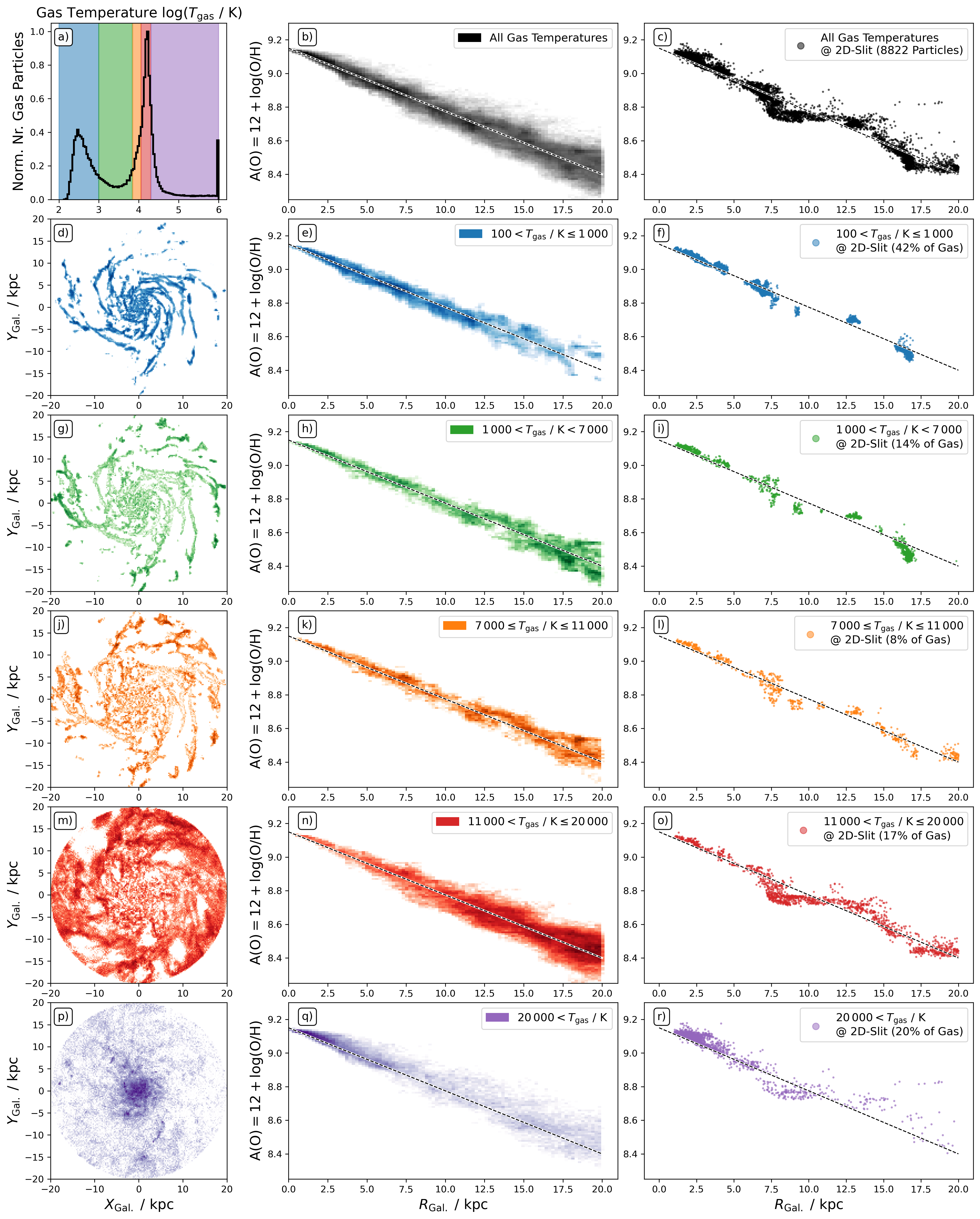}
    \caption{\textbf{Spatial distribution and oxygen abundances of gas with different gas temperatures.} Panel a) shows the gas temperature distribution with different groups (cold and warm neutral medium as well as ionized gas with different temperatures) highlighted. The histogram is saturated at $10^5\,\mathrm{K}$ to better visualize the temperature distribution. The left columns show the spatial distribution of these different groups. The other columns show the radial distribution of oxygen abundances for the full gas disk (middle columns) as well as the same 2-dimensional slit used for Fig.~\ref{fig:region_r_ao_gas_density} (right columns). We show the same dashed line in the $R_\mathrm{Gal.}$-A(O) panels to guide the eye.}
    \label{fig:gas_temperature_tracing_r_ao}
\end{figure*}

\adjusted{To assess the robustness of our analysis of the gas phase oxygen abundances, for example if we would mainly trace abundances through ionised gas like that in HII regions, we examine the temperature distribution of gas within $R_\mathrm{Gal.} < 20\,\mathrm{kpc}$ in Fig.~\ref{fig:gas_temperature_tracing_r_ao}. In Fig.~\ref{fig:gas_temperature_tracing_r_ao}a, see that the gas temperatures of our simulation includes a significant amount of cool and warm neutral gas at the lower temperature end as well as a significant amount of ionised material at the higher end. The distribution peaks around $10^{4.2}\,\mathrm{K} \approx 16\,000\,\mathrm{K}$ -- within the region of $10\,000_{-5\,000}^{+10\,000}\,\mathrm{K}$ tabulated by \citet{Osterbrock1989} for HII regions. We thus divide the gas in different temperature bins of $\log_{10} (T_\mathrm{Gas}~/~\mathrm{K})$ within 2-3 (cold gas, shown in blue), 3-3.85 (warm gas in green), 3.85-4.05 (warm ionised gas in orange), 4.05-4.3 (hot ionised gas in red), and above 4.3 (hottest ionised gas in purple).}

\adjusted{We show the spatial distribution of the gas with these different temperatures in the left columns of Fig.~\ref{fig:gas_temperature_tracing_r_ao}. As expected, we see the cold and warm neutral medium quite tightly constrained on narrow spiral arms. While the warm ionised medium is still following the same pattern (Fig.~\ref{fig:gas_temperature_tracing_r_ao}j), we note that it is more spatially extended around the arms. The hot ionised medium (Fig.~\ref{fig:gas_temperature_tracing_r_ao}m) is well distributed across the plane. The hottest gas (Fig.~\ref{fig:gas_temperature_tracing_r_ao}p) is then again constrained to entirely different regions, including small and localised patches as well as the innermost radii of the galaxy. Most important for HII observations would be Figs.~\ref{fig:gas_temperature_tracing_r_ao}j and \ref{fig:gas_temperature_tracing_r_ao}m, which show quite a significant change in how much gas is distributed away from the spiral arms themselves. This would be especially important, if we see a significantly different behaviour of the gas phase metallicity of oxygen abundance between these two phases.}

\adjusted{We thus use the same temperature bins and plot the radial oxygen abundance gradient in the middle column of Fig.~\ref{fig:gas_temperature_tracing_r_ao}. Here, we firstly note that the warm neutral, and warm ionised medium follow a similar pattern and are also closely aligned with the abundances cold neutral medium (which seems slightly more constrained to the inner part of the galaxy). The most drastic change is again seen when going above temperatures of $10\,000\,\mathrm{K}$ between Figs.~\ref{fig:gas_temperature_tracing_r_ao}k and \ref{fig:gas_temperature_tracing_r_ao}n. Here, the logarithmic color map suggests a stronger increase of scatter. The median abundances agree, however, well with the trend of the cooler gas. This is not the case for the hottest ionised medium (Fig.~\ref{fig:gas_temperature_tracing_r_ao}q), which shows increased abundances with respect to the cooler gas. To confirm this qualitative impression, we have calculated the distribution (16th, 50th, and 84th percentiles) of gas within different temperature and radial bins.}

\adjusted{The oxygen abundance distribution remains consistent between $T_\mathrm{Gas} = 10^2-10^{4.2}\,\mathrm{K}$, with only a mild increase in spread from 0.03 to 0.05\,dex. Above $T_\mathrm{Gas} = 20\,000\,\mathrm{K}$, the scatter increases more significantly (to 0.08-0.11\,dex), but the median trends remain comparable to those at lower temperatures. Therefore, assuming HII regions primarily trace gas with $T_\mathrm{Gas} < 20\,000\,\mathrm{K}$ we confirm that the observed oxygen abundance patterns and their scatter are consistent across these phases. The use of oxygen abundances for all gas particles as a proxy for those of HII regions thus appears justified in this regime, but future studies should take a closer look at the more nuanced difference in spatial distribution and abundances for the different gas temperatures.}

%%%%%%%%%%%%%%%%%%%%%%%%%%%%%%%%%%%%%%%%%%%%%%%%%%
\end{document}